\documentclass{nature}
\usepackage{lipsum}
\usepackage{amsmath}
\usepackage{amssymb}
\usepackage[normalem]{ulem}
\usepackage{lineno}
\usepackage{longtable}
\usepackage{booktabs}
\usepackage{svg}
\usepackage{adjustbox}
\usepackage{ragged2e}
\usepackage{booktabs}
\usepackage{multirow}
\usepackage{tabularx}
\usepackage{pdfpages}

\PassOptionsToPackage{hyphens}{url}\usepackage{hyperref}
\usepackage[vmargin=0.5in]{geometry}

\newcommand\ExtraSep

 \renewcommand{\baselinestretch}{1}
\usepackage{caption}
\usepackage{graphicx}
\makeatletter
\let\saved@includegraphics\includegraphics
\AtBeginDocument{\let\includegraphics\saved@includegraphics}
\makeatother

\title{Unraveling cradle-to-grave disease trajectories from multilayer comorbidity networks}

\author{Elma Dervić$^{1,2}$, Johannes Sorger$^{1}$,  Liuhuaying Yang$^{1}$, Michael Leutner$^{3}$, Alexander Kautzky$^{4}$, Stefan Thurner $^{1,2,5}$, Alexandra Kautzky-Willer$^{3,6}$, Peter Klimek$^{1,2,*}$}

\date{\today} 

\begin{document}

\maketitle

\begin{affiliations} 
\item Complexity Science Hub Vienna, Josefst\"adter Stra\ss e 39, 1080 Vienna, Austria;
\item Medical University of Vienna, Section for Science of Complex Systems, CeMSIIS, Spitalgasse 23, 1090 Vienna, Austria;
\item Medical University of Vienna, Department of Internal Medicine III, Clinical Division of Endocrinology and Metabolism,  W\"ahringer G\"urtel 18–20, A-1090 Vienna, Austria;
\item Medical University of Vienna, Department of Psychiatry and Psychotherapy, W\"ahringer G\"urtel 18-20, A-1090 Vienna, Austria;
\item Santa Fe Institute, 1399 Hyde Park Road, Santa Fe, NM 87501, USA.
\item Gender Institute, A-3571 Gars am Kamp, Austria

\end{affiliations}

$^*$Correspondence: peter.klimek@meduniwien.ac.at\\

\begin{abstract}
We aim to comprehensively identify typical life-spanning trajectories and critical events that impact patients' hospital utilization and mortality.
We use a unique dataset containing 44 million records of almost all inpatient stays from 2003 to 2014 in Austria to investigate disease trajectories.
We develop a new, multilayer disease network approach to quantitatively analyse how cooccurrences of two or more diagnoses form and evolve over the life course of patients. Nodes represent diagnoses in  age groups of ten years; each age group makes up a layer of the comorbidity multilayer network. Inter-layer links encode a significant correlation between diagnoses (p $<$ 0.001, relative risk $>$ 1.5), while intra-layers links encode correlations between diagnoses across different age groups. We use an unsupervised clustering algorithm for detecting typical disease trajectories as overlapping clusters in the multilayer comorbidity network. 

We identify critical events in a patient's career as points where initially overlapping trajectories start to diverge towards different states. We identified 1,260 distinct disease trajectories (618 for females, 642 for males) that on average contain 9 (IQR 2-6) different diagnoses that cover over up to 70 years (mean 23 years). We found 70 pairs of diverging trajectories that share some diagnoses at younger ages but develop into markedly different groups of diagnoses at older ages. 

The disease trajectory framework can help us to identify critical events as specific combinations of risk factors that put patients at high risk for different diagnoses decades later. Our findings enable a data-driven integration of personalized life-course perspectives into clinical decision-making. \\
\end{abstract}

\maketitle

\section*{Introduction}


Multimorbidity, the occurrence of two or more diseases in one patient, is a frequent phenomenon ~~\cite{han2021disease,cezard2021studying}. Today's reality of a 100-year lifespan brings a shifting multimorbidity burden and increased healthcare and long-term care costs ~\cite{whoage, euage}. 
It was estimated that mnore than 50 million people in Europe show more than one chronic condition ~\cite{struckmann2014caring}. In ~\cite{hajat2018global}, authors estimated that 16–57\% of adults in developed countries are diagnosed with more than one chronic disease and predicted a dramatic rise of multimorbidity rates in the next years. The WHO World Report on Ageing and Health emphasizes the importance of research to better understand the dynamics and consequences of ageing ~\cite{world2015world}. Studies on multimorbidity patterns may contribute to successful ageing by the prevention of disease progression by identifying critical events which lead to a rapid deterioration of health ~\cite{rowe1997successful, kudesia2021incidence}.\\

As diseases tend to co-occur and interact with each other (in a way that can worsen the course of both), they cannot be studied separately from each other ~\cite{di2015association}. The analysis of multimorbidity has recently been catalyzed by the massive collection of patient health information on diagnoses, medication, results of laboratory tests in electronic health records (EHR), and other clinical registries. Comorbidity networks have been established as tools to analyse multimorbidity in such datasets ~\cite{strauss2014distinct, fotouhi2018statistical}. Age and sex-specific analyses can further be conducted to address age- and sex-dependent associations between diagnoses ~\cite{jeong2017network, chmiel2014spreading}. These works confirm that patients mainly develop diseases in close network proximity to disorders they already suffer.\\

The concept of disease trajectories has been proposed to formally describe the progression of multimorbidity over time. Disease trajectories are frequently occurring patterns or sequences of 
diagnoses 
at particular times 
and are typically extracted from the medical history of millions of patients. Thus, apart from the pairwise disease associations, uncovering complex disease patterns and assessing their temporal directionality is crucial for estimating disease progression, developing prediction models ~\cite{violan2020five, prados2018cohort}, analysing trajectories ~\cite{siggaard2020disease, jensen2014temporal} and their temporal patterns using clustering algorithms ~\cite{haug2020high, giannoula2018identifying}. Many studies used data on in-hospital stays to construct such trajectories. A summary of applications of machine learning tools to understand the structure and formation of multimorbidity in the population was given in ~\cite{hassaine2020untangling}. However, studies of multimorbidity patterns over the full life span of patients, from cradle to grave, remain scarce, as studies frequently take cross-sectional approaches 
~\cite{cezard2021studying, hsu2015trajectories}.\\

Longitudinal analysis of multimorbidity requires large population-wide disease registries which span over multiple years, if not decades. Such analyses are challenging as they require custom-made methods and that are often computationally challenging ~\cite{siggaard2020disease}. Taken together, a life span perspective on multimorbidities addressing the need for more comprehensive knowledge on disease trajectories and their critical events is largely missing to date ~\cite{vos2015trajectories}.\\

Here, we propose a novel approach to 
dynamical comorbidity networks from longitudinal population-wide healthcare data to comprehensively identify disease trajectories in an entire population. A multilayer comorbidity network is constructed where nodes correspond to diagnoses, layers to age groups, intralayer links to disease co-occurrence relations, and interlayer links encode the directionality of disease pairs (which diagnosis tends to occur first). We identify temporal disease trajectories as communities in this multilayer network.  In some cases, these tightly connected communities share some nodes and be referred to as overlapping communities. \\

The central assumption of our approach is that communities of nodes in the comorbidity network represent patients' diseases trajectories. We identify overlapping communities rather than exclusive clusters as the same diseases (nodes) can naturally be part of different disease trajectories, i.e. sleep disorders in patients with and without obesity and diabetes mellitus type 2. We further try to identify critical events as points along trajectories, where two initially identical trajectories start to diverge 
and will lead to different outcomes in terms of disease burden (hospital utilization) and mortality.  \\

\begin{figure*}[!tb]
    \centering
    \includegraphics[width=12cm]{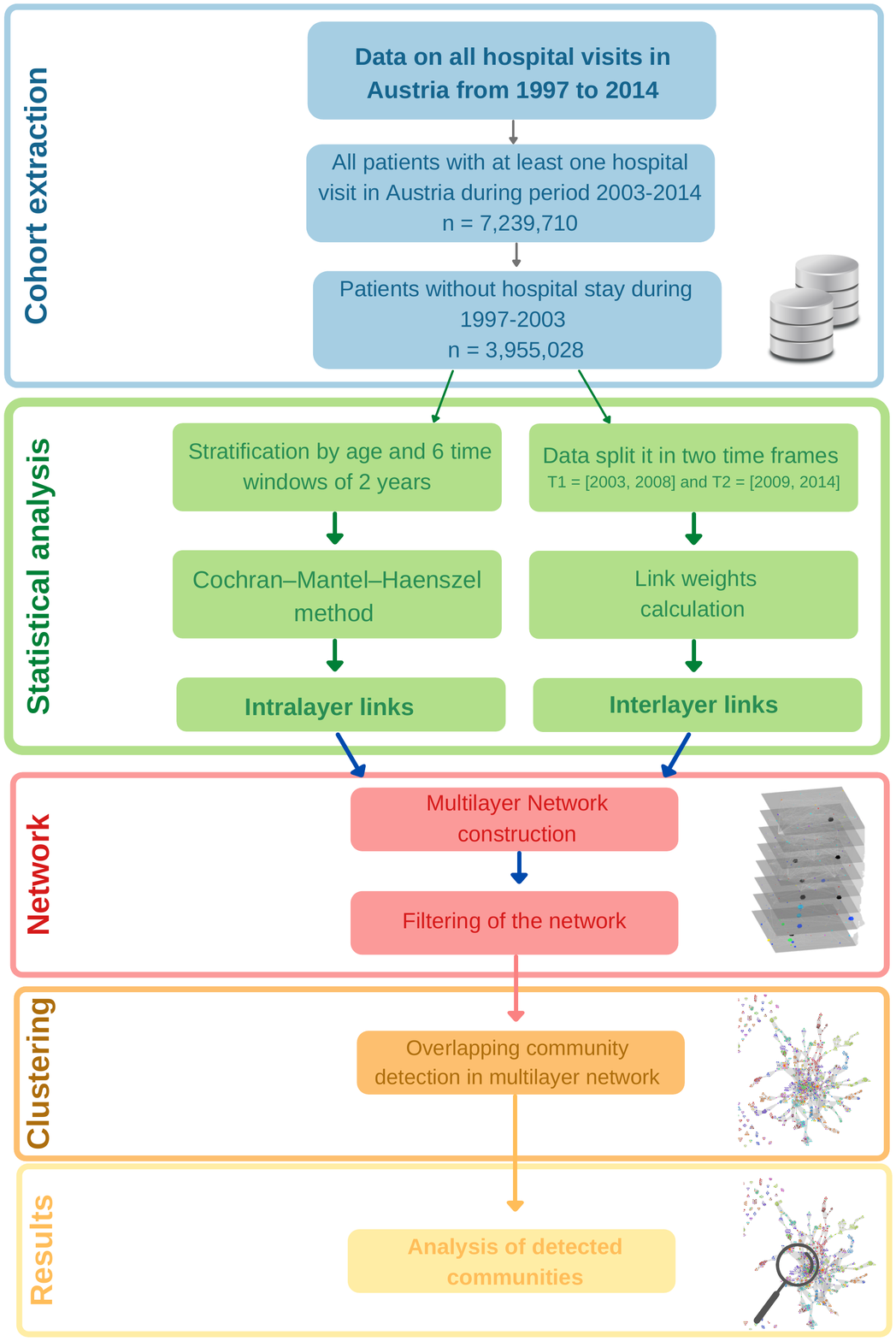}
    \caption{Workflow of the research presented in this article.} 
    \label{fig:Workflow}
\end{figure*}

Figure \ref{fig:Workflow} illustrates the suggested methodology of this large-scale disease trajectory study. We analysed data from an electronic health registry covering almost all 
of 8.9 million  
Austrians with more than 44 million in-hospital stays over 17 years, from 1997-2014.  To ensure the comparability of the health status of our study population, we restricted the analysis to patients who were "healthy" at the beginning of the observed period between 2003 and 2014. Therefore, in the first step of the analysis, we identified as the study population all patients with at least one hospital stay between 1997 and 2002 with a diagnosis from the range A00-N99 (in total 1,081 diagnoses). Moreover, in the early 2000s, Austria transitioned from the previous ICD coding system to ICD-10 2001. It was crucial to avoid combining various classification systems as it would have compromised the reliability of the analysis, Figure \ref{fig:Workflow} (blue box).  \\

In a next step we then constructed a multilayer comorbidity network to explore how different disease conditions co-occur and develop over time. We separated our data into 10-year age groups. For every age group we introduced a layer in the  multilayer comorbidity  network. In this network, two types of link can be found, links that connect nodes in the same layer (intralayer links) and links that connect nodes from different layers (interlayer links).  All identified significant correlations of diagnoses in the same age group are defined as intralayer links, while interlayer links represent the correlation between diagnoses in different age groups, Figure \ref{fig:Workflow} (green box). Nodes without any intralayer links were removed, Figure \ref{fig:Workflow} (red box).\\

We used an algorithm based on the local optimization of a fitness function presented in ~\cite{lancichinetti2009detecting} to identify overlapping communities in the multilayer network, Figure \ref{fig:Workflow} (orange box). Note that the detected communities typically encompass more than one age layer.  We analysed the age structure of the detected overlapping communities and the number of chapters of diagnoses inside the communities. More concrete, we conceptualize disease trajectories as groups of diagnoses that occur at different age groups (layers in the network) and that are more closely connected to other diagnoses in the same community compared to diagnoses outside of the community.\\

As disease trajectories can overlap, this enables us to comprehensively study relationships between disease trajectories across more than one age group. We defined pairs of trajectories as converging if they do not overlap (no shared diagnoses) in younger age groups while they have a nonzero overlap in older age groups. Additionally, diverging pairs of trajectories overlap at the beginning, in younger age groups, but have different pathways in older age groups. 	\\

From this we can identify critical events in patient careers. Critical events are defined as combinations of diagnoses in a specific age group, mainly chronic conditions, that signal that the disease trajectories are about to diverge towards paths that lead to different levels of mortality or lengths of hospital stays in the following age groups. Critical events can be thought of as bifurcation points of disease trajectories that can lead to trajectories associated with strongly varying outcomes. These events can support the identification of patients at risk for more severe multimorbidity trajectories and associated adverse outcomes in the next decade and thereby provide leverage points for targeted preventive actions.

\section*{Data and Methods}
\subsubsection*{Data}
The analysed dataset spans 17 years of nationwide in-hospital data from all hospitals in Austria. Each hospital stay is recorded with primary and secondary diagnoses, age in the resolution of 5 years, sex, admission and release date, release type (i.e., release, transfer, death...). This dataset covers the period from 1997 until 2014 and the vast majority of Austria's population with 8.9 unique patients. 
Diagnoses are coded with the three-digit International Classification of Diseases, 10th Revision (ICD-10) codes. We restricted our analysis to 1,081 codes from A00 to N99, excluding codes describing health encounters that can not be directly related to diseases  (i.e., O00-O9A - Pregnancy, childbirth, and the puerperium, S00-T88 - Injury, poisoning and certain other consequences of external causes...). The data always reports a primary diagnosis as the main reason for hospitalization, along with  a variable number of secondary diagnoses.

In this study, we assigned equal importance to both primary and secondary diagnoses ~\cite{haug2020high, deischinger2020diabetes}. 
To ensure that our study population's health state was comparable at the beginning of the observation period and not in the middle of connected hospitalization episodes, we introduced a wash-out period and limit the analysis to patients who had no hospital visits between 1997 and 2002. Consequently, excluding these patients also ensured that analysed data has only one ICD coding system, as in the early 2000s, Austria updated its ICD coding system to ICD-10 2001 ~\cite{haug2020high, deischinger2020diabetes, dervic2021effect}.

\subsubsection*{Multilayer Comorbidity Network}

Formally, we construct the multilayer comorbidity network given by the tensor $M_{i,j}^{\alpha, \beta}$ where $i$ and $j$ refer to nodes (diagnoses) on layers (age groups) $\alpha$ and $\beta$, respectively. We refer to entries in $M$ with $ \alpha=\beta$ as intralayer links and with  $\alpha \neq \beta$  as interlayer links. The analysis was performed separately for male and female patients.

\subsection{Intralayer links}

Intralayer links give the correlation between diagnoses within the same age group. 
The analysed dataset was stratified by six time windows of two years each, from 2003 to 2014. A contingency table is created for each pair of diagnoses in each stratum (for each sex and age group, the intralayer analysis includes six strata, each covering two calendar years). We used all contingency tables with more than four patients in each subgroup to compute relative risks (RR) and the p-value for rejecting the null hypothesis that the co-occurrence of two analysed diagnoses is statistically independent ~\cite{dervic2021effect}. A weighted average of the estimates of the risk ratios and odds ratios across the stratified data were calculated by using the Cochran–Mantel–Haenszel method ~\cite{kuritz1988general}.

Subsequently, all correlations with RR higher than 1.5 and p-value smaller than 0.05 were extracted and presented as intralayer links ~\cite{chmiel2014spreading}. These links are bidirectional, and we use a normalized RR as the link weight. The normalization of RR was done such that the sum of all total weights of all intralayer links with the same target was one.

\subsection{Interlayer links}
To estimate directionality or time order in pairs of diagnoses, we split the observation period  in two time frames $T1 = [2003, 2008]$ and $T2 = [2009, 2014]$. We investigate if a patient diagnosed with $i$ in $T1$ elevates the risk of being diagnosed with $j$ in $T2$ and compute the interlayer link weight as

\begin{equation}
M_{i,j}^{\alpha \neq \beta} = \frac{P(j_{T2}^{\beta}|i_{T1}^{\alpha})}{P(j_{T2}^{\beta})} \, .
\end{equation}

\subsection{Overlapping community detection in multilayer network}
We deleted all nodes without at least one inbound and one outbound link. Further, we normalized all link weights to range from 0 to 1 by dividing each link's weight by the sum of all links of the same type of a target node

\begin{equation}
M_{ij}^{\alpha \beta} = \frac{M_{ij}^{\alpha \beta}} {\sum_{j} {M_{ij}^{\alpha \beta}}} \, .
\end{equation}

The algorithm for detecting the overlapping and hierarchical community structure in complex networks proposed in ~\cite{lancichinetti2009detecting} was applied. This unsupervised clustering algorithm does not have a predefined number of communities. The detection procedure is initiated starting with a random node, which represents one community by itself.

 Community's fitness is $f_G = \frac{k_{in}^G}{(k_{in}^G + k_{out}^G)^a}$,\\

 where $k_{in}^G$ are the total internal degrees of the nodes in the community $G$ and $k_{out}^G$ are the total external degrees of the nodes in the community $G$. \\

 As long as the $f_G$ improves, neighboring nodes are added, or nodes already community members are removed.

The entailed resolution parameter $a$ enables us to uncover different hierarchical levels of a system, natural choice is $a=1$. Fitness is calculated at each step. Once the fitness cannot be increased anymore by a node removal or addition step, that community is "completed" and "closed." The community detection process ends when all nodes have been assigned to at least one community. To parallelize and optimize this computationally costly process, we identify the community of every node and delete duplicates among the discovered communities.

Identified communities usually consist of diseases in different age groups that tend to co-occur more frequently among themselves than diseases that are not part of the community. Hence, these communities represent typical disease trajectories; we denote a trajectory $X$ as a set of diagnosis-age tuples, 
$X = \{(i_1, \alpha_1), (i_2, \alpha_2), (i_3, \alpha_3)...\}$, where $i$ is an ICD10 code ranging from $[$A00, N99$]$ and $\alpha$ is the age group from $[1,8]$.

We measure the similarity of trajectories by the Jaccard coefficient between two trajectories consisting of tuples with diagnoses $i$ and age groups $\alpha$, $(i,\alpha)$. That is, two trajectories have a non-zero overlap if they share diagnoses within the same age groups. 

\subsection{Identifying converging and diverging trajectories}
\begin{figure*}[h] 
    \centering
    \includegraphics[width=10cm]{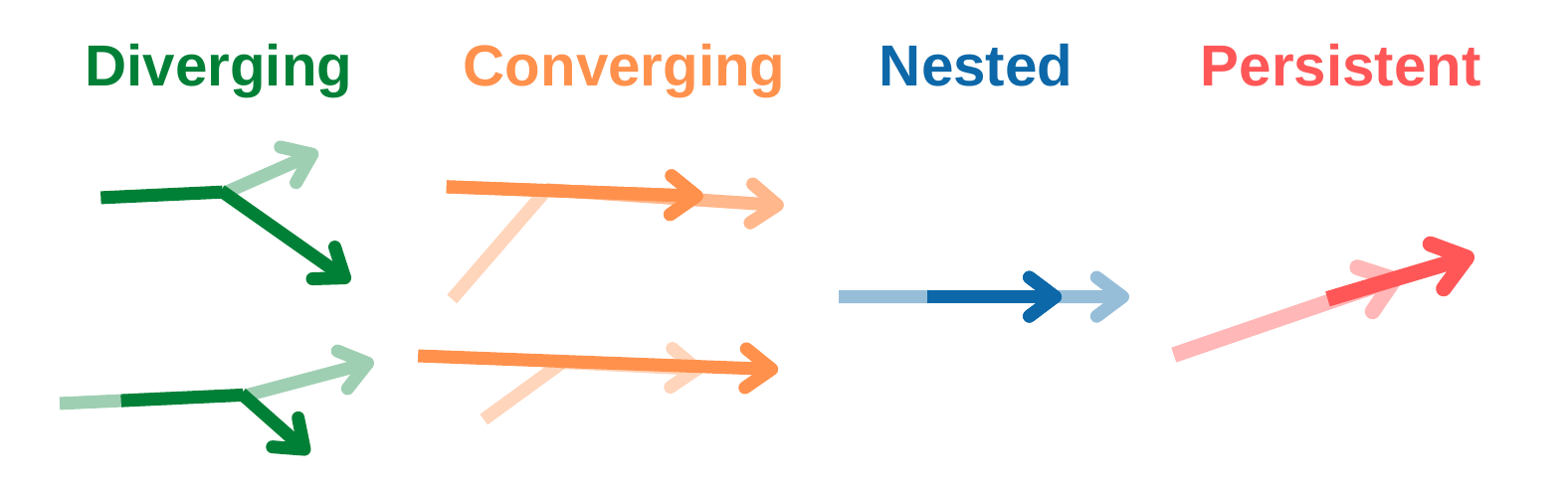}
    \caption{Visual representation of pairwise relations between pair of trajectories} 
    \label{fig:div_con_plot}
\end{figure*}

We performed a comprehensive classification with respect to all pairwise relations between every pair of trajectories. Provided that two trajectories share at least one diagnosis, they can be related in one of four different ways, namely (i)  diverging, (ii) converging, (iii) nested, or (iv) persistent, Figure \ref{fig:div_con_plot}.

Diverging trajectories have some overlapping elements at younger ages, but they develop into markedly different sets of diagnoses at older ages. 

More formally, trajectories $X = \{(i_{11}, \alpha_{11}), (i_{12}, \alpha_{12}), (i_{13}, \alpha_{13})...\}$ and \newline $Y = \{(i_{21}, \alpha_{21}), (i_{22}, \alpha_{22}), (i_{23}, \alpha_{23})...\}$  are diverging if it holds that \newline
\begin{equation}
\begin{aligned}
\Bigl\{ \{ (i_{1i}, \alpha_{1i}) \in X | \alpha_{1i} = \alpha^X_{\min}) \} \cap \{ (i_{2i}, \alpha_{2i}) \in Y | \alpha_{2i} =  \alpha^X_{\min}) \} \Bigl\} \: \cup  \\
\Bigl\{ \{ (i_{1i}, \alpha_{1i}) \in X | \alpha_{1i} = \alpha^Y_{\min}) \} \cap \{ (i_{2i}, \alpha_{2i}) \in Y  | \alpha_{2i} =  \alpha^Y_{\min}) \} \Bigl\} \: \neq  \emptyset \: {\rm and} \\
\{ (i_{1i}, \alpha_{1i}) \in X | \alpha_{1i} >  \alpha^X_{\min}) \}  \: \neq  \: \{ (i_{2i}, \alpha_{2i}) \in Y | \alpha_{2i} >  \alpha^X_{\min}) \} \: {\rm and} \\
\{ (i_{1i}, \alpha_{1i}) \in X | \alpha_{1i} >  \alpha^Y_{\min}) \} \: \neq  \: \{ (i_{2i}, \alpha_{2i}) \in Y | \alpha_{2i} >  \alpha^Y_{\min}) \},
\end{aligned}
\end{equation}

where $ \alpha^X_{\min} = \min_{(i, \alpha)\in X} \: \alpha \: \:, \:  \alpha^Y_{\min}  = \min_{(i, \alpha)\in Y} \:  \alpha$. \\

Converging trajectories overlap at older ages but are clearly different at younger ages. Trajectories $X$ and $Y$ are converging if it holds that
\begin{equation}
\begin{aligned}
\Bigl\{ \{ (i_{1i}, \alpha_{1i}) \in X | \alpha_{1i} = \alpha^X_{\max}) \} \cap \{ (i_{2i}, \alpha_{2i}) \in Y | \alpha_{2i} =  \alpha^X_{\max}) \} \Bigl\} \: \cup  \\
\Bigl\{ \{ (i_{1i}, \alpha_{1i}) \in X | \alpha_{1i} = \alpha^Y_{\max}) \} \cap \{ (i_{2i}, \alpha_{2i}) \in Y  | \alpha_{2i} =  \alpha^Y_{\max}) \} \Bigl\} \: \neq  \emptyset \: {\rm and} \\
\{ (i_{1i}, \alpha_{1i}) \in X | \alpha_{1i} <  \alpha^X_{\max}) \}  \: \neq  \: \{ (i_{2i}, \alpha_{2i}) \in Y | \alpha_{2i} <  \alpha^X_{\max}) \} \: {\rm and} \\
\{ (i_{1i}, \alpha_{1i}) \in X | \alpha_{1i} <  \alpha^Y_{\max}) \} \: \neq  \: \{ (i_{2i}, \alpha_{2i}) \in Y | \alpha_{2i} <  \alpha^Y_{\max}) \},
\end{aligned}
\end{equation}
where  $\alpha^X_{\max} = \max_{(i, \alpha)\in X} \: \alpha \: \:, \:  \alpha^Y_{\max}  = \max_{(i, \alpha)\in Y} \:  \alpha$. \\

Two trajectories are nested if one of them is a subset of another one, $X \subset Y$ or $Y \subset X$. Persistent trajectories $X$ and $Y$ can overlap in the highest age group of $X$ and lowest age group of $Y$, or vice versa.

\subsection{Identifying critical events}
We define critical events by one or a combination of diagnoses and age groups where two trajectories begin to diverge and where one of the diverging trajectories has patients with considerably higher number of diagnoses, higher mortality or more extended hospital stays in the subsequent age group(s) compared to the other diverging trajectory. Mortality of a trajectory for a certain age group is calculated as $M = \sum_i m_i * \prod_{j \neq i} (1- m_j)$, where $m$ is the in-hospital mortality of a diagnosis (defined as the percentage of patient diagnosed with the diagnose in a specific age group who die in-hospital) which is a member of a trajectory. Length of hospital stay of a trajectory in a certain age group is defined as the average number of days spent in hospital for patients who are diagnosed with at least half of all diagnoses from a trajectory.

\section*{Results}
\subsubsection*{Multilayer Comorbidity Network}

\begin{figure*}[h] 
    \centering
    \includegraphics[width=15cm]{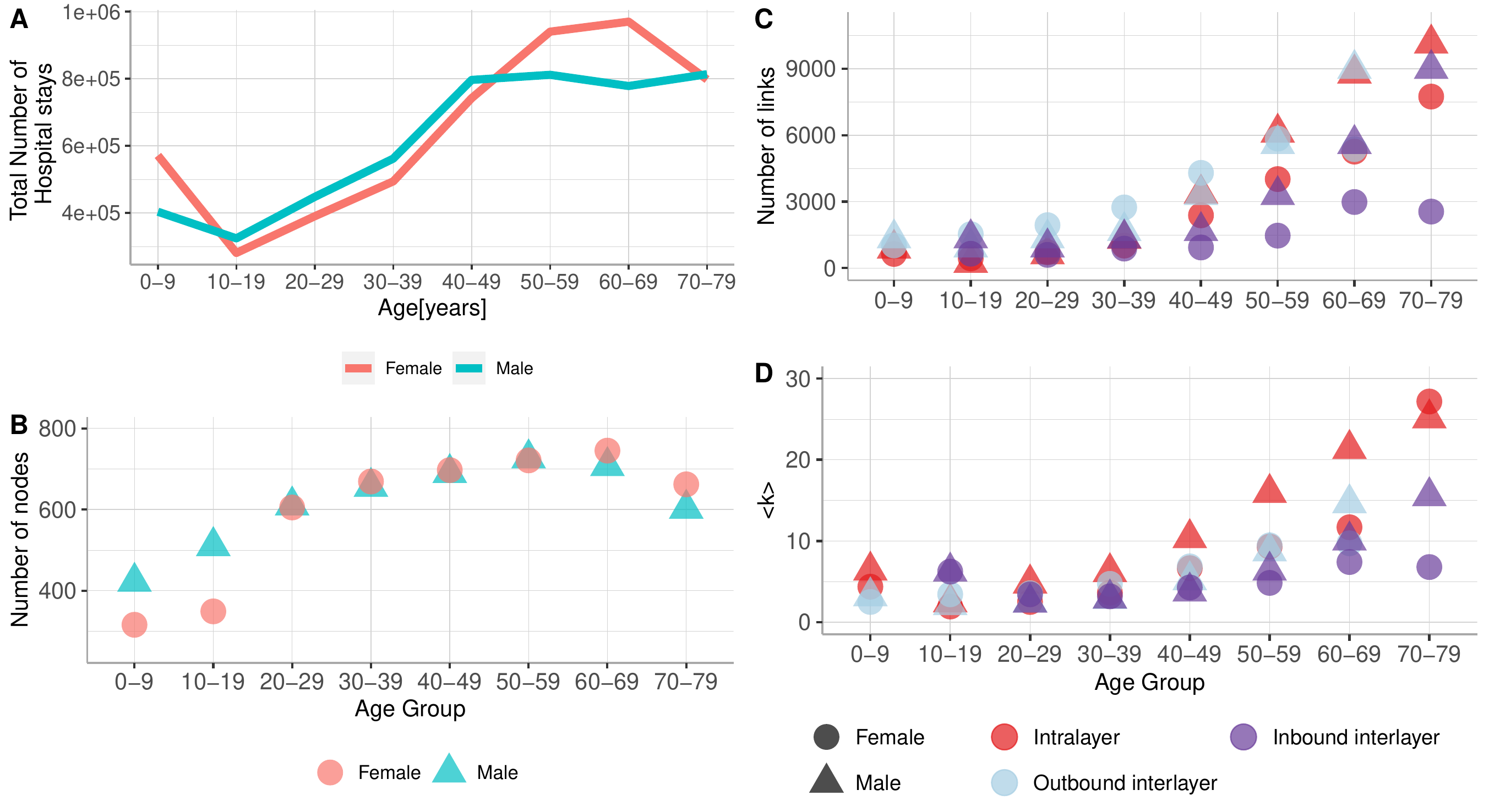}
    \caption{(A) Total number of hospital stays per age group,  Network properties: (B) number of nodes,  (C) number of all inter- and intralayer links and, (D), the average degrees $⟨k⟩$.} 
    \label{fig:Network_properties}
\end{figure*}

We constructed the multilayer comorbidity network based on hospital data, basic characteristics of the database are shown in Figure S1. 
We used all 3-digits ICD10 codes from the range A00-N99 and one more newly introduced code for patients without any diagnosis, in total 1,082 codes. Nodes in the constructed network are ICD10 codes appearing in one of eight different age groups, i.e. E66-0-9, E66-10-19, etc. Hence, we used 8,648 nodes to construct a multilayer comorbidity network with eight layers (one for each ten years age group, 0-9, 10-19,... 70-79 years old). 
We filtered the network by removing nodes without any intralayer links. This reduced the network from 8,648 nodes to 4,923 nodes for males and 4,764 nodes for females. The average degree in the filtered male network is 11.6 SD 39.7, for the female network the average degree is 15.8 SD 46.
The number of hospital stays, Figure \ref{fig:Network_properties}(A), and nodes $N$ Figure \ref{fig:Network_properties}(B) increases with age, reaches a peak at ages 60 to 69 and decreases for older ages. We see similar age trends in Figure \ref{fig:Network_properties}(C) the total number of links and Figure \ref{fig:Network_properties}(D) the average degree for intralayer as well as in- or outbound interlayer links for males and females.

\subsubsection*{Trajectories}

\begin{figure*}[!htt]
  \centering
    \includegraphics[width=11cm]{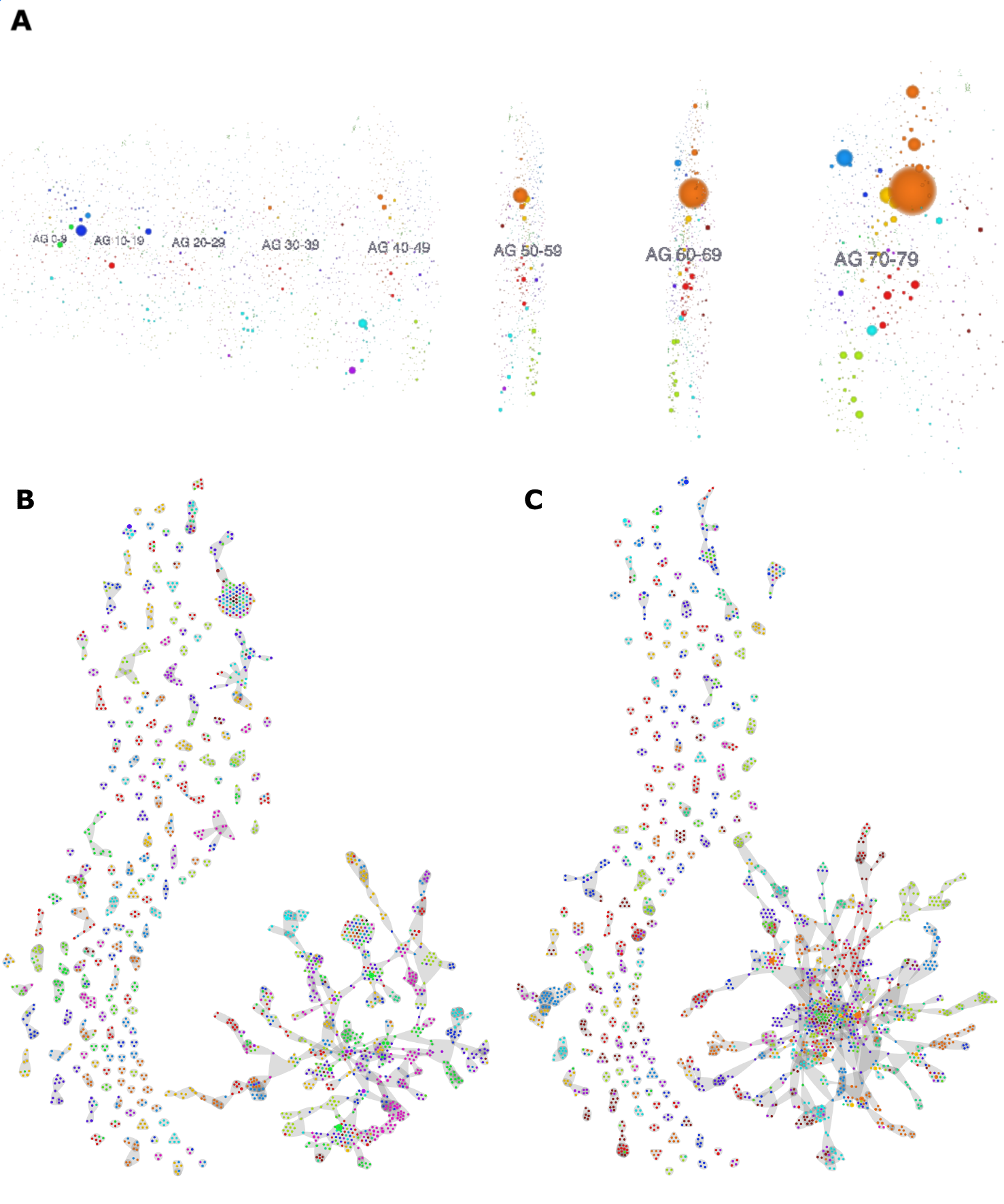}
    \caption{ 
    (A) Multilayer comorbidity network of female patients. Nodes represent diagnoses in age groups of ten years; each age group makes up a layer of the comorbidity multilayer network. The node color indicates the diagnose chapter; the diagnosis prevalence of the node in the network scales node size. 
    All identified trajectories of female (B) and male (C) patients. The members of the trajectories are nodes of a multilayer comorbidity network (diagnose + 10 years age group). The Y-axis approximates the age groups of nodes; the node color indicates the diagnose chapter, and each grey area around nodes is one trajectory. More detailed visualization of these plots can be found in the interactive web application: (A) \url{https://vis.csh.ac.at/netviewer/} and (B) \& (C)  \url{https://vis.csh.ac.at/comorbidity_network_graphics/}} 
    \label{fig:net_com}
\end{figure*}

The unsupervised community detection algorithm discovered 642 distinct disease trajectories in the male and 618 in the females network; they are listed in SI, Tables S1–S2, and shown in Figure \ref{fig:net_com}. These trajectories contain on average 9 (IQR 2-6) different diagnoses that range over up to 7 age groups (mean: 2.3 age groups), meaning that these trajectories range on average over 20-30 years and in some cases over up to 70 years of life. Besides trivial examples like a trajectory with the only diagnosis being K51 (ulcerative colitis) in each age group in males, we also found more complex trajectories spanning 70 years. For instance, for female patients there is a trajectory that starts with personality disorder (F61) at the age of 20-29y. Over the following decades there is an accumulation of mental disorders including depression (F33), post-traumatic stress disorder (F43) and eating disorders (F50) in 50-59y, followed by anxiety disorders (F40) and a few more non-chronic diagnoses in 60-69y.\\

\begin{figure*}[!ht]
    \centering
    \includegraphics[width=15.4cm]{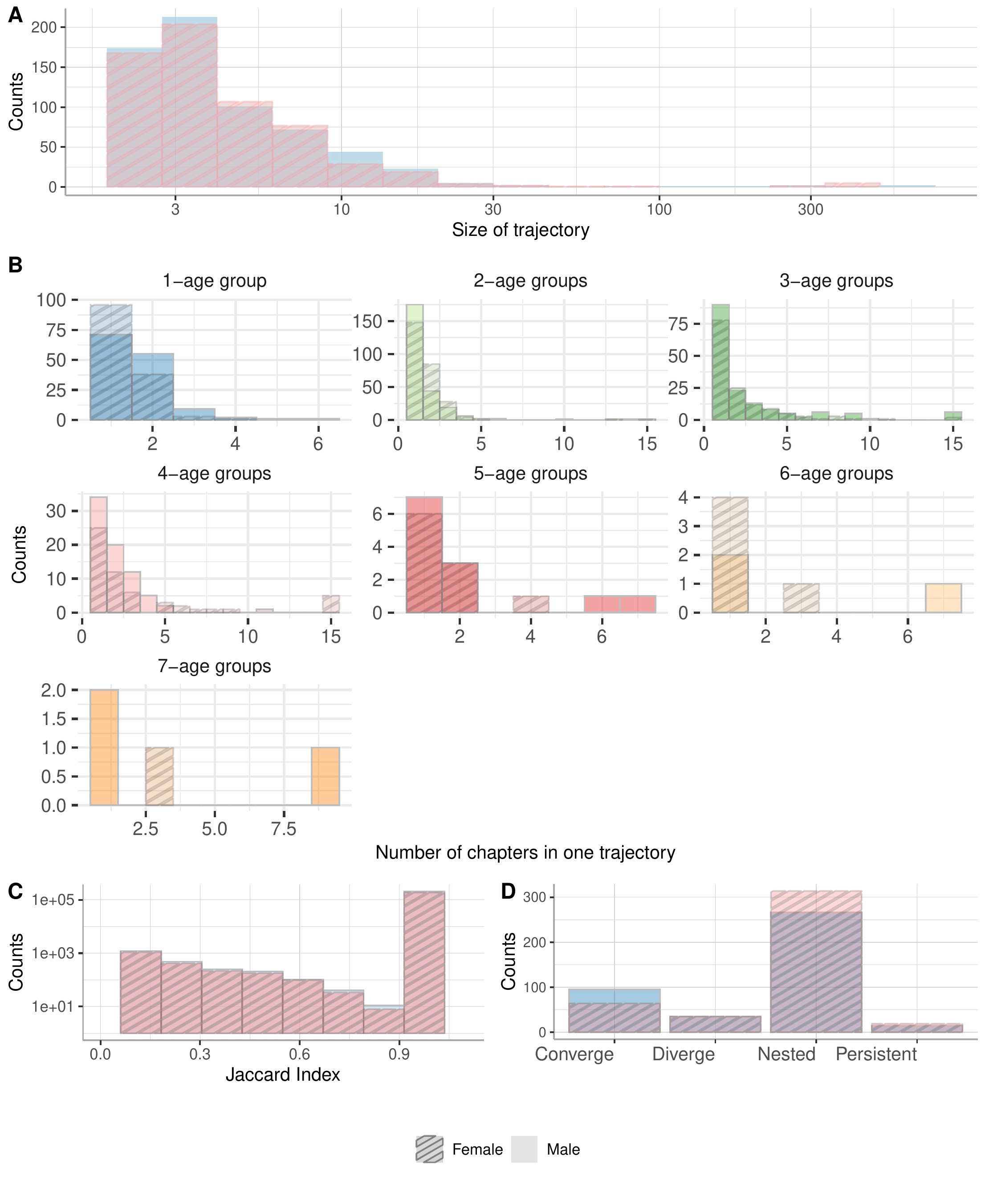}
    \caption{ Properties of trajectories. We show for males and females (A) the distribution of the sizes of the trajectories, (B) the distributions of the number of different ICD chapters per each group of trajectories, i.e. trajectories which span over one age group - first blue plot, C) the histogram of the pairwise Jaccard index and (D) frequency statistics of different types of trajectory pairs.} 
    \label{fig:com_pro}
\end{figure*}

 The distribution of the size of the trajectories (number of diagnoses-age tuples) is presented in Figure \ref{fig:com_pro} (A). Most trajectories contain between 3 and 5 diagnoses-age combinations; while a few trajectories contain more than hundred elements. We split trajectories into seven groups based on the number of age groups in the trajectory and analysed the number of different disease chapters in one trajectory Figure \ref{fig:com_pro} (B). This shows that trajectories typically span heterogeneous chapters of ICD codes, meaning that they often span diagnoses affecting quite different organ systems. 
We calculated the Jaccard index to inspect the pairwise similarity and dissimilarity of trajectories; see the distribution of this index in  Figure \ref{fig:com_pro} (C). Jaccard indices range between zero and one, indicating varying degrees of similarity between two trajectories. The most common relationship among pairs of trajectories is nested, which explains the peak at one in the Jaccard index. Figure  \ref{fig:com_pro} (D) shows frequency statistics of different types of trajectory pairs.  \\

\begin{figure*}[!ht]
    \centering
    \includegraphics[width=14cm]{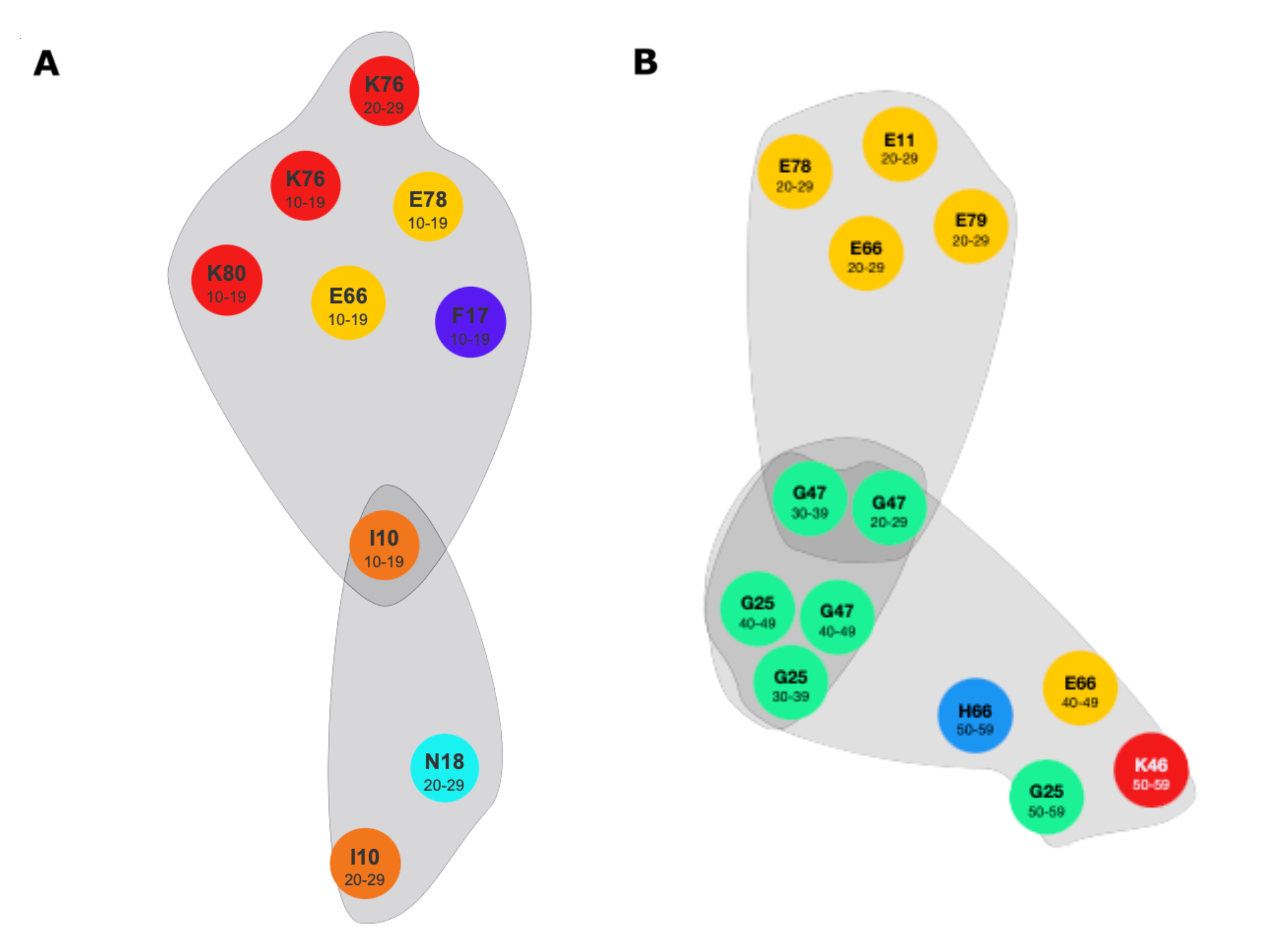}
    \caption{ 
Two examples of diverging trajectories, (A) departing from
 hypertension (I10) at an age of 10-19y in females diverges to the "kidney-trajectory" and the "metabolic trajectory" (B) departing from sleep disorders (G47) at an age of 20-29y in males diverges to metabolic trajectory with diabetes mellitus type 2 (E11), obesity (E66), lipid disorders (E78) and hyperuricemia (E79) and path with movement disorders or otitis media (G25), obesity (E66) and abdominal hernia (K46).
    } 
    \label{fig:example}
\end{figure*}

In Figure \ref{fig:example} we show a more detailed view of some of the trajectories from Figure \ref{fig:net_com}.
We show two examples for trajectories (grey areas) departing from (A)
 hypertension (I10) at an age of 10-19y in females and (B) sleep disorders (G47) at an age of 20-29y in males.
 In both cases, different combinations of other diagnoses appear in subsequent decades.
 The hypertension trajectory diverged  into chronic kidney diseases (2,289 patients) or (1,027 patients) a combination of metabolic (obesity, disorders of lipoprotein metabolism) and digestive disorders (liver diseases, cholelithiasis) with nicotine abuse.
 The sleep disorder trajectory diverged either toward the metabolic syndrome (including obesity and type 2 diabetes) in 115 patients or towards a combination of movement disorders, hernia, obesity and diseases of the middle ear (316 patients).  \\

In total, we identified 35 pairs of such diverging trajectories in females and 35 in males; see Figure \ref{fig:com_pro} D). 
On average, diverging trajectories have 2.9 SD  0.8 age groups,   3.5 SD  1.8 different diagnoses chapters, and 8.1 SD  4.7 different diseases for females, and for males  3.0 SD  1 age groups,   3.5 SD  2.9 different diagnoses chapters, and 11 SD  11 different diseases. 
While there are 64 pairs of converging trajectories in females and 95 in males, converging trajectories in females have: 2.8 SD  0.9 age groups,  4.2 SD  3.2 different diagnoses chapters, and 26  SD  79 different diseases, in males: 3 SD  1 age groups,  3.8 SD  3.5 different diagnoses chapters, and 22 SD  68 different diseases. 
Some of the trajectories are persistent (16 pairs of trajectories in females, 14 in males). These can be combined as they overlap at the end of X trajectory and the beginning of the Y trajectory. \\

The most frequent relationship between trajectories was the complete overlap of shorter and longer trajectories, which we defined as nested. We found 314 pairs of nested trajectories among female trajectories and 266 in males trajectories. \\

We designed and implemented an online visualization tool that allows a user to interactively explore the comorbidity network structure and the underlying diagnose data, \url{https://vis.csh.ac.at/netviewer/}. \\

\subsubsection*{Outcomes of trajectories}
\begin{figure*}[!ht]
    \centering
    \includegraphics[width=13cm]{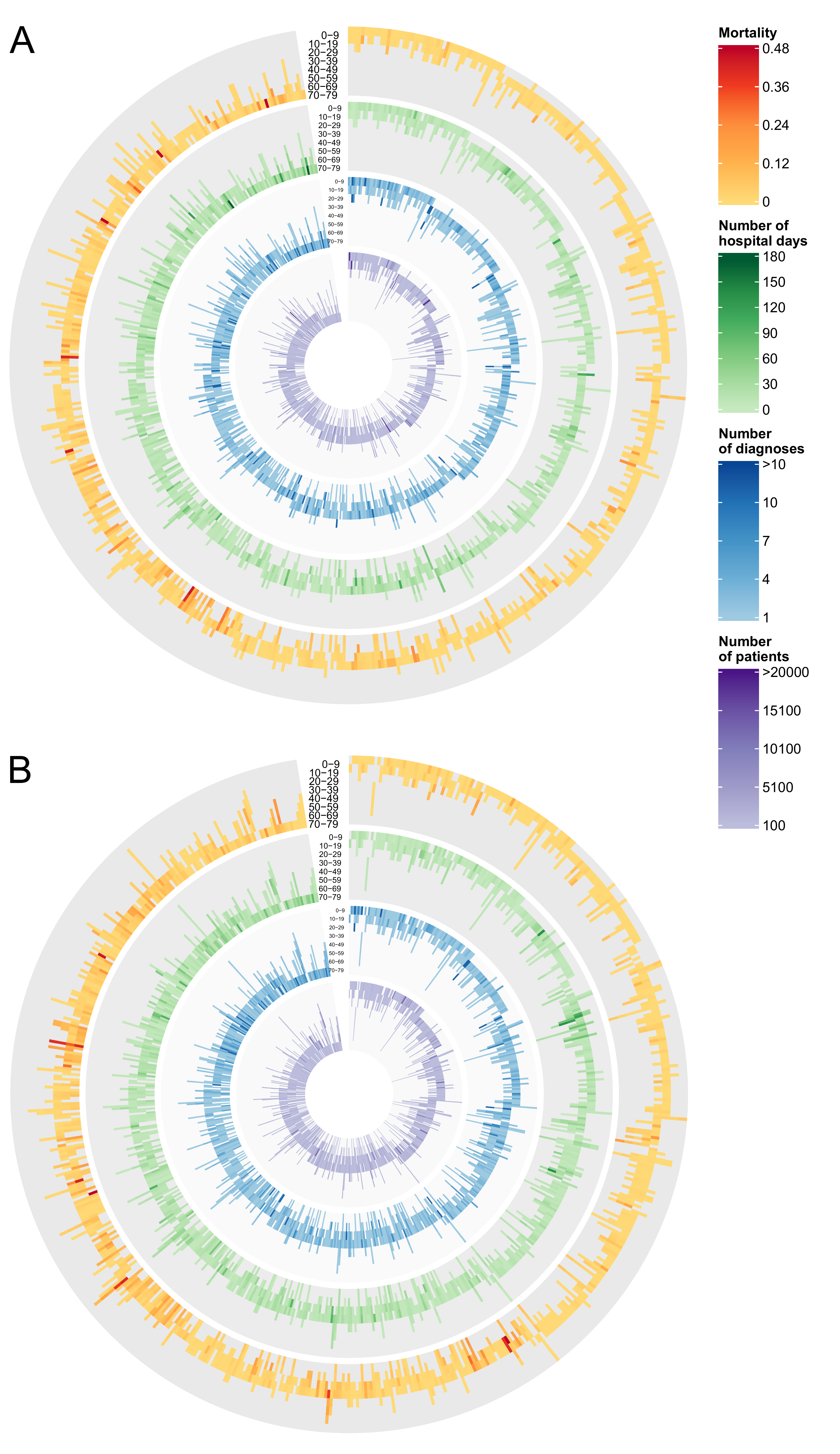}
    \caption{Outcomes of trajectories that span more than one age group (A) Females (B) Males, the outer yellow circle shows mortality of each trajectory in each age group, the green circle presents an estimation of the average number of hospital days of patients of the trajectory, while the blue circle presents an estimation of the average number of diagnoses, the inner purple circle shows a number of patients who are following each trajectory in each age group.Each trajectory is represented by a single line within each circle, which is further divided based on the age groups the trajectory encompasses.} 
    \label{fig:circle}
\end{figure*}

For every trajectory, we calculated (in-hospital) mortality and the number of days spent in the hospital for each age group, Figure \ref{fig:circle}. In-hospital mortality for each trajectory is shown in the yellow outer circle. The analysis reveals notable variations in mortality rates across trajectories, with younger age groups generally exhibiting lower mortality. Moreover, it is evident that certain trajectories undergo significant shifts in mortality as they progress into older age groups. The green circle represents the average duration of hospitalization for trajectories, while the blue circle denotes the number of diagnoses, and the purple inner circle signifies the count of patients who followed at least 50\% of a given trajectory. Notably, the green circle highlights discernible differences in the number of hospital days among different trajectories. Some trajectories have a clearly higher number of hospital days compared to other trajectories; these trajectories mainly consist of mental and behavioral disorders (F chapter) and infectious and parasitic diseases (B chapter) in males, while in females, besides these we see diseases of musculoskeletal and connective tissue (M chapter) and diseases of the nervous system (G chapter). \\

We also compared outcomes of diverging trajectories; some examples are shown in Table \ref{outcome_table} (extended tables in SI, Table S8 and S9).  We calculated an average number of hospital diagnoses, hospital days, and hospital stays for each age group in each trajectory over all patients following these trajectories. We calculated the ratio of each outcome of trajectories in each diverging pair to check if these trajectories develop into different outcomes in terms of disease burden and mortality. For example, both trajectories from the pair starting with N81 in 50s are characterised with a similar average number of hospital diagnoses in the 20s, while in the 30s patients of the second trajectory have, on average, 24\% more hospital diagnoses. In the same example, we see that patients of the first trajectory, on average have more days spent in hospital and hospital stays in the 20s (Ratio of average number of / number of days spent in hospital = 1.547 / hospital stays = 1.548), but in 30s patients of the second trajectory, have remarkably more days spent in hospital and hospital stays (Ratio of average number of / number of days spent in hospital = 0.331 / hospital stays = 0.551), Table \ref{outcome_table}.

\pagebreak
\clearpage

\begin{table*}[ht]
  \centering
  \small 
\begin{tabular} {| p{0.15\textwidth} | p{0.2\textwidth} | p{0.2\textwidth} |   p{0.05\textwidth} | p{0.05\textwidth} | p{0.05\textwidth} | p{0.05\textwidth}| }

\hline
\hline
\scriptsize
\multirow{2}{*} \centering \scriptsize Same diagnoses of trajectories, before they diverge &  
\multirow{2}{*}
                \centering 
                {\scriptsize Exclusive diagnoses of the  first trajectory} & 
                                      
                                      \multirow{2}{*}
                \centering 
                {\scriptsize Exclusive diagnoses of the second trajectory}  & 
                 \relax
                 {\multirow{2}{*} \centering {\scriptsize Age group}}    & \multicolumn{3}{c|}  {\begin{tabular}{l}\scriptsize Ratio of  average number \\ \scriptsize (first vs. second trajectory) \\ \scriptsize  of number of \end{tabular}}  \\ 
 &  &   &  & \scriptsize hospital diagnoses & \scriptsize days spent in hospital & \scriptsize hospital stays         \\ 
\hline
\hline

\multicolumn{7}{c}{\textbf{ \scriptsize FEMALES }}                             \\ 
\hline
\scriptsize
\multirow{3}{*}{ 
\centering 
\begin{tabular}{c}\scriptsize I25-40-49 \\ \scriptsize  I42-40-49 \\ \scriptsize  I50-40-49  \end{tabular}}  & 
         
                \multirow{3}{*}{ 
                \centering 
                \begin{tabular}{c} \scriptsize I21-40-49 I20-50-59 \\ \scriptsize  I21-50-59 I24-50-59 \end{tabular}}   & 
                \multirow{3}{*}{ 
                \centering 
                \begin{tabular}{c}\scriptsize I42-50-59 I44-50-59 \\ \scriptsize  I50-50-59 I51-50-59 \\ \scriptsize  I42-60-69 I44-60-69 \\ \scriptsize  I44-70-79 \end{tabular}} &  
              \scriptsize 40-49   &
               \scriptsize 0.806 & 
               \scriptsize 0.92  & 
               \scriptsize 0.953  \\  
   &   &   &  \scriptsize 50-59   & \scriptsize 0.685    & \scriptsize 0.522  & \scriptsize 0.713 \\ 
 &   &   &  &  &  & \\  
  &   &   &  &  &  & \\  
\toprule
\scriptsize
\multirow{2}{*}{ 
\centering 
\begin{tabular}{c}{\scriptsize N81-50-59}\end{tabular}}  &  \multirow{2}{*}{ 
\centering 
\begin{tabular}{c} \scriptsize N80-50-59 N84-50-59 \\ \scriptsize  N85-50-59 N88-50-59 \\ \scriptsize  N95-50-59 N80-60-69  \end{tabular}}   & \multirow{2}{*}{ \centering 
\begin{tabular}{c}{\scriptsize N81-60-69 N99-60-69}  \end{tabular}}    & \scriptsize 50-59 &  \scriptsize 1.052              & \scriptsize 0.862   & \scriptsize 1.013   \\ 
 &   &  &  \scriptsize 60-69  & \scriptsize 1.258  & \scriptsize 1.255 & \scriptsize 1.377  \\
 &   &   &  &  &  & \\  
\toprule
\multicolumn{7}{c}{\textbf{\scriptsize MALES }}  \\
\hline
\scriptsize
\multirow{2}{*}{
\centering
\begin{tabular}{c}G47-20-29 \\  \scriptsize  G47-30-39\end{tabular}}  &
\multirow{2}{*}{ \begin{tabular}{c}\scriptsize E11-20-29 E66-20-29 \\  \scriptsize  E78-20-29 E79-20-29\end{tabular}}   &
\multirow{2}{*}{ \begin{tabular}{c}\scriptsize G25-30-39 E66-40-49 \\  \scriptsize  G25-40-49 G47-40-49 \\  \scriptsize  G25-50-59 H66-50-59 \\  \scriptsize  K46-50-59\end{tabular}}  &
\scriptsize 20-29 &  \scriptsize 1.021 & \scriptsize 1.547  & \scriptsize 1.548 \\
 &    &  &  \scriptsize 30-39  & \scriptsize 0.76  & \scriptsize 0.331  & \scriptsize 0.551  \\
  &   &    &  &  &  &\\
    &   &    &  &  &  &\\
\toprule
\scriptsize
\multirow{2}{*}{ \begin{tabular}{c}G47-20-29 \\  \scriptsize   G47-30-39\end{tabular}}   &
\multirow{2}{*}{ \begin{tabular}{c}\scriptsize E11-20-29 E66-20-29 \\  \scriptsize  E78-20-29 E79-20-29\end{tabular}}   &
\multirow{2}{*}{ \begin{tabular}{c}\scriptsize G25-30-39 G25-40-49 \\  \scriptsize  G47-40-49

\end{tabular}}   & \scriptsize 20-29  &  \scriptsize 1.021              & \scriptsize 1.547 & \scriptsize 1.548  \\
  &    &   & \scriptsize 30-39  &  \scriptsize 0.76               & \scriptsize 0.331  & \scriptsize 0.551   \\
\toprule
\end{tabular}%
    \caption{Outcomes of trajectories, four examples of diverging pairs of trajectories. Column  same diagnoses of trajectories shows initially overlap of trajectories, exclusive diagnoses of the first and the second trajectory shows to which states trajectories diverge.
    Ratio of average number (first vs. second trajectory) of number of   hospital diagnoses/ days spent in hospital/ hospital stays for each age group are presented for each age group.
    } 
    \label{outcome_table}
\end{table*}


\section*{Discussion and conclusion}

In this work we introduced a novel method to identify life-course disease trajectories, in some cases spanning up to 70 years of life, in terms of sequences and combinations of hospital diagnoses that form and change over time.
Our comprehensive analysis identified 642 disease trajectories in males and 618 in females ranging over the entire diagnostic spectrum (41\% of males and 42\% of female trajectories contained diagnoses from more than one ICD chapter). 
While the most common length of these trajectories was two diagnoses for both sexes, on average they  contained  5.3 SD 5.1 and 5.4 SD 5.5 diagnoses for males and females, respectively, emphasizing the heterogeneous and widespread nature of multimorbidity in the general population.

There is a substantial variation in the number of patients that follow a trajectory. We count patients for each trajectory for each age group if they have at least 50\% of diagnoses from a trajectory. In general, shorter trajectories tend to be followed by more patients (more than 10,000 patients per trajectory per age group) than longer, more specific ones that typically contain approximately a hundred patients. The number of patients in a trajectory typically increases with age.\\
  \pagebreak
 \clearpage
 
The trajectories foster the rapid identification of critical events.
These can take the form of bifurcation points where a trajectory ``splits up'' into multiple diverging trajectories at a specific age group.
More concrete, we found 35 pairs of diverging trajectories for females and 35 pairs for males.
For example, in females diagnosed with arterial hypertension (I10) between 10 and 19 years, two major trajectories were identified by the model. The first trajectory lead to the additional diagnosis of chronic kidney disease (N18) at an age of 20-29 years. This is clinically relevant as the number of pediatric arterial hypertension is increasing worldwide ~\cite{ashraf2020pediatric} and it is well known that aHTN is closely related to chronic kidney disease. So our results point out that from a clinical point of view, a strict monitoring for arterial hypertension should be established especially in children at high risk, such as obese children or children with the metabolic syndrome. Arterial hypertension does not only mean increased risk for chronic kidney disease, but also other complications such as cardiovascular disease. The second trajectory was characterized by patients with the metabolic syndrome; these patients were disproportionally diagnosed with obesity (E66), lipid disorders (E78), steatosis hepatis (K76), cholelithiasis (K80) and nicotine abuse (N17) in their further life. In general, we therefore have two trajectories in females initially diagnosed with arterial hypertension, which are in principal dangerous conditions - the "kidney-trajectory" and the "metabolic trajectory". We found that approximately 2,289 patients follow the "kidney-" and 1,027 patients follow the metabolic trajectory. These trajectories are mostly important as metabolic diseases belong to the most common diseases worldwide and also chronic kidney disease is a disease which is related to multi morbidity and increased mortality rate.\\

In a different example we found that sleeping disorders (G47) in males diagnosed in the age groups between 20-39 years were also followed by a metabolic trajectory which was defined by an over-representation of later diagnoses of diabetes mellitus type 2 (E11), obesity (E66), lipid disorders (E78) and hyperuricemia (E79).  The other trajectory, diverging from sleeping disorders, is characterized by a higher chance of being diagnosed with movement disorders or otitis media (G25), obesity (E66) and abdominal hernia (K46). We found substantial differences in the average number of diagnoses and hospital days between patients of different branches of these diverging trajectories. While patients who followed these two trajectories showed similar average numbers of diagnoses at age 20-29 (3.3 diagnoses in both cases), patients who followed a metabolic trajectory had, on average, 3.9 diagnoses ten years later while patients who followed the other trajectory had, on average, 5.1 diagnoses. The number of sleeping disorders is on the rise and these results show that patients with sleeping disorders have to be monitored for several diseases in different trajectories. 
Our analysis also identified several instances were diverging trajectories differed substantially in their mortality, in some cases of up to 18 times. \\

In terms of mortality we 
identify trajectories that develop into a combination of diagnoses with high mortality in older age groups. For instance, a trajectory consisting of chronic bronchitis and COPD at an age of 40-49y,  bronchiectasis and intraoperative and postprocedural complications at 50-59y and finally in sequelae of tuberculosis, inflammatory polyneuropathy, conjunctivitis,  bronchitis, bronchiectasis, eosinophilia and again intraoperative and postprocedural complications in 60-69y in males had eight times higher mortality in the age group 60-69y compared to its' mortality ten years earlier (mortality increased from $0.089$ in 40-49y to $0.013$ before jumping to $0.11$ in 60-69y). Trajectories with the highest mortality usually contain cancer diagnoses, but cardiovascular or respiratory diseases also feature in the trajectories with high mortality.

\subsubsection*{Strengths and Limitations}

Strengths of this study include its comprehensive population-wide in-hospital database, containing information on about 9 million individuals. Non-systematic errors, such as randomly missing diagnoses, have little impact on our research because of the volume of the data set. However, this study has some limitations caused by data quality and limitations in data availability, in particular, the lack of information on outpatient visits, medication and lifestyle. Consequently, we cannot evaluate the outcomes of outpatient visits, blood tests, examinations, or imaging because primary care diagnoses are not recorded in this dataset; only hospital diagnoses coded with ICD10 codes were available for analysis. \\

Another drawback is that the database was designed for billing purposes, so diagnoses that did not result in financial compensation were frequently not reported. Therefore, we have to point out that some diseases, such as alcohol related disorders or nicotine dependence, are often not recorded correctly in our data. Further, socio-economic indicators for individual patients were also not available in the dataset, leaving it yet to be explored how socio-economic status impacts on these trajectories. An additional constraint associated with the dataset is the exclusive availability of in-hospital mortality data.  On a methodological level, it is also important to bear in mind that a constructed multilayer comorbidity network has two types of links (with normalized links weights); but these types are not distinguishable by the used community detection algorithms. \\

In summary, we presented a novel and statistically grounded way of studying disease progression over time based on a population-wide and decade-spanning data set of hospital diagnoses. We proposed an age multilayer comorbidity network as a base for our modelling approach. We showed that this kind of network is a promising approach for better understanding disease trajectories and their dynamics as patients age. 
While some of the identified trajectories in this study have been described in previously published studies, many novel disease trajectories and their decades-long time dynamics have been revealed.\\
A better understanding of diseases, their correlations and the sequences in they occur has the potential to improve the prevention of focal diseases. Early detection and identification of a patient's projected disease trajectory might enable prompt and timely treatments next to targeted preventive action. Consequently, that will help transition health systems from single-disease models to more effective life-spanning and individualized multimorbidity models ~\cite{zou2022association}. 



\renewcommand{\baselinestretch}{1.5}


\renewcommand{\baselinestretch}{1.5}


\section*{Acknowledgments} 
This study was supported financially by the WWTF "Mathematics and..." Project MA16-045. ED would like to thank Michaela Kaleta, Nina Haug and Rafael Prieto-Curiel for the helpful discussions.

\section*{Author Contributions}
ED and PK conceived the study and devised the analytic methods. ED wrote the manuscript with contributions from PK, ST, ML and JS. ED carried out the analysis and produced the plots and graphics. JS and LY designed and implemented the visualisation too. AK-W, AK and ML contributed medical expertise regarding the medical interpretation of the findings and in developing medical hypotheses. ED and PK researched and prepared the data. All authors reviewed and contributed to the manuscript.
\section*{Competing Interests}
The authors declare no competing interests.

\clearpage

\appendix

\pagebreak
\clearpage


\section*{Supplementary material}

# Unravelling cradle-to-grave disease trajectories from multilayer comorbidity networks

Elma Dervic1,2 , Johannes Sorger1 , Liuhuaying Yang1 , Leutner Michael3 , Kautzky Alexander4 3 , Stefan Thurner 1,2,5 , Alexandra Kautzky-Willer3,6 , Peter Klimek1,2,∗

[1] Section for Science of Complex Systems, CeMSIIS, Medical University of Vienna, Spitalgasse 23, Vienna, Austria

[2] Complexity Science Hub Vienna, Josefstädter Straße 39, 1080 Vienna, Austria

[3] Department of Internal Medicine III, Clinical Division of Endocrinology and Metabolism, Gender Medicine Unit, Medical University of Vienna, Waehringer Guertel 18–20, 1090, Vienna, Austria

[4] Gender Institute, Gars am Kamp, Austria

[5] Department of Psychiatry and Psychotherapy, Medical University of Vienna, Waehringer Guertel 18-20, A-1090 Vienna, Austria

**Suplementary Information**

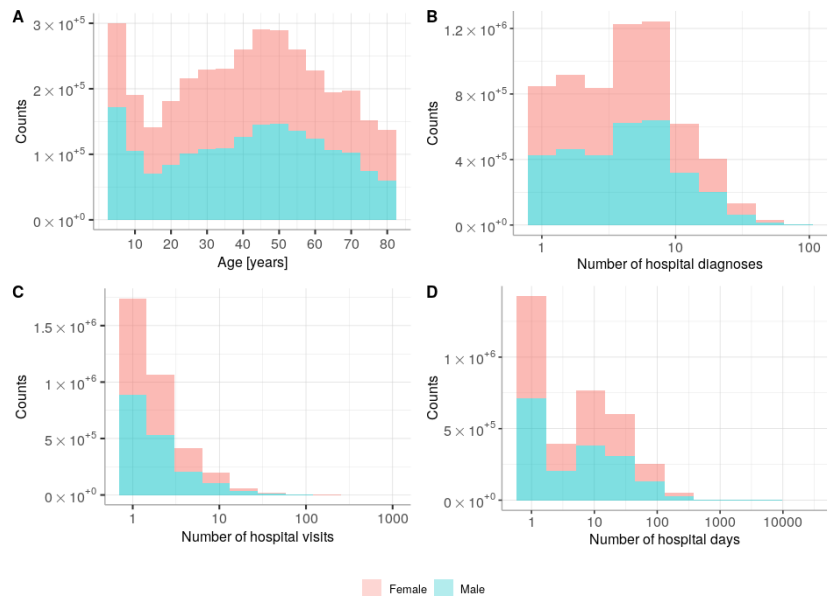

Figure S1: Distribution of (A) age of patients, (B) average number of hospital diagnoses of patients, (C) average number of hospital visits of patients, (D) average number of hospital days of patients

Figure S2 shows an overview of interlayer links aggregated on the level of a chapter (of ICD10 codes) for each layer (age group) for female patients. The size of nodes is proportional to the prevalence of diagnoses from the chapter. Link weight between chapters A and B is the number of links oriented from chapter A- to B divided by the total number of outgoing links from chapter A. More detailed visualization of this plot can be found in the interactive WEB application: https://vis.csh.ac.at/comorbidity_network_graphics/mlcn_chapters.html

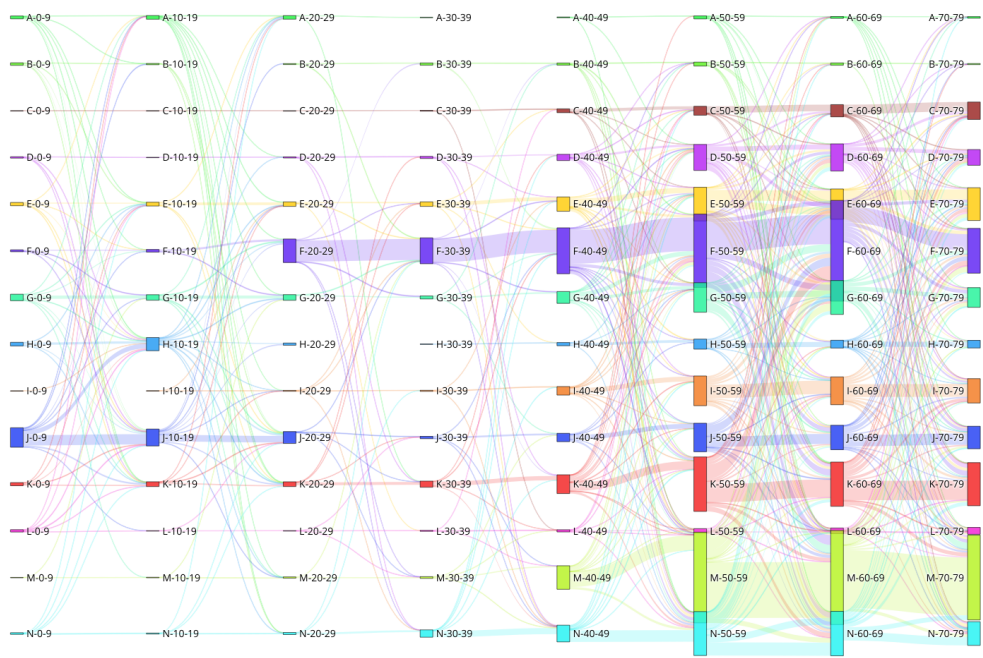

Figure S2: Interlayer links grouped by nodesICD chapters, female patients. More detailed visualization of this plot can be found in the interactive WEB application: https://vis.csh.ac.at/comorbidity_network_graphics/mlcn_chapters.html

**Visualisation tool**

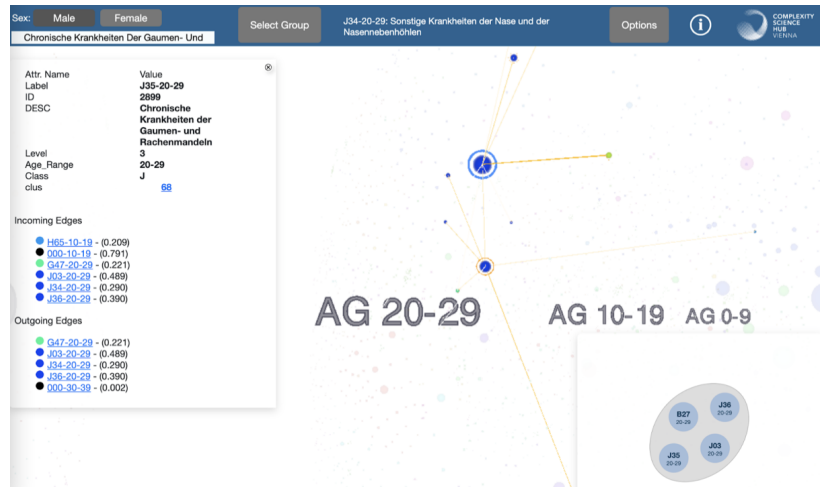

Figure S3: Screenshot from developed visualisation tool

We designed and implemented an online visualization tool that allows a user to interactively explore the comorbidity network structure and the underlying diagnose data, https://vis.csh.ac.at/netviewer/. The tool positions the comorbidity network in a hierarchical fashion - one hierarchy layer for each age group; within each layer the nodes represent diagnoses, labelled by ICD10 code. A layer places nodes on a 2D plane within 3D space and signifies the associated age group by a label. The node color indicates the diagnose chapter; node size is scaled by the diagnosis prevalence of the node in the network. Connections between nodes indicate the likelihood that a pair of diseases is co-occurring in the link color (from white (unlikely) to orange (very likely)). A multitude of options to explore network properties and structure are at the user's disposal. A user can search for diagnoses (top left), or select one of the pre-computed communities (Select Group Button). When selecting a node or a community the camera in the main view automatically transforms to a position that puts all selected network elements into view. A two dimensional representation of the current community is displayed in the bottom right corner. When selecting a node in the main 3D view or in the 2D community view, information on detailed node attributes are displayed in the info box on the left. The info box also displays a list of a node's community memberships as well as incoming and outgoing edges of the currently selected node, along with the edge weights. Selecting a community or an edge automatically transforms the view to the selected element.

With this set of visual information and interactions, the user can easily browse node attributes, inspect community compositions or follow disease paths in the network.

Table S4: All identified communities in males. Communities smaller than 100 members

| Community_ID | Members of community |
|---|---|
| 1 | K35-10-19, K37-10-19, K56-10-19, K65-10-19 |
| 2 | A04__0-9, A08__0-9, B09__0-9, F50__0-9, J10__0-9 |
| 3 | A09__0-9, H73__0-9, K92__0-9 |
| 4 | A41__0-9, C91__0-9, N10__0-9, A41-10-19, C91-10-19 |
| 5 | A49__0-9, B27__0-9, J03__0-9, J06__0-9, J11__0-9, J38__0-9, L04__0-9, L50__0-9, J38-10-19 |
| 6 | B00__0-9, B08__0-9, B34__0-9, B37__0-9, B99__0-9, D50__0-9, H10__0-9, J00__0-9, J02__0-9, J04__0-9, J05__0-9, J12__0-9, J15__0-9, J21__0-9, J22__0-9, J31__0-9, J38__0-9, J40__0-9, K52__0-9, L22__0-9 |
| 7 | L20-30-39, L20-40-49, L20-50-59, L20-60-69 |
| 8 | B27__0-9, J03__0-9, L04__0-9 |
| 9 | D50-10-19, K50-10-19, D50-20-29 |
| 10 | B97__0-9, J20__0-9, J30__0-9, J42__0-9, L30__0-9, D80-10-19 |
| 11 | C71__0-9, C79-10-19, H53-10-19 |
| 12 | D61__0-9, D69__0-9, D70__0-9 |
| 13 | D80__0-9, E45__0-9, E61__0-9, G00__0-9, H02__0-9, H04__0-9, J01__0-9, J16__0-9, K00__0-9, K76__0-9, L27__0-9, M43__0-9, N44__0-9, H65-10-19, J35-10-19, J36-10-19, H65-20-29 |
| 14 | E03__0-9, E03-10-19, E03-20-29 |
| 15 | E10__0-9, E14__0-9, E16__0-9, E10-10-19, E14-10-19, E16-10-19 |
| 16 | E23__0-9, E23-10-19, E23-20-29 |
| 17 | E34__0-9, E34-10-19, E34-20-29 |
| 18 | E66__0-9, E66-10-19, E78-10-19, F17-10-19, I10-10-19, K76-10-19 |
| 19 | E84__0-9, E84-10-19, K86-10-19, E84-20-29, K86-20-29, E84-30-39 |
| 20 | E88__0-9, J96-10-19, J96-20-29 |
| 21 | F41__0-9, K02__0-9, K05__0-9, K10__0-9, L03__0-9 |
| 22 | F43__0-9, F81__0-9, F43-10-19, F70-10-19, F81-10-19, F90-10-19, F93-10-19 |
| 23 | E88__0-9, F79__0-9, G25__0-9, G41__0-9, I47__0-9, J69__0-9, G40-10-19, G41-10-19, J96-10-19, J96-20-29 |
| 24 | F80__0-9, F82__0-9, F80-10-19, F82-10-19 |
| 25 | F90__0-9, F91__0-9, F92__0-9, F93__0-9 |
| 26 | F91__0-9, F91-10-19, F91-20-29 |
| 27 | G40__0-9, G41__0-9, G47__0-9, G81__0-9, G91__0-9, G41-10-19, G47-10-19, G81-10-19, G91-10-19 |
| 28 | G71__0-9, G71-10-19, G71-20-29 |
| 29 | M22-10-19, M23-10-19, M24-10-19, M25-10-19, M65-10-19, M67-10-19 |
| 30 | G93-10-19, G93-20-29, G93-30-39 |
| 31 | H50__0-9, H52__0-9, H53__0-9, H50-10-19, H52-10-19 |
| 32 | H60__0-9, H66__0-9, H70__0-9, H92__0-9, J11__0-9, K04__0-9 |
| 33 | H61__0-9, H65__0-9, H74__0-9, J34__0-9, J39__0-9 |
| 34 | H68__0-9, H69__0-9, J32__0-9, J35__0-9, J36__0-9 |
| 35 | H70__0-9, K04__0-9, K08__0-9, K12__0-9 |

| | |
|---|---|
| 36 | H90__0-9, H91__0-9, H90-10-19, H91-10-19 |
| 37 | J30__0-9, J30-10-19, J30-20-29 |
| 38 | J04__0-9, J05__0-9, J31__0-9, J40__0-9, J31-10-19 |
| 39 | K21__0-9, K21-10-19, K29-10-19, K44-10-19, K52-10-19 |
| 40 | J45__0-9, J46__0-9, J46-10-19 |
| 41 | K40__0-9, K42__0-9, N43__0-9, K40-10-19, N43-10-19 |
| 42 | K59__0-9, K60__0-9, K61__0-9 |
| 43 | K90__0-9, K90-10-19, K90-20-29, K90-30-39 |
| 44 | L01__0-9, L20__0-9, L20-10-19, L20-20-29, L20-30-39, L20-40-49, L20-50-59, L20-60-69 |
| 45 | L90__0-9, N35__0-9, N47__0-9, N48__0-9 |
| 46 | M21__0-9, M21-10-19, M21-20-29 |
| 47 | N12__0-9, N39__0-9, N39-10-19, N39-20-29 |
| 48 | N10__0-9, N13__0-9, N28__0-9, N28-10-19 |
| 49 | N31__0-9, N31-10-19, N31-20-29 |
| 50 | B27-10-19, J03-10-19, J36-10-19 |
| 51 | C71-10-19, C71-20-29, C71-30-39, D43-30-39 |
| 52 | D69-10-19, D69-20-29, D69-30-39 |
| 53 | E84-10-19, E84-20-29, K86-20-29, E84-30-39 |
| 54 | E87-10-19, F10-10-19, F17-10-19 |
| 55 | F11-10-19, F13-10-19, F19-10-19 |
| 56 | F32-10-19, F33-10-19, F41-10-19, F45-10-19 |
| 57 | E87-10-19, F10-10-19, F12-10-19, F17-10-19, F20-10-19, F23-10-19, F60-10-19, F60-20-29, F63-20-29 |
| 58 | F70-10-19, F70-20-29, F70-30-39, F70-40-49 |
| 59 | F91__0-9, F90-10-19, F91-10-19, F93-10-19, F98-10-19, F91-20-29 |
| 60 | G80-10-19, G82-10-19, G80-20-29, G82-20-29, G80-30-39, G80-40-49 |
| 61 | I10-10-19, I10-20-29, I15-30-39, I15-40-49 |
| 62 | F80__0-9, F82__0-9, F83__0-9, F84__0-9, F89__0-9, F80-10-19, F82-10-19, F89-10-19, F89-20-29 |
| 63 | J32-10-19, J33-10-19, J34-10-19, M95-10-19 |
| 64 | K76-10-19, K76-20-29, K76-30-39, K65-40-49 |
| 65 | N44-10-19, N45-10-19, N50-10-19 |
| 66 | A63-20-29, N47-20-29, N48-20-29 |
| 67 | B20-30-39, B24-30-39, B20-40-49, B24-40-49 |
| 68 | B27-20-29, J03-20-29, J35-20-29, J36-20-29 |
| 69 | F40-30-39, F40-40-49, F40-50-59 |
| 70 | C62-20-29, C77-20-29, C78-20-29, D40-20-29 |
| 71 | D16-20-29, L05-20-29, D16-30-39, L05-30-39 |
| 72 | N18__0-9, N18-10-19, N18-20-29, D64-30-39, N19-30-39 |
| 73 | E10__0-9, E14__0-9, E16__0-9, E10-10-19, E14-10-19, E16-10-19, E10-20-29, E14-20-29, E16-20-29 |
| 74 | E11-20-29, E66-20-29, E78-20-29, E79-20-29, G47-20-29, G47-30-39 |
| 75 | E78-20-29, E79-20-29, E79-30-39 |
| 76 | F10-20-29, K70-20-29, K71-20-29, K71-30-39 |

| | |
|---|---|
| 77 | F11-10-19, F13-10-19, F19-10-19, B17-20-29, B18-20-29, B24-20-29, F11-20-29, F13-20-29, F14-20-29, B94-30-39, F15-30-39 |
| 78 | F20-10-19, F23-10-19, F12-20-29, F15-20-29, F20-20-29, F23-20-29, F25-20-29, F31-20-29, F20-30-39, F23-30-39, F25-30-39, F20-40-49, F25-40-49 |
| 79 | B94-20-29, B16-30-39, B17-30-39, B18-30-39, F11-30-39, F12-30-39, F13-30-39, F14-30-39, F19-30-39, K73-30-39, B17-40-49, B18-40-49, B94-40-49, F11-40-49, F12-40-49, F14-40-49, F19-40-49, K73-40-49, F12-50-59, K73-50-59 |
| 80 | F17-20-29, J06-20-29, J93-20-29 |
| 81 | F19-20-29, F90-20-29, B94-30-39 |
| 82 | F20-10-19, F23-10-19, F20-20-29, F23-20-29, F25-20-29, F20-30-39, F23-30-39, F25-30-39 |
| 83 | F31-20-29, F31-30-39, F31-40-49, F31-50-59, F31-60-69, F31-70-79 |
| 84 | F32-20-29, F40-20-29, F63-20-29 |
| 85 | F41-20-29, F45-20-29, F41-30-39, F45-30-39 |
| 86 | F42-20-29, F42-30-39, F42-40-49 |
| 87 | F20-10-19, F23-10-19, F20-20-29, F23-20-29, F25-20-29, F20-30-39, F23-30-39, F25-30-39, F20-40-49, F25-40-49 |
| 88 | G80-20-29, G80-30-39, G80-40-49 |
| 89 | G82-20-29, G82-30-39, G82-40-49 |
| 90 | H83-20-29, H91-20-29, H93-20-29 |
| 91 | I26-20-29, I80-20-29, I26-30-39, I80-30-39 |
| 92 | I42-20-29, I42-30-39, I50-30-39 |
| 93 | J01-20-29, J32-20-29, J33-20-29 |
| 94 | I26-20-29, I80-20-29, J18-20-29, J90-20-29, I26-30-39, I80-30-39 |
| 95 | K01-20-29, K07-20-29, K09-20-29 |
| 96 | K02-20-29, K04-20-29, K08-20-29, K12-20-29 |
| 97 | K03-20-29, K01-30-39, K02-30-39, K04-30-39, K07-30-39, K08-30-39, K09-30-39, K12-30-39 |
| 98 | K51-10-19, K25-20-29, K26-20-29, K29-20-29, K51-20-29, K58-20-29, K59-20-29 |
| 99 | K35-20-29, K37-20-29, K65-20-29 |
| 100 | D50-10-19, K50-10-19, D50-20-29, K50-20-29, K56-20-29, K63-20-29, K56-30-39 |
| 101 | L02-20-29, L03-20-29, L72-20-29 |
| 102 | K74-20-29, I85-30-39, K74-30-39 |
| 103 | K80-20-29, K81-20-29, K85-20-29 |
| 104 | M17-20-29, M23-20-29, M24-20-29, M25-20-29, M65-20-29, M67-20-29, M93-20-29, M94-20-29 |
| 105 | M24-20-29, M25-20-29, M65-20-29, M67-20-29 |
| 106 | M42-20-29, M51-20-29, M53-20-29, M54-20-29 |
| 107 | M81-20-29, M81-30-39, M81-40-49 |
| 108 | N19-20-29, I10-30-39, I11-30-39, I34-30-39, I61-30-39, I67-30-39, I70-30-39, I71-40-49, L97-40-49 |
| 109 | N13-20-29, N20-20-29, N23-20-29, N13-30-39, N20-30-39, N23-30-39 |
| 110 | N18__0-9, N18-10-19, N18-20-29, D64-30-39 |
| 111 | B16-30-39, K73-30-39, B17-40-49, B18-40-49, K73-40-49, B17-50-59, B18-50-59, C22-50-59, K73-50-59, B17-60-69, K73-60-69 |
| 112 | B16-30-39, B17-30-39, B18-30-39, K73-30-39, B17-40-49, B18-40-49, B94-40-49, K73-40-49, B17-50-59, B18-50-59, C22-50-59, K73-50-59, B17-60-69, K73-60-69 |
| 113 | C34-30-39, C34-40-49, D38-40-49 |
| 114 | C62-30-39, C77-30-39, C78-30-39, C79-30-39, D40-30-39, C62-40-49, C77-40-49, C62-50-59, C62-60-69 |
| 115 | C73-30-39, C73-40-49, E89-40-49 |

| | |
|---|---|
| 116 | C81-40-49, C81-50-59, C81-60-69, C81-70-79 |
| 117 | I84-20-29, K62-20-29, D12-30-39, I84-30-39, K62-30-39 |
| 118 | F70-10-19, F70-20-29, D53-30-39, F10-30-39, F34-30-39, F70-30-39, D53-40-49, F05-40-49, F34-40-49, F70-40-49, G31-40-49, H10-40-49, K72-40-49, L21-40-49, D53-50-59 |
| 119 | F70-10-19, F70-20-29, D75-30-39, F70-30-39, D52-40-49, F10-40-49, F22-40-49, F23-40-49, F70-40-49, G31-40-49, L21-40-49, C09-50-59, C10-50-59, C13-50-59, E53-50-59, F22-50-59, F55-50-59, H10-50-59, I86-50-59, N62-50-59, F22-60-69, F22-70-79 |
| 120 | D86-30-39, D86-40-49, D86-50-59 |
| 121 | E04-30-39, E05-30-39, E04-40-49, E05-40-49, E89-50-59 |
| 122 | E10-30-39, E16-30-39, H36-30-39, E10-40-49, E16-40-49, H36-40-49 |
| 123 | E13-30-39, H40-30-39, E11-40-49, E13-40-49, G63-40-49, H25-40-49, H26-40-49, H40-40-49, M86-40-49, N47-40-49, N48-40-49, I79-50-59, L03-50-59, M86-50-59 |
| 124 | I83-30-39, I87-30-39, I83-40-49, I87-40-49, I87-50-59 |
| 125 | K29-30-39, K31-30-39, K51-30-39, K58-30-39, K59-30-39, K59-40-49 |
| 126 | G35-20-29, E78-30-39, E87-30-39, G35-30-39, G45-30-39, H81-30-39, M42-30-39, M47-30-39, M47-40-49 |
| 127 | E88-30-39, E88-40-49, E88-50-59 |
| 128 | F07-20-29, F07-30-39, F06-40-49, F07-40-49 |
| 129 | B94-20-29, F17-30-39, G45-30-39, J06-30-39, J15-30-39, J20-30-39, J93-30-39, L05-40-49 |
| 130 | F63-30-39, F63-40-49, F63-50-59 |
| 131 | F61-40-49, F11-50-59, F13-50-59, F33-50-59, F43-50-59, F60-50-59, F61-50-59, F11-60-69, F60-60-69 |
| 132 | G25-30-39, G25-40-49, G25-50-59 |
| 133 | G40-40-49, G41-40-49, G81-40-49, J69-40-49, F70-50-59, G41-50-59, F70-60-69 |
| 134 | M42-20-29, M51-20-29, M53-20-29, M54-20-29, G54-30-39, M54-30-39, M62-30-39, M99-30-39 |
| 135 | H33-30-39, H35-30-39, H43-40-49, H33-40-49 |
| 136 | H33-30-39, H35-30-39, H36-30-39, H43-40-49, H33-40-49, H34-40-49, H35-40-49, H36-40-49, H43-50-59, H33-50-59, H35-50-59, H36-50-59, H43-60-69, H26-60-69, H33-60-69 |
| 137 | H50-30-39, H52-30-39, H53-30-39, H52-40-49, H53-40-49, H52-50-59, H53-50-59 |
| 138 | H53-30-39, H53-40-49, H53-50-59, H53-60-69, H53-70-79 |
| 139 | H91-30-39, H93-30-39, H90-40-49, H91-40-49, H93-40-49, H90-50-59, H91-50-59, H93-50-59, H90-60-69, H91-60-69, H93-60-69 |
| 140 | I20-30-39, I21-30-39, I25-30-39, I20-40-49, I21-40-49, I24-40-49 |
| 141 | I34-30-39, I07-40-49, I34-40-49, I36-40-49 |
| 142 | I35-30-39, I71-30-39, I35-40-49, I71-40-49, I35-50-59, I71-50-59 |
| 143 | I45-30-39, I45-40-49, I45-50-59 |
| 144 | I73-30-39, I73-40-49, I74-40-49 |
| 145 | J01-30-39, J32-30-39, J33-30-39, J01-40-49, J32-40-49, J33-40-49 |
| 146 | J03-30-39, J35-30-39, J36-30-39 |
| 147 | J18-30-39, J90-30-39, J96-30-39, J15-40-49 |
| 148 | J30-30-39, J30-40-49, J30-50-59, J30-60-69 |
| 149 | J44-30-39, J44-40-49, J98-40-49, J98-50-59 |
| 150 | K01-30-39, K07-30-39, K09-30-39 |
| 151 | K20-30-39, K21-40-49, K25-40-49, K26-40-49, D62-50-59, K20-50-59, K25-50-59, K26-50-59, K20-60-69, K20-70-79 |
| 152 | K21-30-39, K25-30-39, K26-30-39, K31-30-39 |
| 153 | K40-30-39, K42-30-39, K43-30-39 |

| | |
|---|---|
| 154 | D50-30-39, K50-30-39, K56-30-39, D50-40-49, K50-40-49, K56-40-49, K50-50-59, K50-60-69, K50-70-79 |
| 155 | K60-20-29, K61-20-29, L02-20-29, L03-20-29, L72-20-29, K60-30-39, K61-30-39 |
| 156 | K63-30-39, D12-40-49, I84-40-49, K60-40-49, K61-40-49, K62-40-49, K63-40-49, D12-50-59, I84-50-59, K60-50-59, K61-50-59, K62-50-59, K64-50-59, D21-60-69, K64-60-69 |
| 157 | K80-30-39, K81-30-39, K82-30-39, K83-30-39 |
| 158 | K92-30-39, D62-40-49, K92-40-49 |
| 159 | D17-30-39, L02-30-39, L03-30-39, L72-30-39 |
| 160 | M16-30-39, M87-30-39, M16-40-49, M87-40-49 |
| 161 | M17-30-39, M23-30-39, M66-30-39, M71-30-39, M93-30-39 |
| 162 | M75-30-39, M75-40-49, M77-40-49, M76-50-59, M77-50-59 |
| 163 | M25-30-39, M25-40-49, M66-40-49, M87-40-49 |
| 164 | M42-30-39, M47-30-39, M47-40-49 |
| 165 | M43-30-39, M48-30-39, M54-40-49, G54-50-59, H61-50-59 |
| 166 | M50-30-39, M53-30-39, M50-40-49 |
| 167 | M19-30-39, M24-30-39, M25-30-39, M65-30-39, M67-30-39, M75-30-39, M19-40-49, M24-40-49, M65-40-49, M67-40-49, M75-40-49, M77-40-49, M66-50-59, M75-50-59, M76-50-59, M77-50-59 |
| 168 | N18-40-49, B99-50-59, N03-50-59, N25-50-59, N26-50-59, N05-60-69, N26-60-69, N26-70-79 |
| 169 | N31__0-9, N31-10-19, N31-20-29, N31-30-39, N39-30-39, N41-30-39 |
| 170 | A46-30-39, B35-30-39, A46-40-49, B35-40-49, I89-50-59 |
| 171 | B16-40-49, B16-50-59, B16-60-69 |
| 172 | B37-40-49, J18-40-49, J86-40-49 |
| 173 | C15-40-49, C16-40-49, C15-50-59, C16-50-59 |
| 174 | C18-40-49, C19-40-49, C20-40-49 |
| 175 | C21-40-49, C19-50-59, C20-50-59, C21-50-59, K91-50-59, C21-60-69, C21-70-79 |
| 176 | K74-20-29, I85-30-39, K74-30-39, C22-40-49, I85-40-49, K31-40-49, K72-40-49, K74-40-49, B99-50-59, K72-50-59 |
| 177 | C32-40-49, J37-40-49, J38-40-49, C32-50-59, J37-50-59, J38-50-59 |
| 178 | C43-40-49, D22-40-49, C43-50-59, D22-50-59 |
| 179 | C61-40-49, N32-40-49, N40-40-49, C61-50-59, D40-50-59, N32-50-59 |
| 180 | C64-40-49, D41-40-49, C64-50-59, D41-50-59 |
| 181 | C71-40-49, D43-40-49, C71-50-59, D43-50-59 |
| 182 | C82-40-49, C83-40-49, C85-40-49, C82-50-59, C83-50-59, C85-50-59 |
| 183 | C90-40-49, C90-50-59, C90-60-69, C90-70-79 |
| 184 | C92-40-49, C92-50-59, C92-60-69, D46-60-69, C92-70-79, D46-70-79 |
| 185 | D12-40-49, I84-40-49, K60-40-49, K61-40-49, K62-40-49, D12-50-59, I84-50-59, K60-50-59, K61-50-59, K62-50-59, K64-50-59, D21-60-69, K64-60-69 |
| 186 | D13-40-49, K63-50-59, D12-60-69, D13-60-69, D13-70-79 |
| 187 | D17-40-49, D23-40-49, D48-40-49 |
| 188 | D18-40-49, D18-50-59, D18-60-69, D18-70-79 |
| 189 | M45-20-29, E13-30-39, M45-30-39, D35-40-49, I10-40-49, I12-40-49, I72-40-49, I77-40-49, K20-40-49, L98-40-49, M45-40-49, N05-40-49, I12-50-59, I13-50-59, M45-50-59, N05-50-59, M45-60-69, N05-60-69, M45-70-79 |
| 190 | M54-50-59, G54-60-69, L50-60-69, M71-60-69, M76-60-69 |
| 191 | D68-40-49, I26-40-49, I80-40-49, I82-40-49, D68-50-59, I26-50-59, I80-50-59, I82-50-59, D68-60-69 |
| 192 | D75-40-49, D75-50-59, D75-60-69, D75-70-79 |
| 193 | E06-40-49, E06-50-59, E06-60-69, E06-70-79 |

| | |
|---|---|
| 194 | E10-30-39, E16-30-39, H36-30-39, E16-40-49 |
| 195 | E12-40-49, E78-40-49, G45-40-49, G51-40-49, H81-40-49, I72-40-49, I77-40-49, K75-40-49, G45-50-59 |
| 196 | K60-40-49, K61-40-49, K60-50-59, K61-50-59 |
| 197 | E88-30-39, E88-40-49, F54-40-49, H47-40-49, E66-50-59, E88-50-59, E61-60-69, L08-60-69 |
| 198 | G47-20-29, G25-30-39, G47-30-39, E66-40-49, G25-40-49, G47-40-49, G25-50-59, H66-50-59, K46-50-59 |
| 199 | E73-40-49, E73-50-59, E73-60-69, E73-70-79 |
| 200 | F09-40-49, F63-40-49, F03-50-59, F10-50-59, F63-50-59, G31-50-59, L21-50-59, D52-60-69, E51-60-69, E53-60-69, I62-60-69, J39-60-69, K02-60-69, K12-60-69, M46-60-69 |
| 201 | J37-40-49, J38-40-49, J37-50-59, J38-50-59, J37-60-69, J37-70-79 |
| 202 | F22-50-59, F22-60-69, F22-70-79 |
| 203 | F32-40-49, F48-40-49, B37-50-59, F51-50-59, K02-50-59, K04-50-59, K08-50-59 |
| 204 | F34-40-49, F34-50-59, F34-60-69, F34-70-79 |
| 205 | D52-40-49, F54-40-49, F55-40-49, J41-40-49, L71-40-49, F17-50-59, F41-50-59, F42-50-59, J95-50-59, J13-60-69, J39-60-69, K04-60-69, K08-60-69, K12-60-69, L08-60-69, N62-60-69 |
| 206 | F79-50-59, F79-60-69, F79-70-79 |
| 207 | G35-20-29, G35-30-39, G35-40-49, N31-40-49, N31-50-59 |
| 208 | G43-40-49, G43-50-59, G43-60-69 |
| 209 | G46-40-49, I63-40-49, I64-40-49, I66-40-49 |
| 210 | M43-30-39, M48-30-39, G55-40-49, M51-40-49, M54-40-49, M96-40-49, G54-50-59, G55-50-59, H61-50-59, M51-50-59, M96-50-59, M96-60-69 |
| 211 | M96-40-49, M96-50-59, M96-60-69 |
| 212 | F17-40-49, G58-40-49, J01-40-49, J37-40-49, J38-40-49, J42-40-49, J43-40-49, J93-40-49, J02-50-59, J37-50-59, J38-50-59, J37-60-69, J37-70-79 |
| 213 | G91-40-49, I60-40-49, G91-50-59, I60-50-59 |
| 214 | G95-40-49, G95-50-59, G95-60-69, G95-70-79 |
| 215 | H40-30-39, H25-40-49, H26-40-49, H40-40-49 |
| 216 | H91-30-39, H93-30-39, H81-40-49, H90-40-49, H91-40-49, H93-40-49, H81-50-59, H90-50-59, H91-50-59, H93-50-59 |
| 217 | I20-30-39, I21-30-39, I25-30-39, I20-40-49, I21-40-49, I24-40-49, I25-40-49, I20-50-59, I21-50-59, I24-50-59, B00-60-69, I24-60-69 |
| 218 | I12-40-49, I13-40-49, N25-40-49, D44-50-59, E72-50-59, G61-50-59, I10-50-59, I12-50-59, I13-50-59, I30-50-59, I62-50-59, I99-50-59, J22-50-59, J91-50-59, L23-50-59, L27-50-59, L73-50-59, M35-50-59, M46-50-59, M80-50-59, N05-50-59, C31-60-69, E26-60-69, L05-60-69, N05-60-69, N12-60-69 |
| 219 | I25-40-49, I27-40-49, I31-40-49, A04-50-59, I27-50-59 |
| 220 | I35-40-49, I71-40-49, I35-50-59, I71-50-59, I72-50-59, I71-60-69, I72-60-69, I71-70-79, I72-70-79 |
| 221 | I42-20-29, I42-30-39, I50-30-39, I42-40-49, I44-40-49, I50-40-49, I51-40-49, I44-50-59 |
| 222 | G93-10-19, G93-20-29, G93-30-39, G93-40-49, I46-40-49 |
| 223 | G91-40-49, I60-40-49, I61-40-49, G91-50-59, I60-50-59, I61-50-59 |
| 224 | I65-40-49, I65-50-59, G46-60-69, I66-60-69 |
| 225 | D12-40-49, I84-40-49, K60-40-49, K61-40-49, K60-50-59, K61-50-59, K64-50-59 |
| 226 | I98-40-49, I98-50-59, K74-50-59, I86-60-69 |
| 227 | J34-40-49, J35-40-49, J36-40-49, M95-40-49 |
| 228 | J41-40-49, J44-50-59, J47-50-59, J95-50-59, B90-60-69, G61-60-69, H10-60-69, J40-60-69, J47-60-69, J82-60-69, J95-60-69 |
| 229 | J43-40-49, J43-50-59, J93-60-69 |
| 230 | K02-50-59, K04-50-59, K08-50-59 |

28

| | |
|---|---|
| 231 | K02-40-49, K04-40-49, K08-40-49, K12-40-49, K02-50-59, K04-50-59, K08-50-59 |
| 232 | K20-30-39, K22-30-39, K21-40-49, K22-40-49, K25-40-49, K26-40-49, K44-40-49, K20-50-59, K26-50-59, K20-60-69, K20-70-79 |
| 233 | K25-40-49, K26-40-49, K26-50-59 |
| 234 | K40-40-49, K66-40-49, N43-40-49 |
| 235 | K42-40-49, K43-40-49, D17-50-59, K40-50-59, K42-50-59, K43-50-59, K66-50-59, L72-50-59 |
| 236 | K52-30-39, A09-40-49, E86-40-49, K52-40-49 |
| 237 | K58-40-49, K58-50-59, K58-60-69, K58-70-79 |
| 238 | K25-40-49, K26-40-49, K70-40-49, D52-50-59, D62-50-59, F05-50-59, I98-50-59, K25-50-59, K26-50-59 |
| 239 | K76-40-49, A04-50-59, D61-50-59, E55-50-59, K46-50-59, K55-50-59, K75-50-59, K55-60-69 |
| 240 | K80-40-49, K81-40-49, K82-40-49, K83-40-49, K81-50-59 |
| 241 | K90-40-49, K90-50-59, K90-60-69, K90-70-79 |
| 242 | L02-40-49, L03-40-49, L72-40-49 |
| 243 | A46-30-39, B35-30-39, A46-40-49, B35-40-49, A46-50-59, B35-50-59, I89-50-59, L30-50-59, B35-60-69, L30-60-69 |
| 244 | L71-40-49, K70-50-59, D53-60-69, F05-60-69, G31-60-69, G31-70-79 |
| 245 | L89-40-49, L89-50-59, L89-60-69 |
| 246 | L97-40-49, L98-40-49, L97-50-59, L98-50-59 |
| 247 | M11-40-49, M23-40-49, M71-40-49, M93-40-49 |
| 248 | M11-40-49, M17-40-49, M21-40-49, M22-40-49, M23-40-49, M71-40-49, M93-40-49, M94-40-49, M21-50-59, M22-50-59, M94-50-59 |
| 249 | M20-40-49, M20-50-59, M20-60-69, M20-70-79 |
| 250 | M31-40-49, M31-50-59, M31-60-69, M31-70-79 |
| 251 | M42-30-39, M47-30-39, M41-40-49, M42-40-49, M43-40-49, M47-40-49, M48-40-49, M41-50-59, M42-50-59, M43-50-59, M48-50-59 |
| 252 | M50-30-39, M53-30-39, M50-40-49, M53-40-49, M62-40-49, M62-50-59 |
| 253 | M70-40-49, E79-50-59, M16-50-59, M70-50-59, M87-50-59, M87-60-69 |
| 254 | M72-40-49, M72-50-59, M72-60-69 |
| 255 | M77-40-49, M76-50-59, M77-50-59 |
| 256 | L98-40-49, M86-40-49, L98-50-59, M86-50-59 |
| 257 | N03-40-49, A49-50-59, A69-50-59, E78-50-59, F54-50-59, G46-50-59, G58-50-59, H49-50-59, H81-50-59, I05-50-59, I66-50-59, I99-50-59, L27-50-59, M15-50-59, F23-60-69, F44-60-69, J03-60-69 |
| 258 | N04-40-49, N04-50-59, N04-60-69 |
| 259 | N47-60-69, N48-60-69, N47-70-79, N48-70-79 |
| 260 | N13-40-49, N20-40-49, N23-40-49, N13-50-59, N20-50-59, N23-50-59, N20-60-69, N21-60-69, N23-60-69 |
| 261 | N25-40-49, N26-40-49, N18-50-59, N25-50-59, N26-50-59, A04-60-69, B02-60-69, N03-60-69, N26-60-69, N26-70-79 |
| 262 | F54-40-49, N26-40-49, I25-50-59, I97-50-59, I13-60-69, I30-60-69, I33-60-69, K04-60-69 |
| 263 | N28-40-49, N28-50-59, N28-60-69 |
| 264 | N39-50-59, N45-50-59, A04-60-69, N30-60-69, N45-60-69 |
| 265 | N32-40-49, N35-40-49, N32-50-59, N35-50-59 |
| 266 | N39-40-49, N45-40-49, I95-50-59 |
| 267 | N41-40-49, N41-50-59, N42-50-59 |
| 268 | A08-50-59, I25-60-69, I97-60-69, I99-60-69, J22-60-69, B34-70-79, C75-70-79, D14-70-79, F51-70-79, F52-70-79, G12-70-79, G98-70-79, J85-70-79, K02-70-79 |
| 269 | A41-50-59, J80-50-59, K65-50-59 |

| | |
|---|---|
| 270 | A49-50-59, K29-50-59, K50-50-59, K51-50-59, K75-50-59, D73-60-69, E61-60-69, K50-60-69, L50-60-69, N62-60-69, K50-70-79 |
| 271 | B02-50-59, J18-60-69, J86-60-69, B00-70-79, B96-70-79, C79-70-79, D72-70-79, D73-70-79, M84-70-79 |
| 272 | B16-30-39, K73-30-39, B17-40-49, B18-40-49, K73-40-49, B17-50-59, B18-50-59, C22-50-59, F19-50-59, K73-50-59, B17-60-69, F19-60-69, K73-60-69 |
| 273 | B37-50-59, C34-50-59, C38-50-59, C71-50-59, D43-50-59, J82-50-59, J93-50-59 |
| 274 | B90-50-59, J47-50-59, J44-60-69, J47-60-69, A16-70-79, B44-70-79, B90-70-79, D52-70-79, G58-70-79, J04-70-79, J21-70-79, J39-70-79, L27-70-79, L28-70-79, M00-70-79 |
| 275 | C04-50-59, C04-60-69, C04-70-79 |
| 276 | C09-50-59, C09-60-69, C10-60-69, C09-70-79, C10-70-79 |
| 277 | C18-40-49, C19-40-49, C20-40-49, C18-50-59, C19-50-59 |
| 278 | C21-50-59, C21-60-69, C21-70-79 |
| 279 | C21-50-59, C18-60-69, C19-60-69, C20-60-69, C21-60-69, C48-60-69, K36-60-69, K91-60-69, C19-70-79, C20-70-79, C21-70-79, C48-70-79, K36-70-79, K91-70-79 |
| 280 | C49-50-59, C49-60-69, C49-70-79 |
| 281 | C67-40-49, C67-50-59, C68-50-59, C67-60-69, C68-60-69, D41-60-69, N30-60-69, C65-70-79, C68-70-79 |
| 282 | C71-50-59, D43-50-59, C71-60-69, D43-60-69, C71-70-79, D43-70-79 |
| 283 | C73-50-59, E89-50-59, C73-60-69, E89-60-69, C73-70-79, E89-70-79 |
| 284 | D17-50-59, K40-50-59, K66-50-59, L72-50-59 |
| 285 | D32-50-59, D32-60-69, D32-70-79 |
| 286 | D35-50-59, D35-60-69, D35-70-79 |
| 287 | D47-50-59, D47-60-69, D47-70-79 |
| 288 | D59-50-59, D59-60-69, D59-70-79 |
| 289 | E83-40-49, D64-50-59, E03-50-59, E21-50-59, E83-50-59, A49-60-69, E83-60-69, E83-70-79 |
| 290 | K51-30-39, K51-40-49, D73-50-59, J04-50-59, K51-50-59, B96-60-69, K29-60-69, K51-60-69, A69-70-79, D16-70-79, D21-70-79, K51-70-79, L50-70-79, N10-70-79, N62-70-79 |
| 291 | L89-40-49, N08-40-49, E11-50-59, E12-50-59, E55-50-59, L89-50-59, G61-60-69, I13-60-69, I33-60-69, L28-60-69, L89-60-69 |
| 292 | E10-50-59, E13-50-59, E16-50-59, E16-60-69 |
| 293 | E83-40-49, E83-50-59, E83-60-69, E83-70-79 |
| 294 | G93-10-19, G93-20-29, G93-30-39, G93-40-49, I46-40-49, G93-50-59, I46-50-59 |
| 295 | F07-20-29, F07-30-39, F06-40-49, F07-40-49, F06-50-59, F07-50-59 |
| 296 | H53-30-39, H52-40-49, H53-40-49, H52-50-59, H53-50-59, H43-60-69, H02-60-69, H26-60-69, H33-60-69, H52-60-69, H53-60-69, H43-70-79, H02-70-79, H26-70-79, H33-70-79, H52-70-79, H53-70-79 |
| 297 | G35-60-69, N47-60-69, N48-60-69, G35-70-79, N47-70-79, N48-70-79 |
| 298 | G44-30-39, F45-40-49, G44-40-49, F45-50-59, G44-50-59 |
| 299 | G47-50-59, J30-50-59, F51-60-69, J30-60-69 |
| 300 | G91-40-49, I60-40-49, G91-50-59, I60-50-59, G91-60-69, G91-70-79 |
| 301 | G20-50-59, F02-60-69, G20-60-69, G21-60-69, G21-70-79 |
| 302 | G30-50-59, F00-60-69, G30-60-69 |
| 303 | G40-50-59, G41-50-59, F01-60-69, F71-60-69, G41-60-69 |
| 304 | G51-50-59, G51-60-69, G51-70-79 |
| 305 | M43-30-39, M48-30-39, M51-40-49, M54-40-49, M96-40-49, G54-50-59, H61-50-59, M96-50-59, M96-60-69 |
| 306 | G82-20-29, G82-30-39, G82-40-49, G82-50-59, G83-60-69 |

| | |
|---|---|
| 307 | G99-50-59, J31-50-59, J94-50-59, K45-50-59, E66-60-69, G62-60-69, G61-70-79, H16-70-79, L71-70-79, M93-70-79 |
| 308 | H02-50-59, H25-50-59, H26-50-59, H27-50-59, H34-50-59, H34-60-69, H34-70-79 |
| 309 | H34-50-59, H34-60-69, H34-70-79 |
| 310 | H40-50-59, H47-50-59, H40-60-69, H47-60-69, H40-70-79, H47-70-79 |
| 311 | H44-50-59, H44-60-69, H44-70-79 |
| 312 | H54-50-59, H54-60-69, H54-70-79 |
| 313 | H90-50-59, H91-50-59, H93-50-59, H90-60-69, H91-60-69, H93-60-69, H90-70-79, H91-70-79, H93-70-79 |
| 314 | I07-50-59, I34-50-59, I36-50-59, I07-60-69, I08-60-69, I34-60-69, I36-60-69, I07-70-79, I34-70-79, I36-70-79 |
| 315 | I20-50-59, B00-60-69, K04-60-69 |
| 316 | I31-50-59, I31-60-69, I31-70-79 |
| 317 | I34-50-59, I36-50-59, I08-60-69 |
| 318 | I48-20-29, I48-30-39, I42-40-49, I44-40-49, I48-40-49, I07-50-59, I34-50-59, I36-50-59, I44-50-59, I08-60-69 |
| 319 | I48-50-59, I05-60-69, I05-70-79 |
| 320 | I50-50-59, J81-50-59, B99-60-69, G93-60-69, I46-60-69 |
| 321 | I77-50-59, I77-60-69, I77-70-79 |
| 322 | I71-50-59, I72-50-59, I71-60-69, I72-60-69, I71-70-79, I72-70-79 |
| 323 | K60-40-49, K61-40-49, I84-50-59, K60-50-59, K61-50-59, D21-60-69 |
| 324 | J01-50-59, J32-50-59, J33-50-59, J01-60-69 |
| 325 | B37-40-49, J18-50-59, J40-50-59, J86-50-59, J93-50-59, B02-60-69, L72-60-69 |
| 326 | J32-60-69, J33-60-69, J34-60-69, J32-70-79, J33-70-79, J34-70-79 |
| 327 | J34-50-59, J35-50-59, M95-50-59 |
| 328 | K20-50-59, K20-60-69, K20-70-79 |
| 329 | C44-40-49, C44-50-59, L57-50-59, C44-60-69, D23-60-69, L57-60-69, L57-70-79 |
| 330 | D12-60-69, D13-60-69, D13-70-79 |
| 331 | K55-50-59, K55-60-69, K55-70-79 |
| 332 | K59-30-39, K59-40-49, K56-50-59, K59-50-59, K65-50-59, K91-50-59 |
| 333 | K80-50-59, K81-50-59, K82-50-59 |
| 334 | L28-50-59, K76-60-69, B96-70-79, D73-70-79, E51-70-79, G58-70-79, M89-70-79 |
| 335 | L57-50-59, L57-60-69, L57-70-79 |
| 336 | M07-50-59, L40-60-69, L40-70-79 |
| 337 | M15-50-59, M15-60-69, M15-70-79 |
| 338 | M22-70-79, M23-70-79, M24-70-79, M67-70-79, M94-70-79 |
| 339 | M35-50-59, M35-60-69, M35-70-79 |
| 340 | M42-30-39, M47-30-39, M42-40-49, M43-40-49, M47-40-49, M48-40-49, M43-50-59, M48-50-59 |
| 341 | M45-20-29, M45-30-39, M45-40-49, M45-50-59, M45-60-69, M45-70-79 |
| 342 | M81-50-59, M80-60-69, M85-60-69, M80-70-79 |
| 343 | M89-50-59, F32-60-69, F60-70-79, G83-70-79 |
| 344 | M17-40-49, M21-40-49, M22-40-49, M17-50-59, M21-50-59, M22-50-59 |
| 345 | N08-50-59, N08-60-69, N08-70-79 |
| 346 | N10-50-59, N26-50-59, N18-60-69, N26-60-69, B02-70-79, B25-70-79, H10-70-79, I13-70-79, N05-70-79, N26-70-79 |
| 347 | N13-50-59, N20-50-59, N23-50-59, N13-60-69, N20-60-69, N21-60-69, N23-60-69, N21-70-79 |

| | |
|---|---|
| 348 | N32-40-49, N40-40-49, N41-40-49, N21-50-59, N40-50-59, N41-50-59, N42-50-59, N41-60-69, N42-60-69 |
| 349 | N26-50-59, N26-60-69, N26-70-79 |
| 350 | N43-50-59, N43-60-69, N43-70-79 |
| 351 | A02-60-69, A18-60-69, A26-60-69, A40-60-69, A48-60-69, A69-60-69, A84-60-69, B07-60-69, B36-60-69, B44-60-69, B49-60-69, B86-60-69, B91-60-69, C00-60-69, C06-60-69, C11-60-69, C12-60-69, C14-60-69, C26-60-69, C39-60-69, C41-60-69, C45-60-69, C50-60-69, C60-60-69, C66-60-69, C69-60-69, C76-60-69, C84-60-69, C88-60-69, D01-60-69, D03-60-69, D04-60-69, D07-60-69, D11-60-69, D15-60-69, D31-60-69, D34-60-69, D51-60-69, D89-60-69, E07-60-69, E22-60-69, E23-60-69, E27-60-69, E41-60-69, E46-60-69, E74-60-69, E85-60-69, F52-60-69, G06-60-69, G24-60-69, G52-60-69, G90-60-69, G92-60-69, H21-60-69, H31-60-69, H55-60-69, H61-60-69, H70-60-69, J02-60-69, J10-60-69, J11-60-69, J12-60-69, J21-60-69, J61-60-69, J85-60-69, K11-60-69, K28-60-69, K41-60-69, L25-60-69, L58-60-69, L60-60-69, L80-60-69, L81-60-69, M12-60-69, M60-60-69, N11-60-69, N34-60-69, N36-60-69, N50-60-69, O00-70-79 |
| 352 | K57-30-39, K57-40-49, K57-50-59, A08-60-69 |
| 353 | G20-50-59, F02-60-69, G20-60-69, G21-60-69, F02-70-79, G20-70-79, G21-70-79 |
| 354 | A41-60-69, B37-60-69, J80-60-69 |
| 355 | A46-60-69, B35-60-69, A46-70-79, B35-70-79 |
| 356 | B18-60-69, C22-60-69, C24-60-69, K73-70-79 |
| 357 | C01-50-59, C02-50-59, C01-60-69, C02-60-69, C01-70-79, C02-70-79 |
| 358 | C01-60-69, C02-60-69, C01-70-79, C02-70-79 |
| 359 | C13-60-69, C32-60-69, J38-60-69, C13-70-79, C32-70-79, J38-70-79 |
| 360 | C15-40-49, C16-40-49, C15-50-59, C16-50-59, C15-60-69, C16-60-69 |
| 361 | C25-40-49, C25-50-59, C24-60-69, C25-60-69, C25-70-79 |
| 362 | C71-60-69, D43-60-69, C71-70-79, D43-70-79 |
| 363 | C65-60-69, D30-60-69, C66-70-79, C67-70-79, C68-70-79 |
| 364 | C65-60-69, C67-60-69, C68-60-69, D30-60-69, N30-60-69, C65-70-79, C66-70-79, C67-70-79, C68-70-79 |
| 365 | G82-50-59, C79-60-69, G82-60-69, G83-60-69, M84-60-69, G82-70-79 |
| 366 | C91-40-49, C91-50-59, C82-60-69, C83-60-69, C85-60-69, C91-60-69, D72-60-69, C82-70-79, C83-70-79, C85-70-79, C91-70-79, D80-70-79 |
| 367 | C91-40-49, C91-50-59, C91-60-69, D80-70-79 |
| 368 | D17-60-69, K40-60-69, K46-60-69 |
| 369 | N43-50-59, D29-60-69, D30-60-69, N43-60-69, N45-60-69, D30-70-79, N02-70-79, N39-70-79, N40-70-79, N43-70-79, N45-70-79, N99-70-79 |
| 370 | D72-60-69, C83-70-79, C85-70-79, C91-70-79 |
| 371 | D86-30-39, D86-40-49, D86-50-59, D86-60-69, D86-70-79 |
| 372 | E12-50-59, E12-60-69, E12-70-79 |
| 373 | E10-50-59, E13-50-59, E16-50-59, E10-60-69, E13-60-69, E16-60-69, I79-60-69, E16-70-79, I79-70-79 |
| 374 | E68-60-69, J16-60-69, L95-60-69, C45-70-79, D89-70-79, E01-70-79, E72-70-79, G90-70-79, I10-70-79, I22-70-79, I33-70-79, I60-70-79, I78-70-79, J04-70-79, J13-70-79, J16-70-79, K27-70-79, L08-70-79, L20-70-79, L21-70-79, L23-70-79, L29-70-79, L50-70-79, L71-70-79, L85-70-79, M05-70-79, M18-70-79, M31-70-79, M45-70-79, M85-70-79, M96-70-79, N62-70-79 |
| 375 | E61-50-59, G57-50-59, E78-60-69, F34-60-69, G43-60-69, G44-60-69, G57-60-69, G58-60-69, G83-60-69, H49-60-69, I08-60-69, I22-60-69, I99-60-69, C75-70-79, D11-70-79, F40-70-79, G44-70-79, G71-70-79, H00-70-79, J85-70-79, K13-70-79 |
| 376 | M70-50-59, E79-60-69, L82-60-69, M18-60-69, M70-60-69, H61-70-79, L21-70-79, M54-70-79, M70-70-79, M79-70-79 |
| 377 | F06-60-69, F07-60-69, F06-70-79, F07-70-79 |
| 378 | F10-60-69, F13-60-69, F31-60-69, F60-60-69, G31-60-69, D53-70-79, E51-70-79, F13-70-79, F31-70-79, G31-70-79, K08-70-79 |
| 379 | F13-60-69, F33-60-69, F41-60-69, F43-60-69, F45-60-69, F13-70-79, F32-70-79, F33-70-79, F41-70-79, F43-70-79, F45-70-79 |

| | |
|---|---|
| 380 | H43-60-69, H26-60-69, H33-60-69, H43-70-79, H26-70-79, H33-70-79 |
| 381 | J04-50-59, F17-60-69, D51-70-79, H49-70-79, J82-70-79, K04-70-79, K12-70-79 |
| 382 | F32-50-59, G20-50-59, G35-50-59, F48-60-69, F61-60-69, G35-60-69, K02-60-69, N47-60-69, N48-60-69, G35-70-79, N47-70-79, N48-70-79 |
| 383 | F54-60-69, G60-60-69, H57-60-69, I38-60-69, F09-70-79, I25-70-79, I97-70-79, I99-70-79, J01-70-79, J41-70-79, M46-70-79 |
| 384 | F01-50-59, G40-60-69, G41-60-69, I62-60-69, G40-70-79, G41-70-79 |
| 385 | G82-50-59, G82-60-69, G83-60-69, G82-70-79 |
| 386 | H90-50-59, H91-50-59, H93-50-59, H00-60-69, H16-60-69, H81-60-69, H90-60-69, H91-60-69, H93-60-69, I38-60-69, K30-60-69, M40-60-69, E78-70-79, F34-70-79, G57-70-79, H10-70-79, H81-70-79, H90-70-79, H91-70-79, H93-70-79, I06-70-79 |
| 387 | H02-60-69, H02-70-79, H26-70-79 |
| 388 | H44-50-59, H11-60-69, H17-60-69, H18-60-69, H25-60-69, H27-60-69, H44-60-69, H50-60-69, H11-70-79, H17-70-79, H18-70-79, H21-70-79, H27-70-79, H44-70-79 |
| 389 | E10-70-79, E13-70-79, E16-70-79, H43-70-79, H26-70-79, H33-70-79, H35-70-79, H36-70-79 |
| 390 | G91-40-49, I60-40-49, D73-50-59, F52-50-59, G30-50-59, G91-50-59, G99-50-59, I60-50-59, L28-50-59, E11-60-69, E55-60-69, F00-60-69, F05-60-69, G30-60-69, G91-60-69, H49-60-69, A26-70-79, A48-70-79, C76-70-79, D34-70-79, G91-70-79, H59-70-79, H60-70-79, L08-70-79, L28-70-79, M00-70-79, N04-70-79 |
| 391 | H53-30-39, H52-40-49, H53-40-49, H52-50-59, H53-50-59, H52-60-69, H53-60-69, H52-70-79, H53-70-79 |
| 392 | I77-50-59, H57-60-69, I77-60-69, I70-70-79, I73-70-79, I74-70-79, I77-70-79, L02-70-79, L03-70-79 |
| 393 | H90-60-69, H93-60-69, H90-70-79, H91-70-79, H93-70-79 |
| 394 | B36-60-69, D34-60-69, E72-60-69, F09-60-69, G90-60-69, I10-60-69, I30-60-69, I33-60-69, I60-60-69, I78-60-69, J13-60-69, J30-60-69, K73-60-69, L08-60-69, L21-60-69, L23-60-69, L27-60-69, L29-60-69, L71-60-69, L85-60-69, N02-60-69, N04-60-69, C05-70-79, C74-70-79, D15-70-79, E26-70-79, E29-70-79, F55-70-79, G37-70-79, G96-70-79, H72-70-79, I68-70-79, J35-70-79, K09-70-79, L70-70-79, M14-70-79, N03-70-79, N49-70-79 |
| 395 | I26-60-69, I80-60-69, I82-60-69, I26-70-79, I80-70-79, I82-70-79 |
| 396 | J84-40-49, J84-50-59, I07-60-69, I27-60-69, I34-60-69, I36-60-69, J84-60-69, I07-70-79, I27-70-79, I34-70-79, I36-70-79, J84-70-79 |
| 397 | I48-60-69, A48-70-79, C76-70-79, H18-70-79, H61-70-79, I33-70-79, J30-70-79, K04-70-79 |
| 398 | I08-50-59, I31-50-59, F54-60-69, G60-60-69, H57-60-69, I31-60-69, I38-60-69, I50-60-69, D52-70-79, F09-70-79, I13-70-79, I25-70-79, I31-70-79, I97-70-79, I99-70-79, J01-70-79, J41-70-79, M46-70-79 |
| 399 | K60-60-69, K61-60-69, K60-70-79, K61-70-79 |
| 400 | C15-60-69, C16-60-69, C15-70-79, C16-70-79 |
| 401 | F09-60-69, J62-60-69, J92-60-69, J44-70-79, J47-70-79, J62-70-79, K20-70-79 |
| 402 | C34-60-69, C38-60-69, C71-60-69, D43-60-69, J82-60-69, C71-70-79, D43-70-79 |
| 403 | J84-40-49, J84-50-59, J84-60-69, J84-70-79 |
| 404 | J90-60-69, J91-60-69, J94-60-69 |
| 405 | K83-60-69, B18-70-79, C22-70-79, C24-70-79, C25-70-79, K80-70-79, K81-70-79, K82-70-79, K83-70-79, K85-70-79, K86-70-79 |
| 406 | C21-50-59, C19-60-69, C20-60-69, C21-60-69, K36-60-69, C19-70-79, C20-70-79, C21-70-79, K36-70-79 |
| 407 | K43-60-69, K56-60-69, K66-60-69, K91-60-69, A08-70-79, K91-70-79 |
| 408 | K83-60-69, K85-60-69, K86-60-69, B18-70-79, C22-70-79, C24-70-79, C25-70-79, K80-70-79, K81-70-79, K82-70-79, K83-70-79, K85-70-79, K86-70-79 |
| 409 | K51-10-19, K51-20-29, K51-30-39, K51-40-49, K51-50-59, K51-60-69, K51-70-79 |
| 410 | I98-60-69, I85-70-79, I98-70-79, K70-70-79, K72-70-79, K74-70-79 |
| 411 | G31-60-69, F10-70-79, F43-70-79, G31-70-79, K71-70-79 |
| 412 | K80-60-69, K81-60-69, K82-60-69 |
| 413 | K83-60-69, B18-70-79, C22-70-79, C24-70-79, K83-70-79 |

| | |
|---|---|
| 414 | K83-60-69, K85-60-69, K86-60-69, B18-70-79, C22-70-79, C24-70-79, C25-70-79, K83-70-79, K85-70-79, K86-70-79 |
| 415 | L98-40-49, M86-40-49, L97-50-59, L98-50-59, M86-50-59, L97-60-69, L98-60-69, M86-60-69, L97-70-79, L98-70-79, M86-70-79 |
| 416 | L82-60-69, M18-60-69, M54-70-79, M70-70-79, M79-70-79 |
| 417 | N26-50-59, L84-60-69, N25-60-69, N26-60-69, N18-70-79, N25-70-79, N26-70-79 |
| 418 | M05-40-49, M05-50-59, M05-60-69, M05-70-79 |
| 419 | M17-60-69, M21-60-69, M22-60-69, M17-70-79, M21-70-79, M22-70-79 |
| 420 | M17-70-79, M21-70-79, M22-70-79 |
| 421 | M66-50-59, M22-60-69, M23-60-69, M24-60-69, M65-60-69, M67-60-69, M71-60-69, M93-60-69, M94-60-69, M24-70-79 |
| 422 | M16-50-59, M87-50-59, M16-60-69, M87-60-69 |
| 423 | M16-50-59, M87-50-59, M16-60-69, M41-60-69, M42-60-69, M43-60-69, M62-60-69, M87-60-69, G54-70-79 |
| 424 | M41-60-69, M42-60-69, M43-60-69, M50-60-69, M62-60-69, M99-60-69, G54-70-79, G55-70-79, M41-70-79, M42-70-79, M43-70-79, M47-70-79, M48-70-79, M50-70-79, M51-70-79, M53-70-79, M99-70-79 |
| 425 | M70-50-59, M54-60-69, M70-60-69, B00-70-79, G43-70-79, G83-70-79, M76-70-79, M93-70-79 |
| 426 | M66-60-69, M75-60-69, M77-60-69 |
| 427 | N30-50-59, N43-50-59, D29-60-69, D30-60-69, N10-60-69, N39-60-69, N43-60-69, N45-60-69, D30-70-79, H10-70-79, N02-70-79, N39-70-79, N40-70-79, N41-70-79, N42-70-79, N43-70-79, N45-70-79, N99-70-79 |
| 428 | G35-40-49, N31-40-49, G35-50-59, N31-50-59, G35-60-69, G35-70-79 |
| 429 | N32-60-69, N35-60-69, N99-60-69, N32-70-79, N35-70-79 |
| 430 | D29-60-69, N40-60-69, N50-70-79 |
| 431 | I65-60-69, N39-70-79, N45-70-79, N99-70-79 |
| 432 | C21-50-59, C21-60-69, C20-70-79, C21-70-79, K36-70-79 |
| 433 | B18-70-79, C22-70-79, C24-70-79 |
| 434 | B37-70-79, C34-70-79, C38-70-79, J82-70-79 |
| 435 | C43-40-49, D22-40-49, C43-50-59, D22-50-59, C43-60-69, D22-60-69, C43-70-79 |
| 436 | H02-60-69, C44-70-79, D22-70-79, D23-70-79, H02-70-79, L57-70-79, L82-70-79 |
| 437 | C61-70-79, C64-70-79, D40-70-79, D41-70-79, I89-70-79 |
| 438 | B37-70-79, C34-70-79, C38-70-79, C77-70-79, C79-70-79, C80-70-79, D38-70-79, D48-70-79, D63-70-79, D70-70-79, J82-70-79, M84-70-79 |
| 439 | D17-70-79, K40-70-79, K46-70-79 |
| 440 | H44-50-59, H44-60-69, D31-70-79, H04-70-79, H25-70-79, H34-70-79, H44-70-79, H50-70-79 |
| 441 | N30-50-59, K63-60-69, C71-70-79, D36-70-79, D43-70-79 |
| 442 | D50-70-79, D62-70-79, K25-70-79, K26-70-79 |
| 443 | D75-40-49, D75-50-59, D75-60-69, D64-70-79, D68-70-79, D75-70-79, E21-70-79, E83-70-79 |
| 444 | E10-70-79, E13-70-79, E16-70-79, H43-70-79, H26-70-79, H33-70-79 |
| 445 | G91-40-49, I60-40-49, G91-50-59, I60-50-59, G91-60-69, G99-60-69, H00-60-69, H04-60-69, H21-60-69, H59-60-69, E11-70-79, E12-70-79, E53-70-79, E55-70-79, G82-70-79, G91-70-79, N08-70-79 |
| 446 | I73-30-39, I70-40-49, I73-40-49, I74-40-49, I70-50-59, I73-50-59, I74-50-59, I77-50-59, I74-60-69, I77-60-69 |
| 447 | G47-20-29, G25-30-39, G47-30-39, G25-40-49, G47-40-49 |
| 448 | F32-70-79, F33-70-79, F41-70-79, F43-70-79, F45-70-79 |
| 449 | G56-70-79, M65-70-79, M72-70-79 |
| 450 | H36-60-69, E10-70-79, E13-70-79, E16-70-79, H43-70-79, H26-70-79, H33-70-79, H35-70-79, H36-70-79 |
| 451 | I31-50-59, I31-60-69, I20-70-79, I21-70-79, I24-70-79, I31-70-79 |

| | |
|---|---|
| 452 | J90-70-79, J91-70-79, J93-70-79, J94-70-79 |
| 453 | I61-40-49, I61-50-59, I61-60-69, I61-70-79, I62-70-79 |
| 454 | G40-70-79, G41-70-79, G81-70-79, I61-70-79, I62-70-79 |
| 455 | J32-70-79, J33-70-79, J34-70-79 |
| 456 | L97-70-79, L98-70-79, M86-70-79 |
| 457 | M16-70-79, M41-70-79, M42-70-79, M43-70-79, M87-70-79 |
| 458 | M19-70-79, M25-70-79, M75-70-79 |
| 459 | M99-60-69, G55-70-79, M41-70-79, M42-70-79, M43-70-79, M47-70-79, M48-70-79, M50-70-79, M51-70-79, M53-70-79, M99-70-79 |
| 460 | N13-50-59, N20-50-59, N23-50-59, N13-60-69, N20-60-69, N21-60-69, N23-60-69, N20-70-79, N21-70-79, N23-70-79 |
| 461 | N41-40-49, D40-50-59, N41-50-59, N42-50-59, D40-60-69, N41-60-69, C61-70-79, C64-70-79, D40-70-79, N41-70-79, N42-70-79 |

Table S5: All identified trajectories in females. trajectories smaller than 100 members

| Community_ID | Members of community |
|---|---|
| 1 | I88__0-9, K56__0-9, K59__0-9, K60__0-9, K56-10-19 |
| 2 | A08__0-9, A38-10-19, B01-10-19, H61-10-19, J38-10-19 |
| 3 | A38__0-9, B97__0-9, E45__0-9, G47__0-9, H04__0-9, J10__0-9, J35-10-19, J36-10-19, B07-20-29, B34-20-29, G51-20-29, H53-20-29, H72-20-29, H74-20-29, J15-20-29 |
| 4 | A41__0-9, C91__0-9, C91-10-19 |
| 5 | B00__0-9, B99__0-9, J05__0-9, J31__0-9, J40__0-9, K52__0-9, B99-10-19, J31-10-19, K08-10-19 |
| 6 | B27__0-9, J03__0-9, L04__0-9 |
| 7 | B00__0-9, B08__0-9, B37__0-9, B99__0-9, H10__0-9, H66__0-9, H70__0-9, H92__0-9, J02__0-9, J22__0-9, K21__0-9 |
| 8 | B96__0-9, N10__0-9, N12__0-9, N10-10-19 |
| 9 | B97__0-9, J10__0-9, J20__0-9, J10-10-19, J98-10-19, L22-10-19 |
| 10 | C71__0-9, C71-10-19, C71-20-29 |
| 11 | D50__0-9, J15__0-9, L22__0-9, J15-10-19 |
| 12 | D64__0-9, D69__0-9, D70__0-9 |
| 13 | D68__0-9, H61__0-9, H65__0-9, H74__0-9, J34__0-9, J35__0-9, J39__0-9, H72-10-19 |
| 14 | E10__0-9, E16__0-9, E87__0-9 |
| 15 | E45__0-9, E66__0-9, G47__0-9, H65-10-19, H90-10-19, D27-20-29, D39-20-29, F81-20-29, F93-20-29, H65-20-29, H66-20-29, H90-20-29, I88-20-29, J00-20-29, N83-20-29 |
| 16 | F41__0-9, K02__0-9, K04__0-9, K08__0-9, K10__0-9, K12__0-9, L03__0-9 |
| 17 | F80__0-9, F82__0-9, F83__0-9, F80-10-19, F82-10-19 |
| 18 | F84__0-9, G40__0-9, G41__0-9, G81__0-9, G91__0-9, F84-10-19, G41-10-19, G81-10-19, F84-20-29, G81-20-29 |
| 19 | F89__0-9, F89-10-19, F89-20-29 |
| 20 | F98__0-9, J06__0-9, J11__0-9, J38__0-9, L50__0-9, J11-10-19, J38-10-19 |
| 21 | G80__0-9, G82__0-9, M62__0-9, G80-10-19, G82-10-19, M62-10-19, M62-20-29 |
| 22 | G81__0-9, G81-10-19, G81-20-29 |
| 23 | H50__0-9, H52__0-9, H53__0-9, H50-10-19, H52-10-19, H50-20-29 |
| 24 | H66__0-9, H70__0-9, H92__0-9, J22__0-9 |
| 25 | H73__0-9, J00__0-9, J04__0-9, J12__0-9, J38__0-9, L50__0-9 |
| 26 | H90__0-9, H91__0-9, H90-10-19, H90-20-29 |

| | |
|---|---|
| 27 | J05__0-9, J31__0-9, K52__0-9, J31-10-19, K08-10-19 |
| 28 | J30__0-9, J45__0-9, J46__0-9 |
| 29 | J05__0-9, J31__0-9, J40__0-9, B99-10-19, K08-10-19 |
| 30 | K90__0-9, K90-10-19, K90-20-29 |
| 31 | L20__0-9, L20-10-19, L20-20-29, L20-30-39, L20-40-49, L20-50-59 |
| 32 | N11__0-9, N30__0-9, N35__0-9, N39__0-9, N76__0-9 |
| 33 | N13__0-9, N28__0-9, N31__0-9 |
| 34 | 000__0-9, A02-10-19, A16-10-19, A38-10-19, A68-10-19, A69-10-19, A87-10-19, B01-10-19, B07-10-19, B09-10-19, B35-10-19, B80-10-19, B95-10-19, C74-10-19, D10-10-19, D57-10-19, D59-10-19, D61-10-19, D70-10-19, D80-10-19, E30-10-19, E45-10-19, F95-10-19, G03-10-19, G04-10-19, G47-10-19, G71-10-19, H04-10-19, H05-10-19, H10-10-19, H35-10-19, H49-10-19, H51-10-19, H60-10-19, H68-10-19, H69-10-19, H70-10-19, H73-10-19, H92-10-19, I78-10-19, J02-10-19, J05-10-19, J09-10-19, J12-10-19, J13-10-19, J16-10-19, J21-10-19, J22-10-19, J39-10-19, J40-10-19, J42-10-19, J46-10-19, J90-10-19, J91-10-19, K00-10-19, K10-10-19, K12-10-19, K13-10-19, K14-10-19, K40-10-19, K42-10-19, K43-10-19, L01-10-19, L03-10-19, L04-10-19, L92-10-19, L95-10-19, M02-10-19, M12-10-19, M13-10-19, M60-10-19, N04-10-19, N11-10-19, N32-10-19, N35-10-19 |
| 35 | B00-10-19, J31-20-29, J35-20-29, K01-30-39, K07-30-39, K07-40-49 |
| 36 | B27-10-19, J03-10-19, J36-10-19 |
| 37 | D27-10-19, N70-10-19, N73-10-19, N83-10-19 |
| 38 | D69-10-19, D69-20-29, D69-30-39 |
| 39 | E10__0-9, E16__0-9, E06-10-19, E10-10-19, E14-10-19, E16-10-19, E06-20-29, E16-20-29 |
| 40 | E66-10-19, E78-10-19, F17-10-19, I10-10-19, K76-10-19, K80-10-19, K76-20-29 |
| 41 | E73-10-19, E74-10-19, E73-20-29, E74-20-29 |
| 42 | E84-10-19, E84-20-29, K86-20-29, E84-30-39 |
| 43 | E87-10-19, F10-10-19, F12-10-19 |
| 44 | F11-10-19, F13-10-19, F19-10-19 |
| 45 | F23-10-19, F43-10-19, F70-10-19, F81-10-19, F90-10-19, F93-10-19, F90-20-29 |
| 46 | F32-10-19, F33-10-19, F41-10-19, F44-10-19, F45-10-19, F93-10-19, L98-10-19 |
| 47 | F41-10-19, F45-10-19, F41-20-29 |
| 48 | F60-10-19, F91-10-19, F92-10-19, F92-20-29 |
| 49 | J96__0-9, G40-10-19, G41-10-19, J96-10-19, G41-20-29, J96-20-29 |
| 50 | H83-10-19, H91-10-19, H93-10-19 |
| 51 | I10-10-19, I10-20-29, N18-20-29 |
| 52 | J20-10-19, J00-20-29, J20-20-29 |
| 53 | J30__0-9, J45__0-9, J46__0-9, L20__0-9, J30-10-19, J45-10-19, L20-10-19 |
| 54 | J32-10-19, J34-10-19, M95-10-19 |
| 55 | K02-10-19, K04-10-19, I88-20-29 |
| 56 | K21-10-19, K29-10-19, K44-10-19, K92-10-19 |
| 57 | D50-10-19, K50-10-19, K52-10-19, N92-10-19 |
| 58 | M21-10-19, M21-20-29, M21-30-39 |
| 59 | M22-10-19, M23-10-19, M25-10-19, M65-10-19, M67-10-19 |
| 60 | M41-10-19, M51-10-19, M54-10-19 |
| 61 | N12-10-19, N30-10-19, N39-10-19, N70-10-19, N73-10-19, N76-10-19 |
| 62 | N10-10-19, N13-10-19, N20-10-19, N23-10-19 |
| 63 | A08-10-19, E86-10-19, A08-20-29, A09-20-29, E86-20-29 |
| 64 | A63-20-29, D06-20-29, N87-20-29 |

| | |
|---|---|
| 65 | B17-20-29, B18-20-29, F10-20-29, F11-20-29, F12-20-29, F13-20-29, F14-20-29, F19-20-29 |
| 66 | C73-20-29, E04-20-29, E05-20-29, E89-20-29 |
| 67 | D24-20-29, D48-20-29, N60-20-29 |
| 68 | D25-20-29, N84-20-29, N85-20-29, N92-20-29 |
| 69 | D27-20-29, K66-20-29, N73-20-29, N80-20-29, N94-20-29 |
| 70 | D27-20-29, D39-20-29, K66-20-29, N73-20-29, N80-20-29, N83-20-29, N94-20-29 |
| 71 | E03__0-9, E03-10-19, D50-20-29, D64-20-29, E03-20-29, E03-30-39, E06-30-39, E06-40-49 |
| 72 | D68-20-29, D68-30-39, D68-40-49 |
| 73 | E03__0-9, E03-10-19, E03-20-29, E03-30-39, E06-30-39, E06-40-49 |
| 74 | E06-10-19, E10-10-19, E14-10-19, E16-10-19, E06-20-29, E10-20-29, E14-20-29, E16-20-29 |
| 75 | E11-20-29, E66-20-29, G47-20-29, M51-20-29, E65-30-39, L02-30-39 |
| 76 | F20-20-29, F23-20-29, F25-20-29 |
| 77 | F31-20-29, F31-30-39, F31-40-49 |
| 78 | F32-20-29, F42-20-29, I95-20-29, I95-30-39, I95-40-49 |
| 79 | F61-20-29, F61-30-39, F61-40-49, F61-50-59 |
| 80 | F70-30-39, F70-40-49, F70-50-59 |
| 81 | F44-20-29, G40-20-29, F44-30-39, G40-30-39, F44-40-49, J15-40-49, F44-50-59 |
| 82 | F44-10-19, F45-10-19, F45-20-29, F45-30-39, F45-40-49 |
| 83 | F50-20-29, M81-20-29, M81-30-39, N91-30-39, M81-40-49 |
| 84 | G43-10-19, G44-10-19, G43-20-29, G44-20-29 |
| 85 | G82-20-29, G82-30-39, G82-40-49, G82-50-59 |
| 86 | I26-20-29, I80-20-29, I82-20-29 |
| 87 | I95-20-29, I95-30-39, I95-40-49, I95-50-59 |
| 88 | J01-20-29, J32-20-29, J33-20-29 |
| 89 | G35-20-29, H46-20-29, F32-30-39, F34-30-39, F48-30-39, G35-30-39, M50-30-39 |
| 90 | L20-30-39, L20-40-49, L20-50-59 |
| 91 | K01-20-29, K02-20-29, K04-20-29, K02-30-39 |
| 92 | K21-10-19, K44-10-19, K25-20-29, K29-20-29, K31-20-29, K58-20-29, K63-20-29 |
| 93 | D27-20-29, K66-20-29, N70-20-29, N71-20-29, N73-20-29, N76-20-29, N80-20-29, N94-20-29 |
| 94 | K51-20-29, K51-30-39, K51-40-49, K51-50-59, K51-60-69, K51-70-79 |
| 95 | B37-10-19, K59-20-29, K59-30-39, K59-40-49 |
| 96 | I84-20-29, K60-20-29, K61-20-29, K62-20-29 |
| 97 | K80-20-29, K81-20-29, K82-20-29, K83-20-29, K85-20-29 |
| 98 | L93-20-29, M32-20-29, M32-30-39 |
| 99 | M22-20-29, M23-20-29, M24-20-29, M65-20-29, M94-20-29 |
| 100 | M51-20-29, G55-30-39, M43-30-39, M48-30-39, M51-30-39 |
| 101 | N10-20-29, N13-20-29, N20-20-29, N23-20-29 |
| 102 | C53-30-39, D06-30-39, N87-30-39 |
| 103 | B17-20-29, B18-20-29, F10-20-29, F11-20-29, F12-20-29, F13-20-29, F14-20-29, F19-20-29, B18-30-39, F11-30-39, F12-30-39, F13-30-39, F19-30-39, F12-40-49 |
| 104 | B24-30-39, B24-40-49, B24-50-59 |
| 105 | D18-30-39, K25-30-39, K26-30-39, K29-30-39, K31-30-39, K57-30-39, K58-30-39 |

| | |
|---|---|
| 106 | C50-30-39, C77-30-39, C78-30-39, C79-30-39, D05-30-39 |
| 107 | C71__0-9, C71-10-19, C71-20-29, C71-30-39, D43-30-39 |
| 108 | C73-30-39, E04-30-39, E05-30-39, E89-30-39 |
| 109 | C77-30-39, C78-30-39, C79-30-39 |
| 110 | D21-30-39, D25-30-39, N84-30-39, N85-30-39, N88-30-39, N92-30-39, N93-30-39 |
| 111 | D23-30-39, I83-30-39, I87-30-39 |
| 112 | D24-30-39, D48-30-39, N60-30-39 |
| 113 | D35-30-39, D35-40-49, D35-50-59 |
| 114 | N80-30-39, N84-30-39, N85-30-39, N88-30-39, N99-30-39, N99-40-49 |
| 115 | D50-20-29, D64-20-29, E03-20-29, K51-20-29, D50-30-39, D64-30-39, E03-30-39, E06-30-39, K51-30-39 |
| 116 | D86-30-39, D86-40-49, D86-50-59 |
| 117 | E10-20-29, E14-20-29, E10-30-39, E11-30-39, E14-30-39 |
| 118 | E73-30-39, E74-30-39, E73-40-49, E74-40-49 |
| 119 | E78-30-39, E79-30-39, I21-30-39, I25-30-39, I63-30-39 |
| 120 | F31-20-29, F10-30-39, F31-30-39, K70-30-39, K71-30-39, K71-40-49 |
| 121 | F14-30-39, F11-40-49, F11-50-59 |
| 122 | F17-30-39, J20-30-39, J44-30-39, J43-40-49, J44-40-49, J96-40-49 |
| 123 | F33-30-39, F40-30-39, F42-30-39, F42-40-49, F42-50-59 |
| 124 | F42-30-39, F42-40-49, F42-50-59 |
| 125 | E87-20-29, E87-30-39, F50-30-39, E87-40-49 |
| 126 | F07-40-49, G40-40-49, G93-40-49, G41-50-59, G93-50-59 |
| 127 | G43-10-19, G44-10-19, G43-20-29, G44-20-29, G43-30-39, G44-30-39, G43-40-49, G44-40-49, G45-40-49, H81-40-49 |
| 128 | E66-30-39, G47-30-39, K42-30-39, K43-30-39, E65-40-49 |
| 129 | M53-30-39, M62-30-39, M99-30-39 |
| 130 | H33-30-39, H50-30-39, H52-30-39 |
| 131 | H90-30-39, H93-30-39, H93-40-49 |
| 132 | I10-30-39, I11-30-39, I20-30-39, I47-30-39, I63-30-39 |
| 133 | I47-30-39, I47-40-49, I49-40-49, I47-50-59, I49-50-59 |
| 134 | I73-30-39, I70-40-49, I73-40-49, M34-40-49, M34-50-59 |
| 135 | I84-30-39, K60-30-39, K61-30-39, K62-30-39 |
| 136 | J01-30-39, J32-30-39, J33-30-39, J01-40-49, J32-40-49, J33-40-49 |
| 137 | J03-30-39, J35-30-39, J36-30-39 |
| 138 | J18-30-39, J90-30-39, J96-30-39 |
| 139 | J44-30-39, J45-30-39, J43-40-49, J44-40-49, J45-40-49, J96-40-49, J06-50-59 |
| 140 | J44-30-39, J43-40-49, J44-40-49, J96-40-49 |
| 141 | K01-30-39, K07-30-39, K07-40-49 |
| 142 | K21-30-39, K22-30-39, K44-30-39, K21-40-49, K22-40-49, K44-40-49 |
| 143 | K66-30-39, N70-30-39, N71-30-39, N76-30-39 |
| 144 | N84-40-49, N85-40-49, N88-40-49, N93-40-49, N95-40-49 |
| 145 | K44-30-39, D37-40-49, K44-40-49 |
| 146 | K50-30-39, K56-30-39, K63-30-39, D37-40-49, K56-40-49, K63-40-49, K65-40-49 |

| | |
|---|---|
| 147 | K76-30-39, K76-40-49, N28-40-49, B37-50-59 |
| 148 | K80-30-39, K81-30-39, K82-30-39, K83-30-39 |
| 149 | K85-40-49, K86-40-49, K85-50-59, K86-50-59 |
| 150 | L40-30-39, L40-40-49, L40-50-59 |
| 151 | L93-30-39, L93-40-49, L93-50-59, L93-60-69, M32-60-69, L93-70-79, M32-70-79 |
| 152 | M06-30-39, M05-40-49, M06-40-49, M13-50-59 |
| 153 | M17-30-39, M22-30-39, M23-30-39, M24-30-39, M65-30-39, M67-30-39, M94-30-39 |
| 154 | M19-30-39, M65-30-39, M75-30-39, M65-40-49, M65-50-59, M72-50-59 |
| 155 | N10-20-29, N13-20-29, N20-20-29, N23-20-29, N10-30-39, N13-30-39, N20-30-39, N23-30-39, N10-40-49 |
| 156 | E21-30-39, N18-30-39, N19-30-39, E21-40-49, I15-40-49, N18-40-49, N19-40-49, I15-50-59, N17-50-59, N19-50-59 |
| 157 | N80-30-39, N84-30-39, N85-30-39, N88-30-39, N92-30-39, N93-30-39, N99-30-39, N99-40-49 |
| 158 | A46-40-49, A46-50-59, B35-50-59 |
| 159 | B16-40-49, B17-50-59, B18-50-59, B17-60-69, B18-60-69, K73-60-69, B18-70-79 |
| 160 | G83-50-59, H10-50-59, M54-60-69, M70-60-69, M76-60-69, J04-70-79, J31-70-79, M76-70-79 |
| 161 | C18-40-49, C18-50-59, C19-50-59 |
| 162 | C43-40-49, C43-50-59, C43-60-69, C43-70-79 |
| 163 | C48-40-49, C56-40-49, D39-40-49 |
| 164 | C50-40-49, D05-40-49, D17-40-49, D24-40-49, D48-40-49, N60-40-49 |
| 165 | C73-40-49, E89-40-49, C73-50-59, E89-50-59 |
| 166 | C82-40-49, C83-40-49, C85-40-49 |
| 167 | C53-30-39, D06-30-39, C53-40-49, C54-40-49, C55-40-49, D06-40-49, N87-40-49 |
| 168 | D12-40-49, I84-40-49, K64-50-59 |
| 169 | D17-40-49, D24-40-49, D48-40-49, N60-40-49 |
| 170 | D21-40-49, D25-40-49, D27-40-49, D62-40-49, K36-40-49 |
| 171 | C43-40-49, D22-40-49, C43-50-59, D22-50-59 |
| 172 | I83-40-49, I87-40-49, I83-50-59, I87-50-59 |
| 173 | D26-40-49, N84-40-49, N85-40-49, N88-40-49, N92-40-49, N93-40-49, N95-40-49 |
| 174 | D34-40-49, D44-40-49, E04-40-49, E55-40-49, E83-40-49, J38-40-49 |
| 175 | K51-20-29, K51-30-39, D50-40-49, E61-40-49, K51-40-49, E61-50-59, K51-50-59, K51-60-69, K51-70-79 |
| 176 | D53-40-49, F10-40-49, G62-40-49, I85-40-49, K71-40-49, K74-40-49, D53-50-59, F12-50-59, G62-50-59, K71-50-59 |
| 177 | D69-10-19, D69-20-29, D69-30-39, D69-40-49, D70-40-49 |
| 178 | H25-40-49, H26-40-49, H33-40-49, H35-40-49, H33-50-59 |
| 179 | E13-40-49, E11-50-59, E12-50-59, E13-50-59, G63-50-59, L03-50-59, G63-60-69, M86-60-69, N08-60-69 |
| 180 | E65-40-49, L90-40-49, N62-40-49 |
| 181 | E66-40-49, I89-40-49, K42-40-49, K43-40-49, M16-40-49, E65-50-59, K42-50-59, K43-50-59 |
| 182 | E73-40-49, E74-40-49, E73-50-59, E74-50-59 |
| 183 | M42-40-49, M43-40-49, M48-40-49 |
| 184 | F48-40-49, G25-40-49, G47-40-49, F48-50-59, G25-50-59, G47-50-59 |
| 185 | F17-40-49, G45-40-49, J38-40-49, J40-40-49, J43-40-49, J98-40-49 |
| 186 | F20-20-29, F23-20-29, F25-20-29, F20-30-39, F23-30-39, F25-30-39, F20-40-49, F25-40-49 |
| 187 | F20-50-59, F25-50-59, F20-60-69, F25-60-69, F20-70-79, F25-70-79 |

| | |
|---|---|
| 188 | F79-50-59, F79-60-69, F79-70-79 |
| 189 | N31-30-39, B00-40-49, G35-40-49, N31-40-49, B00-50-59, G35-50-59, N31-50-59, N31-60-69 |
| 190 | G25-40-49, G47-40-49, G25-50-59, G47-50-59 |
| 191 | M51-20-29, G55-30-39, M43-30-39, M48-30-39, M51-30-39, G55-40-49, G57-40-49, M51-40-49, M96-40-49 |
| 192 | G55-40-49, M42-40-49, M43-40-49, M48-40-49, M50-40-49 |
| 193 | M51-20-29, G55-30-39, M43-30-39, M48-30-39, M51-30-39, G54-40-49, G55-40-49, G57-40-49, M40-40-49, M51-40-49, M54-40-49, M96-40-49, G54-50-59, M96-50-59, M96-60-69 |
| 194 | G58-40-49, M40-50-59, M54-50-59, M70-50-59, M76-50-59, M80-50-59, M93-50-59, D16-60-69, D36-60-69, F54-60-69, G83-60-69, H65-60-69, K04-60-69, M40-60-69 |
| 195 | G70-40-49, G70-50-59, G70-60-69 |
| 196 | G80-40-49, G80-50-59, G80-60-69 |
| 197 | G91-40-49, I60-40-49, I61-40-49 |
| 198 | H02-40-49, H52-40-49, H53-40-49 |
| 199 | H80-40-49, H90-40-49, H80-50-59 |
| 200 | H81-40-49, H91-40-49, H93-40-49, H81-50-59, H90-50-59, H91-50-59, H93-50-59, H81-60-69, H90-60-69, H91-60-69, H93-60-69, H91-70-79, H93-70-79 |
| 201 | M35-30-39, I10-40-49, I11-40-49, I15-40-49, I27-40-49, I35-40-49, I44-40-49, I51-40-49, I74-40-49, I77-40-49, J15-40-49, L30-40-49, M13-40-49, M35-40-49 |
| 202 | I21-40-49, I25-40-49, I42-40-49, I50-40-49, I20-50-59, I21-50-59, I24-50-59 |
| 203 | I20-40-49, I21-40-49, I25-40-49, I42-40-49, I50-40-49, I20-50-59, I21-50-59, I24-50-59, I25-50-59, I42-50-59, I44-50-59, I50-50-59, I51-50-59 |
| 204 | I25-40-49, I42-40-49, I50-40-49, I42-50-59, I44-50-59, I50-50-59, I51-50-59, I42-60-69, I44-60-69, I44-70-79 |
| 205 | I26-40-49, I80-40-49, I82-40-49 |
| 206 | I42-40-49, I50-40-49, I42-50-59, I44-50-59, I44-60-69, I44-70-79 |
| 207 | I45-40-49, I45-50-59, I45-60-69, I45-70-79 |
| 208 | E78-40-49, E79-40-49, G45-40-49, I20-40-49, I21-40-49, I44-40-49, I51-40-49, M41-40-49, E79-50-59, M41-50-59 |
| 209 | I63-40-49, I64-40-49, I65-40-49, G46-50-59, I63-50-59, I64-50-59, I65-50-59 |
| 210 | I67-40-49, I72-40-49, I67-50-59 |
| 211 | I69-40-49, G91-50-59, I60-50-59, I61-50-59, I69-50-59, G91-60-69, I60-60-69 |
| 212 | I85-40-49, K74-40-49, I85-50-59, K74-50-59 |
| 213 | A41-40-49, J18-40-49, J90-40-49, J96-40-49, N17-40-49 |
| 214 | J34-40-49, J35-40-49, M95-40-49 |
| 215 | K51-40-49, K51-50-59, K51-60-69, K51-70-79 |
| 216 | K31-40-49, K31-50-59, K31-60-69 |
| 217 | F60-50-59, F60-60-69, F61-60-69, F60-70-79 |
| 218 | K42-40-49, K43-40-49, K42-50-59, K43-50-59 |
| 219 | D21-30-39, D25-30-39, D27-30-39, K66-30-39, N70-30-39, N71-30-39, N73-30-39, N80-30-39, N83-30-39, N84-30-39, N85-30-39, N88-30-39, N92-30-39, N93-30-39, N94-30-39, N99-30-39, K66-40-49, N70-40-49, N73-40-49 |
| 220 | K58-40-49, K58-50-59, K58-60-69, K58-70-79 |
| 221 | K60-40-49, K61-40-49, K60-50-59, K61-50-59 |
| 222 | K80-40-49, K81-40-49, K82-40-49, K83-40-49 |
| 223 | K83-40-49, C25-50-59, K83-50-59, K83-60-69, K83-70-79 |
| 224 | L02-40-49, L02-50-59, L02-60-69 |
| 225 | D21-50-59, D25-50-59, D27-50-59, D62-50-59, N70-50-59, N73-50-59, N83-50-59, A69-60-69, D21-60-69, D27-60-69, G58-60-69, K35-60-69, N35-60-69, N70-60-69, N73-60-69, N83-60-69 |

| # | Codes |
|---|---|
| 226 | C79-60-69, D61-60-69, D70-60-69, M84-60-69 |
| 227 | M16-40-49, M16-50-59, M87-50-59 |
| 228 | M17-40-49, M21-40-49, M22-40-49, M23-40-49, M71-40-49, M93-40-49, M94-40-49, M94-50-59 |
| 229 | M20-30-39, M77-30-39, M20-40-49, M21-40-49, G57-50-59, M20-50-59 |
| 230 | M22-40-49, M23-40-49, M71-40-49, M93-40-49, M94-40-49, M22-50-59, M23-50-59, M71-50-59, M93-50-59, M94-50-59 |
| 231 | M23-40-49, M71-40-49, M93-40-49 |
| 232 | M65-30-39, M24-40-49, M65-40-49, M67-40-49, M24-50-59, M65-50-59, M67-50-59, M72-50-59 |
| 233 | M34-40-49, M34-50-59, M34-60-69, M34-70-79 |
| 234 | M45-40-49, M45-50-59, M45-60-69 |
| 235 | G54-40-49, G55-40-49, M41-40-49, M42-40-49, M43-40-49, M47-40-49, M48-40-49, M50-40-49, M53-40-49, M62-40-49, M99-40-49, E55-50-59, M41-50-59, M42-50-59, M43-50-59, M47-50-59, M48-50-59, M62-50-59, M99-50-59 |
| 236 | N04-40-49, A69-50-59, B99-50-59, E07-50-59, I10-50-59, I31-50-59, I46-50-59, I71-50-59, J03-50-59, J22-50-59, J30-50-59, J41-50-59, J42-50-59, J81-50-59, J84-50-59, K55-50-59, K75-50-59, L27-50-59, L50-50-59, L98-50-59, M86-50-59, N30-50-59, E26-60-69, N03-60-69 |
| 237 | N35-40-49, N39-40-49, N76-40-49, N81-40-49, N10-50-59, N30-50-59 |
| 238 | N13-40-49, N20-40-49, N23-40-49, N13-50-59, N20-50-59, N23-50-59 |
| 239 | E21-30-39, E21-40-49, N18-40-49, N19-40-49, E21-50-59, N17-50-59, N19-50-59 |
| 240 | K66-30-39, N70-30-39, N71-30-39, N73-30-39, N76-30-39, K66-40-49, N70-40-49, N73-40-49 |
| 241 | N83-40-49, N99-40-49, N76-50-59, N99-50-59 |
| 242 | N97-40-49, D13-50-59, K80-50-59, K81-50-59, K82-50-59, N97-50-59, K91-60-69 |
| 243 | A04-50-59, K80-60-69, K81-60-69, K82-60-69, D72-70-79 |
| 244 | A08-50-59, J31-50-59, M46-50-59, A49-60-69, B96-60-69, K29-60-69, K55-60-69, D22-70-79, K20-70-79, L28-70-79, N60-70-79 |
| 245 | B02-50-59, D13-50-59, E78-50-59, E83-50-59, G51-50-59, G58-50-59, I07-50-59, I15-50-59, I71-50-59, I72-50-59, I72-60-69 |
| 246 | B17-50-59, B17-60-69, B18-60-69, B18-70-79 |
| 247 | C16-50-59, C16-60-69, C16-70-79 |
| 248 | C18-40-49, C18-50-59, C19-50-59, C20-50-59, C21-50-59 |
| 249 | C22-50-59, C38-50-59, C49-50-59, C78-50-59, D61-50-59, J91-50-59 |
| 250 | C34-40-49, D38-40-49, C34-50-59, D38-50-59, J82-50-59 |
| 251 | C48-40-49, C56-40-49, D39-40-49, C48-50-59, C56-50-59, C57-50-59 |
| 252 | C50-50-59, D05-50-59, D24-50-59, N60-50-59, N62-50-59, N64-50-59, N60-60-69 |
| 253 | C53-70-79, C54-70-79, C55-70-79 |
| 254 | C53-50-59, C54-50-59, C55-50-59, C53-60-69, C54-60-69, C55-60-69, C53-70-79 |
| 255 | C67-50-59, C67-60-69, C67-70-79, C68-70-79 |
| 256 | C71-40-49, D43-40-49, C71-50-59, D43-50-59 |
| 257 | C82-50-59, C82-60-69, C82-70-79 |
| 258 | C83-50-59, C85-50-59, C85-60-69 |
| 259 | C90-50-59, C90-60-69, C90-70-79 |
| 260 | C91-50-59, C91-60-69, C91-70-79 |
| 261 | C92-50-59, C92-60-69, C92-70-79 |
| 262 | D13-50-59, K80-50-59, K81-50-59, K82-50-59, K91-60-69 |
| 263 | D17-50-59, D17-60-69, D17-70-79 |

| | |
|---|---|
| 264 | D32-50-59, D32-60-69, D32-70-79 |
| 265 | M47-40-49, D34-50-59, D44-50-59, E04-50-59, E21-50-59, E55-50-59, E83-50-59, J38-50-59, E07-60-69 |
| 266 | D47-50-59, D47-60-69, D47-70-79 |
| 267 | D17-50-59, D48-50-59, D17-60-69, D48-60-69, D17-70-79 |
| 268 | D50-50-59, D62-60-69, E61-60-69 |
| 269 | A04-70-79, A09-70-79, A41-70-79, A46-70-79, B35-70-79, K65-70-79 |
| 270 | C50-60-69, D05-60-69, D24-60-69, N64-60-69, I97-70-79 |
| 271 | D68-50-59, D68-60-69, D68-70-79 |
| 272 | D75-50-59, J37-50-59, J44-60-69, J93-60-69, B96-70-79, J93-70-79 |
| 273 | E03-50-59, E10-50-59, E13-60-69, E16-60-69, E13-70-79 |
| 274 | E16-50-59, E10-60-69, E10-70-79, E13-70-79, E16-70-79, G63-70-79 |
| 275 | I99-50-59, A04-60-69, E78-60-69, E83-60-69, G46-60-69, G51-60-69, G58-60-69, I08-60-69, J22-60-69, M10-60-69, M85-60-69, E83-70-79, F50-70-79, N04-70-79, N05-70-79, N62-70-79, N87-70-79 |
| 276 | M16-40-49, I89-50-59, M16-50-59, M87-50-59, I89-60-69, I89-70-79 |
| 277 | F06-50-59, F07-50-59, F06-60-69, F07-60-69 |
| 278 | F10-50-59, F19-50-59, K71-50-59, D52-60-69, E86-60-69, F19-60-69, G93-60-69, I46-60-69, K26-60-69, M80-60-69, G93-70-79, K26-70-79, M80-70-79 |
| 279 | F06-40-49, F48-40-49, F32-50-59, F40-50-59, F44-50-59, F48-50-59, D16-60-69, F11-60-69, F22-60-69, F23-60-69, F44-60-69, F51-60-69, F55-60-69, J00-60-69, J37-60-69, L28-60-69, N89-60-69 |
| 280 | F61-20-29, F61-30-39, F61-40-49, F33-50-59, F50-50-59, F61-50-59, F50-60-69 |
| 281 | G43-30-39, G44-30-39, G43-40-49, G44-40-49, G45-40-49, H81-40-49, G43-50-59, G44-50-59, H81-50-59 |
| 282 | G35-60-69, G35-70-79, N31-70-79 |
| 283 | G20-50-59, G20-60-69, G21-70-79 |
| 284 | F07-40-49, G93-40-49, G40-50-59, G93-50-59, G41-60-69 |
| 285 | G45-50-59, G45-60-69, G45-70-79 |
| 286 | G50-50-59, G50-60-69, G50-70-79 |
| 287 | G56-30-39, G56-40-49, G56-50-59, M18-50-59, M72-50-59 |
| 288 | G81-50-59, G81-60-69, G81-70-79 |
| 289 | G91-50-59, I60-50-59, I61-50-59, G91-60-69, I60-60-69 |
| 290 | H25-40-49, H26-40-49, H33-40-49, H35-40-49, H25-50-59, H26-50-59, H33-50-59 |
| 291 | H02-40-49, H52-40-49, H53-40-49, H02-50-59, H52-50-59, H53-50-59 |
| 292 | H90-50-59, H91-50-59, H93-50-59, H81-60-69, H90-60-69, H91-60-69, H93-60-69, H91-70-79, H93-70-79 |
| 293 | I05-50-59, I05-60-69, I05-70-79 |
| 294 | I07-40-49, I34-40-49, I07-50-59, I34-50-59, I35-50-59, I36-50-59 |
| 295 | I11-50-59, I11-60-69, I11-70-79 |
| 296 | I12-50-59, N18-60-69, N25-60-69, N26-60-69, N18-70-79, N25-70-79, N26-70-79 |
| 297 | I26-40-49, I80-40-49, I82-40-49, I26-50-59, I80-50-59 |
| 298 | I35-50-59, I35-60-69, I71-60-69, I35-70-79, I71-70-79 |
| 299 | I44-50-59, I44-60-69, I44-70-79 |
| 300 | I70-50-59, I77-50-59, I77-60-69 |
| 301 | I82-50-59, I82-60-69, I82-70-79 |
| 302 | I89-50-59, I89-60-69, I89-70-79 |
| 303 | J01-50-59, J32-50-59, J33-50-59, J01-60-69, J32-60-69, J33-60-69, J34-60-69 |

| | |
|---|---|
| 304 | J98-50-59, J98-60-69, J98-70-79 |
| 305 | K22-50-59, K22-60-69, K22-70-79 |
| 306 | J38-50-59, J38-60-69, J38-70-79 |
| 307 | J43-50-59, J96-50-59, J43-60-69, J96-60-69, J15-70-79, J43-70-79, J96-70-79 |
| 308 | J84-50-59, J84-60-69, J84-70-79 |
| 309 | K21-50-59, K22-50-59, K26-50-59, K22-60-69, K22-70-79 |
| 310 | K42-60-69, K43-60-69, K66-60-69, K42-70-79, K43-70-79, K66-70-79 |
| 311 | C48-50-59, K56-50-59, K65-50-59, K66-50-59 |
| 312 | K59-30-39, E65-40-49, K59-40-49, L90-40-49, N62-40-49, K59-50-59, L90-50-59 |
| 313 | K63-50-59, K65-50-59, D13-60-69 |
| 314 | K90-50-59, K90-60-69, K90-70-79 |
| 315 | M16-40-49, E66-50-59, I89-50-59, L30-50-59, M16-50-59, M87-50-59, E65-60-69, I89-60-69, K45-60-69, L27-60-69, N62-60-69, I89-70-79 |
| 316 | K59-60-69, L90-60-69, L90-70-79 |
| 317 | L93-50-59, L93-60-69, M32-60-69, L93-70-79, M32-70-79 |
| 318 | L97-50-59, L97-60-69, L97-70-79, L98-70-79 |
| 319 | M17-40-49, M21-40-49, M22-40-49, M23-40-49, M71-40-49, M93-40-49, M94-40-49, M17-50-59, M22-50-59, M23-50-59, M71-50-59, M93-50-59, M94-50-59 |
| 320 | M31-50-59, M31-60-69, M31-70-79 |
| 321 | M32-50-59, M32-60-69, M32-70-79 |
| 322 | M35-30-39, M35-40-49, M35-50-59, M35-60-69, M35-70-79 |
| 323 | M81-30-39, M81-40-49, M81-50-59, M85-60-69 |
| 324 | N00-50-59, D30-60-69, D46-60-69, E53-60-69, F05-60-69, G31-60-69, H49-60-69, I10-60-69, I12-60-69, I13-60-69, I62-60-69, I97-60-69, I99-60-69, J02-60-69, J04-60-69, J22-60-69, J30-60-69, J41-60-69, J42-60-69, J69-60-69, J94-60-69, J95-60-69, K20-60-69, K73-60-69, K75-60-69, L29-60-69, M10-60-69, M40-60-69, M46-60-69, N10-60-69, N31-60-69, D26-70-79, E24-70-79, L73-70-79 |
| 325 | E10-40-49, E13-40-49, E14-40-49, E10-50-59, E11-50-59, E12-50-59, E13-50-59, E14-50-59, H43-50-59, G63-50-59, L03-50-59, N08-50-59, A04-60-69, E11-60-69, E12-60-69, E13-60-69, E16-60-69, F01-60-69, H43-60-69, G51-60-69, G63-60-69, I79-60-69, L03-60-69, L89-60-69, L98-60-69, M86-60-69, N08-60-69 |
| 326 | N28-50-59, N28-60-69, N28-70-79 |
| 327 | N35-50-59, N39-50-59, N76-50-59, G83-60-69, N30-60-69 |
| 328 | N70-50-59, N73-50-59, N73-60-69, N83-60-69 |
| 329 | N80-50-59, N81-50-59, N84-50-59, N85-50-59, N88-50-59, N95-50-59, N80-60-69 |
| 330 | N81-50-59, N81-60-69, N99-60-69 |
| 331 | N70-50-59, N73-50-59, N83-50-59, D27-60-69, G58-60-69, N73-60-69, N83-60-69 |
| 332 | N84-50-59, N88-50-59, N95-50-59 |
| 333 | C53-50-59, C54-50-59, C55-50-59, N95-50-59 |
| 334 | N39-60-69, D72-70-79, N32-70-79, N35-70-79 |
| 335 | A18-60-69, A40-60-69, B36-60-69, B90-60-69, C15-60-69, C23-60-69, C32-60-69, C51-60-69, C52-60-69, C69-60-69, C76-60-69, D03-60-69, D04-60-69, D30-60-69, D31-60-69, D46-60-69, D59-60-69, D89-60-69, E20-60-69, E41-60-69, E46-60-69, E53-60-69, E72-60-69, F05-60-69, F09-60-69, G12-60-69, G24-60-69, G31-60-69, G90-60-69, H01-60-69, H17-60-69, H21-60-69, H31-60-69, H49-60-69, H57-60-69, H82-60-69, I06-60-69, I22-60-69, I33-60-69, I81-60-69, I98-60-69, J11-60-69, J12-60-69, J21-60-69, K06-60-69, K14-60-69, K28-60-69, K41-60-69, K46-60-69, L08-60-69, L24-60-69, L53-60-69, L82-60-69, L84-60-69, L85-60-69, M00-60-69, M11-60-69, M82-60-69, N32-60-69, N36-60-69, O00-70-79 |
| 336 | A46-40-49, A46-50-59, B35-50-59, A46-60-69, B35-60-69 |
| 337 | M51-50-59, M96-50-59, A69-60-69, G54-60-69, G57-60-69, M96-60-69, G57-70-79 |
| 338 | D12-40-49, D12-50-59, C18-60-69, C19-60-69, D12-60-69 |

| | |
|---|---|
| 339 | C18-40-49, C18-50-59, C19-50-59, C20-50-59, C21-50-59, C20-60-69, C21-60-69 |
| 340 | C34-60-69, D38-60-69, J82-60-69, C34-70-79, D38-70-79 |
| 341 | C38-60-69, J90-60-69, J91-60-69 |
| 342 | C43-60-69, C44-60-69, L57-60-69, C43-70-79, C44-70-79, L57-70-79 |
| 343 | C48-60-69, C48-70-79, C56-70-79, C57-70-79, D39-70-79 |
| 344 | C38-60-69, C78-60-69, D61-60-69, D70-60-69, J90-60-69, J91-60-69, C80-70-79, D70-70-79 |
| 345 | C53-60-69, C54-60-69, C55-60-69, C53-70-79, C54-70-79, C55-70-79 |
| 346 | C48-60-69, C56-60-69, C57-60-69, D39-60-69, C48-70-79, C56-70-79, C57-70-79, D39-70-79 |
| 347 | C64-60-69, D41-60-69, C64-70-79, D41-70-79 |
| 348 | C71-60-69, D43-60-69, C71-70-79, D43-70-79 |
| 349 | C73-40-49, E89-40-49, C73-50-59, E89-50-59, C73-60-69, E89-60-69, C73-70-79 |
| 350 | C38-60-69, C77-60-69, C78-60-69, C79-60-69, C80-60-69, D61-60-69, D63-60-69, D70-60-69, J90-60-69, J91-60-69, M84-60-69, C23-70-79, C77-70-79, C78-70-79, C79-70-79, C80-70-79, D48-70-79, D63-70-79, D70-70-79, M84-70-79 |
| 351 | D63-50-59, D69-50-59, D70-50-59, D61-60-69, D63-60-69, D69-60-69, D70-60-69, D46-70-79, D63-70-79, D69-70-79 |
| 352 | K50-60-69, A02-70-79, B17-70-79, B34-70-79, D52-70-79, D53-70-79, E07-70-79, E46-70-79, E72-70-79, F09-70-79, G60-70-79, G70-70-79, G83-70-79, G90-70-79, H49-70-79, H61-70-79, I10-70-79, I30-70-79, I99-70-79, J02-70-79, J04-70-79, J11-70-79, J30-70-79, J41-70-79, J94-70-79, K20-70-79, K27-70-79, K28-70-79, K30-70-79, K46-70-79, K50-70-79, K73-70-79, K75-70-79, L08-70-79, L28-70-79, L29-70-79, L50-70-79, L82-70-79, L85-70-79, M00-70-79, M34-70-79, M40-70-79, M46-70-79, N12-70-79 |
| 353 | D18-40-49, D18-50-59, D18-60-69, D18-70-79 |
| 354 | L72-40-49, N92-50-59, D21-60-69, D23-60-69, D72-60-69, E65-60-69, K35-60-69, K91-60-69, L50-60-69, L72-60-69, M84-60-69, N35-60-69, N87-60-69, N94-60-69 |
| 355 | C53-60-69, C54-60-69, C55-60-69, D25-60-69, C53-70-79, C54-70-79, C55-70-79, K81-70-79 |
| 356 | D35-30-39, D35-40-49, D35-50-59, D35-60-69, D35-70-79 |
| 357 | D38-60-69, C34-70-79, D38-70-79 |
| 358 | L25-60-69, D53-70-79, N39-70-79, N76-70-79 |
| 359 | D61-60-69, D69-60-69, D70-60-69, D46-70-79, D61-70-79, D69-70-79 |
| 360 | E21-50-59, E21-60-69, E21-70-79 |
| 361 | D69-60-69, D46-70-79, D61-70-79, D69-70-79 |
| 362 | D86-30-39, D86-40-49, D86-50-59, D86-60-69, D86-70-79 |
| 363 | E13-60-69, E10-70-79, E13-70-79, E16-70-79 |
| 364 | E14-60-69, E12-70-79, E14-70-79, E16-70-79, I79-70-79, L03-70-79, M86-70-79, N08-70-79 |
| 365 | D34-50-59, E01-50-59, J38-50-59, E01-60-69, E04-60-69, E55-60-69, J38-60-69, J38-70-79, L72-70-79, N35-70-79 |
| 366 | E73-60-69, E74-60-69, E73-70-79, E74-70-79 |
| 367 | F00-60-69, F01-60-69, F03-60-69, G30-60-69, F00-70-79, F05-70-79, G30-70-79 |
| 368 | F01-60-69, F03-60-69, G30-60-69 |
| 369 | H43-60-69, H33-60-69, H43-70-79, H18-70-79, H26-70-79, H33-70-79 |
| 370 | F17-60-69, A04-70-79, A09-70-79, A41-70-79, A46-70-79, B35-70-79, B37-70-79, I24-70-79, I77-70-79, J41-70-79, K65-70-79, L30-70-79 |
| 371 | F31-20-29, F31-30-39, F31-40-49, F31-50-59, F31-60-69, F31-70-79 |
| 372 | F32-60-69, F48-60-69, F07-70-79, F19-70-79, F22-70-79, F40-70-79, F44-70-79, F48-70-79, F51-70-79, G58-70-79, L29-70-79, L82-70-79 |
| 373 | H90-50-59, H91-50-59, H93-50-59, H90-60-69, H91-60-69, H93-60-69, H91-70-79, H93-70-79 |
| 374 | H10-50-59, F13-60-69, F33-60-69, F41-60-69, F43-60-69, F45-60-69, F13-70-79, F33-70-79, F41-70-79, F43-70-79, F45-70-79, K59-70-79 |

| | |
|---|---|
| 375 | K20-40-49, K51-40-49, A49-50-59, B96-50-59, K20-50-59, K29-50-59, K51-50-59, H65-60-69, J37-60-69, K51-60-69, N89-60-69, N93-60-69, K51-70-79 |
| 376 | G21-60-69, F02-70-79, G20-70-79 |
| 377 | G81-50-59, G40-60-69, G41-60-69, G81-60-69, G40-70-79, G41-70-79, G81-70-79 |
| 378 | I72-50-59, I67-60-69, I72-60-69, I67-70-79 |
| 379 | M96-50-59, G55-60-69, M51-60-69, M96-60-69 |
| 380 | G56-70-79, M15-70-79, M18-70-79 |
| 381 | G82-30-39, G82-40-49, G82-50-59, G82-60-69, G82-70-79 |
| 382 | G93-60-69, I46-60-69, G93-70-79 |
| 383 | G95-60-69, M48-70-79, M50-70-79 |
| 384 | H10-60-69, H11-60-69, H18-60-69, H25-60-69, H27-60-69, H34-60-69, H44-60-69, H34-70-79 |
| 385 | H40-40-49, H25-50-59, H26-50-59, H40-50-59, H26-60-69 |
| 386 | H34-60-69, H54-60-69, D31-70-79, H43-70-79, H04-70-79, H11-70-79, H16-70-79, H17-70-79, H18-70-79, H25-70-79, H26-70-79, H27-70-79, H33-70-79, H34-70-79, H35-70-79, H44-70-79, H54-70-79 |
| 387 | H40-60-69, H47-60-69, H40-70-79, H47-70-79 |
| 388 | H50-60-69, H52-60-69, H53-60-69, H50-70-79, H52-70-79, H53-70-79 |
| 389 | H90-50-59, H91-50-59, H93-50-59, H90-60-69 |
| 390 | I26-40-49, I80-40-49, I82-40-49, I26-50-59, I80-50-59, I82-50-59, I26-60-69, I80-60-69, I82-60-69, I26-70-79, I82-70-79 |
| 391 | I07-40-49, I34-40-49, I07-50-59, I34-50-59, I35-50-59, I36-50-59, I07-60-69, I34-60-69, I35-60-69, I36-60-69 |
| 392 | I49-40-49, I49-50-59, I49-60-69, I49-70-79 |
| 393 | G46-60-69, I63-60-69, I66-60-69, I63-70-79 |
| 394 | I73-30-39, I70-40-49, I73-40-49, M34-40-49, I73-50-59, I74-50-59, M34-50-59, I74-60-69 |
| 395 | I84-60-69, K60-60-69, K60-70-79, K64-70-79 |
| 396 | I83-40-49, I87-40-49, I83-50-59, I87-50-59, I87-60-69, I87-70-79 |
| 397 | F61-20-29, F61-30-39, F61-40-49, F33-50-59, F43-50-59, F50-50-59, F60-50-59, F61-50-59, B37-60-69, F40-60-69, F50-60-69, F60-60-69, F61-60-69, L30-60-69, F60-70-79 |
| 398 | J32-60-69, J33-60-69, J34-60-69 |
| 399 | J20-50-59, J20-60-69, J20-70-79 |
| 400 | K42-40-49, K43-40-49, K42-50-59, K43-50-59, K42-60-69, K43-60-69, K42-70-79, K43-70-79 |
| 401 | J98-50-59, J45-60-69, J98-60-69, J45-70-79, J96-70-79, J98-70-79 |
| 402 | K59-30-39, E65-40-49, K59-40-49, L90-40-49, N62-40-49, K59-50-59, L90-50-59, K59-60-69, L90-60-69, L90-70-79 |
| 403 | I85-60-69, K70-60-69, K72-60-69, K74-60-69, I85-70-79, K70-70-79, K74-70-79 |
| 404 | K85-60-69, K86-60-69, K85-70-79, K86-70-79 |
| 405 | L23-40-49, L23-50-59, L23-60-69, L23-70-79 |
| 406 | G50-50-59, G50-60-69, E78-70-79, G43-70-79, G50-70-79, G51-70-79, G55-70-79, G57-70-79, G58-70-79, G91-70-79, G95-70-79, H10-70-79, H90-70-79, I06-70-79, I60-70-79, I99-70-79, M40-70-79 |
| 407 | M06-30-39, M05-40-49, M06-40-49, M05-50-59, M06-50-59, M13-50-59, M05-60-69, M06-60-69, M13-60-69, M05-70-79, M06-70-79, M13-70-79 |
| 408 | M17-70-79, M22-70-79, M23-70-79, M24-70-79, M65-70-79, M67-70-79, M71-70-79, M87-70-79, M94-70-79 |
| 409 | G57-60-69, M20-60-69, M77-60-69, G57-70-79 |
| 410 | M17-60-69, M21-60-69, M22-60-69, M23-60-69, M24-60-69, M67-60-69, M71-60-69, M94-60-69, M00-70-79, M17-70-79, M21-70-79, M22-70-79 |
| 411 | G95-60-69, M41-60-69, M42-60-69, M43-60-69, M48-60-69, M50-60-69, M62-60-69, M99-60-69, G54-70-79, G95-70-79, M41-70-79, M42-70-79, M43-70-79, M47-70-79, M48-70-79, M50-70-79, M51-70-79, M53-70-79, M54-70-79, M62-70-79, M70-70-79, M93-70-79, M96-70-79, M99-70-79 |

| | |
|---|---|
| 412 | G95-60-69, M48-60-69, M50-60-69, G95-70-79, M48-70-79, M50-70-79 |
| 413 | N13-60-69, N20-60-69, N23-60-69, N13-70-79, N20-70-79, N23-70-79 |
| 414 | N26-40-49, N26-50-59, N26-60-69, N26-70-79 |
| 415 | C50-70-79, D05-70-79, D24-70-79, L90-70-79, N64-70-79 |
| 416 | D17-50-59, N84-60-69, N88-60-69 |
| 417 | C48-60-69, C18-70-79, C19-70-79, C20-70-79, C21-70-79, C48-70-79, D12-70-79 |
| 418 | C19-70-79, C20-70-79, C21-70-79 |
| 419 | C77-60-69, C80-60-69, D63-60-69, D70-60-69, C23-70-79, C77-70-79, C78-70-79, C79-70-79, C80-70-79, D48-70-79, D63-70-79, D70-70-79, M84-70-79 |
| 420 | C43-70-79, C44-70-79, L57-70-79 |
| 421 | C83-50-59, C85-50-59, C83-60-69, C85-60-69, C83-70-79, C85-70-79 |
| 422 | D25-70-79, N81-70-79, N99-70-79 |
| 423 | H34-60-69, D31-70-79, H04-70-79, H11-70-79, H16-70-79, H17-70-79, H25-70-79, H27-70-79, H34-70-79, H44-70-79 |
| 424 | D50-70-79, D62-70-79, E61-70-79 |
| 425 | F31-60-69, A49-70-79, E11-70-79, E12-70-79, E53-70-79, F31-70-79, G51-70-79, I12-70-79, I13-70-79, I62-70-79, K71-70-79, K72-70-79, L02-70-79, N08-70-79, N10-70-79 |
| 426 | E79-40-49, E79-50-59, E79-60-69, E79-70-79, M10-70-79 |
| 427 | E88-40-49, E88-50-59, E88-60-69, E88-70-79 |
| 428 | F03-70-79, F13-70-79, F20-70-79, F22-70-79, G31-70-79 |
| 429 | H43-70-79, H18-70-79, H26-70-79, H33-70-79 |
| 430 | G40-70-79, G41-70-79, G81-70-79 |
| 431 | M96-50-59, G55-60-69, G95-60-69, M48-60-69, M50-60-69, M51-60-69, M96-60-69, G54-70-79, G55-70-79, G95-70-79, M47-70-79, M48-70-79, M50-70-79, M51-70-79, M54-70-79, M70-70-79, M93-70-79, M96-70-79 |
| 432 | H02-60-69, H02-70-79, H18-70-79, H26-70-79 |
| 433 | I21-70-79, I24-70-79, I46-70-79 |
| 434 | A49-70-79, I25-70-79, I31-70-79, I48-70-79, I50-70-79, I62-70-79, I72-70-79, I97-70-79, J06-70-79, J22-70-79, J42-70-79, J81-70-79, J90-70-79, J91-70-79, J95-70-79, L27-70-79, M10-70-79 |
| 435 | I07-40-49, I34-40-49, I07-50-59, I34-50-59, I35-50-59, I36-50-59, I07-60-69, I34-60-69, I35-60-69, I36-60-69, I07-70-79, I34-70-79, I36-70-79 |
| 436 | I50-60-69, J81-60-69, I50-70-79, J22-70-79, J90-70-79, J91-70-79, J95-70-79 |
| 437 | J44-70-79, J47-70-79, J93-70-79 |
| 438 | K85-60-69, K86-60-69, C23-70-79, K80-70-79, K81-70-79, K82-70-79, K85-70-79, K86-70-79 |
| 439 | L40-30-39, L40-40-49, L40-50-59, L40-60-69, L40-70-79 |
| 440 | G55-60-69, G95-60-69, M41-60-69, M43-60-69, M48-60-69, M50-60-69, M51-60-69, M53-60-69, M62-60-69, M99-60-69, G54-70-79, G55-70-79, G95-70-79, M16-70-79, M25-70-79, M41-70-79, M42-70-79, M43-70-79, M47-70-79, M48-70-79, M50-70-79, M51-70-79, M53-70-79, M54-70-79, M62-70-79, M70-70-79, M87-70-79, M93-70-79, M96-70-79, M99-70-79 |
| 441 | M19-70-79, M20-70-79, M77-70-79 |
| 442 | G95-60-69, M47-70-79, M48-70-79, M50-70-79, M93-70-79 |
| 443 | G35-50-59, N31-50-59, G35-60-69, N31-60-69, G35-70-79, N31-70-79 |
| 444 | D25-70-79, D27-70-79, N81-70-79, N83-70-79, N84-70-79, N85-70-79, N99-70-79 |

Table S6: All identified diverging trajectories in males.

| Community_1 | Community_2 | |
|---|---|---|
| N32-40-49, N35-40-49, N32-50-59, N35-50-59 | N32-40-49, N40-40-49, N41-40-49, N21-50-59, N40-50-59, N41-50-59, N42-50-59, N41-60-69, N42-60-69 | Diverge |

| | | |
|---|---|---|
| D52-40-49, F54-40-49, F55-40-49, J41-40-49, L71-40-49, F17-50-59, J41-50-59, J42-50-59, J95-50-59, J13-60-69, J39-60-69, K04-60-69, K08-60-69, K12-60-69, L08-60-69, N62-60-69 | L71-40-49, K70-50-59, D53-60-69, F05-60-69, G31-60-69, G31-70-79 | Diverge |
| N25-40-49, N26-40-49, N18-50-59, N25-50-59, N26-50-59, A04-60-69, B02-60-69, N03-60-69, N26-60-69, N26-70-79 | F54-40-49, N26-40-49, I25-50-59, I97-50-59, I13-60-69, I30-60-69, I33-60-69, K04-60-69 | Diverge |
| A02-60-69, A18-60-69, A26-60-69, A40-60-69, A48-60-69, A69-60-69, A84-60-69, B07-60-69, B36-60-69, B44-60-69, B49-60-69, B86-60-69, B91-60-69, C00-60-69, C06-60-69, C11-60-69, C12-60-69, C14-60-69, C26-60-69, C39-60-69, C41-60-69, C45-60-69, C50-60-69, C60-60-69, C66-60-69, C69-60-69, C76-60-69, C84-60-69, C88-60-69, D01-60-69, D03-60-69, D04-60-69, D07-60-69, D11-60-69, D15-60-69, D31-60-69, D34-60-69, D51-60-69, D89-60-69, E07-60-69, E22-60-69, E23-60-69, E27-60-69, E41-60-69, E46-60-69, E74-60-69, E85-60-69, F52-60-69, G06-60-69, G24-60-69, G52-60-69, G90-60-69, G92-60-69, H21-60-69, H31-60-69, H55-60-69, H61-60-69, H70-60-69, J02-60-69, J10-60-69, J11-60-69, J12-60-69, J21-60-69, J61-60-69, J85-60-69, K11-60-69, K28-60-69, K41-60-69, L25-60-69, L58-60-69, L60-60-69, L80-60-69, L81-60-69, M12-60-69, M60-60-69, N11-60-69, N34-60-69, N36-60-69, N50-60-69, O00-70-79 | B36-60-69, D34-60-69, E72-60-69, F09-60-69, G90-60-69, I10-60-69, I30-60-69, I33-60-69, I60-60-69, I78-60-69, J13-60-69, J30-60-69, K73-60-69, L08-60-69, L21-60-69, L23-60-69, L27-60-69, L29-60-69, L71-60-69, L85-60-69, N02-60-69, N04-60-69, C05-70-79, C74-70-79, D15-70-79, E26-70-79, E29-70-79, F55-70-79, G37-70-79, G96-70-79, H72-70-79, I68-70-79, J35-70-79, K09-70-79, L70-70-79, M14-70-79, N03-70-79, N49-70-79 | Diverge |
| E87-10-19, F10-10-19, F12-10-19, F17-10-19, F20-10-19, F23-10-19, F60-10-19, F60-20-29, F63-20-29 | F20-10-19, F23-10-19, F12-20-29, F15-20-29, F20-20-29, F23-20-29, F25-20-29, F31-20-29, F20-30-39, F23-30-39, F25-30-39, F20-40-49, F25-40-49 | Diverge |
| K29-30-39, K31-30-39, K51-30-39, K58-30-39, K59-30-39, K59-40-49 | K51-30-39, K51-40-49, D73-50-59, J04-50-59, K51-50-59, B96-60-69, K29-60-69, K51-60-69, A69-70-79, D16-70-79, D21-70-79, K51-70-79, L50-70-79, N10-70-79, N62-70-79 | Diverge |
| C61-40-49, N32-40-49, N40-40-49, C61-50-59, D40-50-59, N32-50-59 | N32-40-49, N40-40-49, N41-40-49, N21-50-59, N40-50-59, N41-50-59, N42-50-59, N41-60-69, N42-60-69 | Diverge |
| A41__0-9, C91__0-9, N10__0-9, A41-10-19, C91-10-19 | N10__0-9, N13__0-9, N28__0-9, N28-10-19 | Diverge |
| G35-20-29, E78-30-39, E87-30-39, G35-30-39, G45-30-39, H81-30-39, M42-30-39, M47-30-39, M47-40-49 | G35-20-29, G35-30-39, G35-40-49, N31-40-49, N31-50-59 | Diverge |
| E10-30-39, E16-30-39, H36-30-39, E10-40-49, E16-40-49, H36-40-49 | H33-30-39, H35-30-39, H36-30-39, H43-40-49, H33-40-49, H34-40-49, H35-40-49, H36-40-49, H43-50-59, H33-50-59, H35-50-59, H36-50-59, H43-60-69, H26-60-69, H33-60-69 | Diverge |
| E11-20-29, E66-20-29, E78-20-29, E79-20-29, G47-20-29, G47-30-39 | G47-20-29, G25-30-39, G47-30-39, E66-40-49, G25-40-49, G47-40-49, G25-50-59, H66-50-59, K46-50-59 | Diverge |
| G20-50-59, F02-60-69, G20-60-69, G21-60-69, G21-70-79 | F32-50-59, G20-50-59, G35-50-59, F48-60-69, F61-60-69, G35-60-69, K02-60-69, N47-60-69, N48-60-69, G35-70-79, N47-70-79, N48-70-79 | Diverge |
| E11-20-29, E66-20-29, E78-20-29, E79-20-29, G47-20-29, G47-30-39 | E78-20-29, E79-20-29, E79-30-39 | Diverge |
| B36-60-69, D34-60-69, E72-60-69, F09-60-69, G90-60-69, I10-60-69, I30-60-69, I33-60-69, I60-60-69, I78-60-69, J13-60-69, J30-60-69, K73-60-69, L08-60-69, L21-60-69, L23-60-69, L27-60-69, L29-60-69, L71-60-69, L85-60-69, N02-60-69, N04-60-69, C05-70-79, C74-70-79, D15-70-79, E26-70-79, E29-70-79, F55-70-79, G37-70-79, G96-70-79, H72-70-79, I68-70-79, J35-70-79, K09-70-79, L70-70-79, M14-70-79, N03-70-79, N49-70-79 | F09-60-69, J62-60-69, J92-60-69, J44-70-79, J47-70-79, J62-70-79, K20-70-79 | Diverge |
| D52-40-49, F54-40-49, F55-40-49, J41-40-49, L71-40-49, F17-50-59, J41-50-59, J42-50-59, J95-50-59, J13-60-69, J39-60-69, K04-60-69, K08-60-69, K12-60-69, L08-60-69, N62-60-69 | J41-40-49, J44-50-59, J47-50-59, J95-50-59, B90-60-69, G61-60-69, H10-60-69, J40-60-69, J47-60-69, J82-60-69, J95-60-69 | Diverge |
| N30-50-59, N43-50-59, D29-60-69, D30-60-69, N10-60-69, N39-60-69, N43-60-69, N45-60-69, D30-70-79, H10-70-79, N02-70-79, N39-70-79, N40-70-79, N41-70-79, N42-70-79, N43-70-79, N45-70-79, N99-70-79 | N30-50-59, K63-60-69, C71-70-79, D36-70-79, D43-70-79 | Diverge |

| | | |
|---|---|---|
| L97-40-49, L98-40-49, L97-50-59, L98-50-59 | L98-40-49, M86-40-49, L97-50-59, L98-50-59, M86-50-59, L97-60-69, L98-60-69, M86-60-69, L97-70-79, L98-70-79, M86-70-79 | Diverge |
| M25-30-39, M25-40-49, M66-40-49, M87-40-49 | M19-30-39, M24-30-39, M25-30-39, M65-30-39, M67-30-39, M75-30-39, M19-40-49, M24-40-49, M65-40-49, M67-40-49, M75-40-49, M77-40-49, M66-50-59, M75-50-59, M76-50-59, M77-50-59 | Diverge |
| G40-40-49, G41-40-49, G81-40-49, J69-40-49, F70-50-59, G41-50-59, F70-60-69 | G81-40-49, G81-50-59 | Diverge |
| E87-10-19, F10-10-19, F12-10-19, F17-10-19, F20-10-19, F23-10-19, F60-10-19, F60-20-29, F63-20-29 | F20-10-19, F23-10-19, F20-20-29, F23-20-29, F25-20-29, F20-30-39, F23-30-39, F25-30-39 | Diverge |
| G20-50-59, F02-60-69, G20-60-69, G21-60-69, F02-70-79, G20-70-79, G21-70-79 | F32-50-59, G20-50-59, G35-50-59, F48-60-69, F61-60-69, G35-60-69, K02-60-69, N47-60-69, N48-60-69, G35-70-79, N47-70-79, N48-70-79 | Diverge |
| F17-40-49, G58-40-49, J01-40-49, J37-40-49, J38-40-49, J42-40-49, J43-40-49, J93-40-49, J02-50-59, J37-50-59, J38-50-59, J37-60-69, J37-70-79 | J43-40-49, J43-50-59, J93-60-69 | Diverge |
| F70-10-19, F70-20-29, D53-30-39, F10-30-39, F34-30-39, F70-30-39, D53-40-49, F05-40-49, F34-40-49, F70-40-49, G31-40-49, H10-40-49, K72-40-49, L21-40-49, D53-50-59 | F70-10-19, F70-20-29, D75-30-39, F70-30-39, D52-40-49, F10-40-49, F22-40-49, F23-40-49, F70-40-49, G31-40-49, L21-40-49, C09-50-59, C10-50-59, C13-50-59, E53-50-59, F22-50-59, F55-50-59, H10-50-59, I86-50-59, N62-50-59, F22-60-69, F22-70-79 | Diverge |
| B94-20-29, B16-30-39, B17-30-39, B18-30-39, F11-30-39, F12-30-39, F13-30-39, F14-30-39, F19-30-39, K73-30-39, B17-40-49, B18-40-49, B94-40-49, F11-40-49, F12-40-49, F14-40-49, F19-40-49, K73-40-49, F12-50-59, K73-50-59 | B94-20-29, F17-30-39, G45-30-39, J06-30-39, J15-30-39, J20-30-39, J93-30-39, L05-40-49 | Diverge |
| H33-30-39, H35-30-39, H36-30-39, H43-40-49, H33-40-49, H34-40-49, H35-40-49, H36-40-49, H43-50-59, H33-50-59, H35-50-59, H36-50-59, H43-60-69, H26-60-69, H33-60-69 | E10-30-39, E16-30-39, H36-30-39, E16-40-49 | Diverge |
| E88__0-9, F79__0-9, G25__0-9, G41__0-9, I47__0-9, J69__0-9, G40-10-19, G41-10-19, J96-10-19, J96-20-29 | G40__0-9, G41__0-9, G47__0-9, G81__0-9, G91__0-9, G41-10-19, G47-10-19, G81-10-19, G91-10-19 | Diverge |
| N32-40-49, N40-40-49, N41-40-49, N21-50-59, N40-50-59, N41-50-59, N42-50-59, N41-60-69, N42-60-69 | N41-40-49, D40-50-59, N41-50-59, N42-50-59, D40-60-69, N41-60-69, C61-70-79, C64-70-79, D40-70-79, N41-70-79, N42-70-79 | Diverge |
| B97__0-9, J20__0-9, J30__0-9, J42__0-9, L30__0-9, D80-10-19 | J30__0-9, J30-10-19, J30-20-29 | Diverge |
| E11-20-29, E66-20-29, E78-20-29, E79-20-29, G47-20-29, G47-30-39 | G47-20-29, G25-30-39, G47-30-39, G25-40-49, G47-40-49 | Diverge |
| D12-40-49, I84-40-49, K60-40-49, K61-40-49, K60-50-59, K61-50-59, K64-50-59 | K60-40-49, K61-40-49, I84-50-59, K60-50-59, K61-50-59, D21-60-69 | Diverge |
| I12-40-49, I13-40-49, N25-40-49, D44-50-59, E72-50-59, G61-50-59, I10-50-59, I12-50-59, I13-50-59, I30-50-59, I62-50-59, I99-50-59, J22-50-59, J91-50-59, L23-50-59, L27-50-59, L73-50-59, M35-50-59, M46-50-59, M80-50-59, N05-50-59, C31-60-69, E26-60-69, L05-60-69, N05-60-69, N12-60-69 | N25-40-49, N26-40-49, N18-50-59, N25-50-59, N26-50-59, A04-60-69, B02-60-69, N03-60-69, N26-60-69, N26-70-79 | Diverge |
| G91-40-49, I60-40-49, I61-40-49, G91-50-59, I60-50-59, I61-50-59 | G91-40-49, I60-40-49, D73-50-59, F52-50-59, G30-50-59, G91-50-59, G99-50-59, I60-50-59, L28-50-59, E11-60-69, E55-60-69, F00-60-69, F05-60-69, G30-60-69, G91-60-69, H49-60-69, A26-70-79, A48-70-79, C76-70-79, D34-70-79, G91-70-79, H59-70-79, H60-70-79, L08-70-79, L28-70-79, M00-70-79, N04-70-79 | Diverge |
| E87-10-19, F10-10-19, F12-10-19, F17-10-19, F20-10-19, F23-10-19, F60-10-19, F60-20-29, F63-20-29 | F20-10-19, F23-10-19, F20-20-29, F23-20-29, F25-20-29, F20-30-39, F23-30-39, F25-30-39, F20-40-49, F25-40-49 | Diverge |
| C32-40-49, J37-40-49, J38-40-49, C32-50-59, J37-50-59, J38-50-59 | F17-40-49, G58-40-49, J01-40-49, J37-40-49, J38-40-49, J42-40-49, J43-40-49, J93-40-49, J02-50-59, J37-50-59, J38-50-59, J37-60-69, J37-70-79 | Diverge |

48

| Community_1 | Community_2 | |
|---|---|---|
| M70-50-59, E79-60-69, L82-60-69, M18-60-69, M70-60-69, H61-70-79, L21-70-79, M54-70-79, M70-70-79, M79-70-79 | M70-50-59, M54-60-69, M70-60-69, B00-70-79, G43-70-79, G83-70-79, M76-70-79, M93-70-79 | Diverge |

Table S7: All identified diverging trajectories in females.

| Community_1 | Community_2 | |
|---|---|---|
| F48-40-49, G25-40-49, G47-40-49, F48-50-59, G25-50-59, G47-50-59 | F06-40-49, F48-40-49, F32-50-59, F40-50-59, F44-50-59, F48-50-59, D16-60-69, F11-60-69, F22-60-69, F23-60-69, F44-60-69, F51-60-69, F55-60-69, J00-60-69, J37-60-69, L28-60-69, N89-60-69 | Diverge |
| A38__0-9, B97__0-9, E45__0-9, G47__0-9, H04__0-9, J10__0-9, J35-10-19, J36-10-19, B07-20-29, B34-20-29, G51-20-29, H53-20-29, H72-20-29, H74-20-29, J15-20-29 | B97__0-9, J10__0-9, J20__0-9, J10-10-19, J98-10-19, L22-10-19 | Diverge |
| G83-50-59, H10-50-59, M54-60-69, M70-60-69, M76-60-69, J04-70-79, J31-70-79, M76-70-79 | H10-50-59, F13-60-69, F33-60-69, F41-60-69, F43-60-69, F45-60-69, F13-70-79, F33-70-79, F41-70-79, F43-70-79, F45-70-79, K59-70-79 | Diverge |
| N80-30-39, N84-30-39, N85-30-39, N88-30-39, N99-30-39, N99-40-49 | D21-30-39, D25-30-39, D27-30-39, K66-30-39, N70-30-39, N71-30-39, N73-30-39, N80-30-39, N83-30-39, N84-30-39, N85-30-39, N88-30-39, N92-30-39, N93-30-39, N94-30-39, N99-30-39, K66-40-49, N70-40-49, N73-40-49 | Diverge |
| F41-10-19, F45-10-19, F41-20-29 | F44-10-19, F45-10-19, F45-20-29, F45-30-39, F45-40-49 | Diverge |
| I21-40-49, I25-40-49, I42-40-49, I50-40-49, I20-50-59, I21-50-59, I24-50-59 | I42-40-49, I50-40-49, I42-50-59, I44-50-59, I44-60-69, I44-70-79 | Diverge |
| E66-40-49, I89-40-49, K42-40-49, K43-40-49, M16-40-49, E65-50-59, K42-50-59, K43-50-59 | M16-40-49, E66-50-59, I89-50-59, L30-50-59, M16-50-59, M87-50-59, E65-60-69, I89-60-69, K45-60-69, L27-60-69, N62-60-69, I89-70-79 | Diverge |
| M51-50-59, M96-50-59, A69-60-69, G54-60-69, G57-60-69, M96-60-69, G57-70-79 | M96-50-59, G55-60-69, M51-60-69, M96-60-69 | Diverge |
| B02-50-59, D13-50-59, E78-50-59, E83-50-59, G51-50-59, G58-50-59, I07-50-59, I15-50-59, I71-50-59, I72-50-59, I72-60-69 | D13-50-59, K80-50-59, K81-50-59, K82-50-59, K91-60-69 | Diverge |
| D17-50-59, D17-60-69, D17-70-79 | D17-50-59, N84-60-69, N88-60-69 | Diverge |
| C38-60-69, C78-60-69, D61-60-69, D70-60-69, J90-60-69, J91-60-69, C80-70-79, D70-70-79 | D61-60-69, D69-60-69, D70-60-69, D46-70-79, D61-70-79, D69-70-79 | Diverge |
| D61-60-69, D69-60-69, D70-60-69, D46-70-79, D61-70-79, D69-70-79 | C77-60-69, C80-60-69, D63-60-69, D70-60-69, C23-70-79, C77-70-79, C78-70-79, C79-70-79, C80-70-79, D48-70-79, D63-70-79, D70-70-79, M84-70-79 | Diverge |
| F31-20-29, F31-30-39, F31-40-49 | F31-20-29, F10-30-39, F31-30-39, K70-30-39, K71-30-39, K71-40-49 | Diverge |
| G54-40-49, G55-40-49, M41-40-49, M42-40-49, M43-40-49, M47-40-49, M48-40-49, M50-40-49, M53-40-49, M62-40-49, M99-40-49, E55-50-59, M41-50-59, M42-50-59, M43-50-59, M47-50-59, M48-50-59, M62-50-59, M99-50-59 | M47-40-49, D34-50-59, D44-50-59, E04-50-59, E21-50-59, E55-50-59, E83-50-59, J38-50-59, E07-60-69 | Diverge |
| A38__0-9, B97__0-9, E45__0-9, G47__0-9, H04__0-9, J10__0-9, J35-10-19, J36-10-19, B07-20-29, B34-20-29, G51-20-29, H53-20-29, H72-20-29, H74-20-29, J15-20-29 | E45__0-9, E66__0-9, G47__0-9, H65-10-19, H90-10-19, D27-20-29, D39-20-29, F81-20-29, F93-20-29, H65-20-29, H66-20-29, H90-20-29, I88-20-29, J00-20-29, N83-20-29 | Diverge |
| E66-40-49, I89-40-49, K42-40-49, K43-40-49, M16-40-49, E65-50-59, K42-50-59, K43-50-59 | M16-40-49, I89-50-59, M16-50-59, M87-50-59, I89-60-69, I89-70-79 | Diverge |
| E66-10-19, E78-10-19, F17-10-19, I10-10-19, K76-10-19, K80-10-19, K76-20-29 | I10-10-19, I10-20-29, N18-20-29 | Diverge |

| | | |
|---|---|---|
| N80-30-39, N84-30-39, N85-30-39, N88-30-39, N92-30-39, N93-30-39, N99-30-39, N99-40-49 | D21-30-39, D25-30-39, D27-30-39, K66-30-39, N70-30-39, N71-30-39, N73-30-39, N80-30-39, N83-30-39, N84-30-39, N85-30-39, N88-30-39, N92-30-39, N93-30-39, N94-30-39, N99-30-39, K66-40-49, N70-40-49, N73-40-49 | Diverge |
| F31-20-29, F10-30-39, F31-30-39, K70-30-39, K71-30-39, K71-40-49 | F31-20-29, F31-30-39, F31-40-49, F31-50-59, F31-60-69, F31-70-79 | Diverge |
| F17-30-39, J20-30-39, J44-30-39, J43-40-49, J44-40-49, J96-40-49 | J44-30-39, J45-30-39, J43-40-49, J44-40-49, J45-40-49, J96-40-49, J06-50-59 | Diverge |
| C38-60-69, C77-60-69, C78-60-69, C79-60-69, C80-60-69, D61-60-69, D63-60-69, D70-60-69, J90-60-69, J91-60-69, M84-60-69, C23-70-79, C77-70-79, C78-70-79, C79-70-79, C80-70-79, D48-70-79, D63-70-79, D70-70-79, M84-70-79 | D61-60-69, D69-60-69, D70-60-69, D46-70-79, D61-70-79, D69-70-79 | Diverge |
| D53-40-49, F10-40-49, G62-40-49, I85-40-49, K71-40-49, K74-40-49, D53-50-59, F12-50-59, G62-50-59, K71-50-59 | I85-40-49, K74-40-49, I85-50-59, K74-50-59 | Diverge |
| F07-40-49, G40-40-49, G93-40-49, G41-50-59, G93-50-59 | F07-40-49, G93-40-49, G40-50-59, G93-50-59, G41-60-69 | Diverge |
| E11-20-29, E66-20-29, G47-20-29, M51-20-29, E65-30-39, L02-30-39 | M51-20-29, G55-30-39, M43-30-39, M48-30-39, M51-30-39 | Diverge |
| I21-40-49, I25-40-49, I42-40-49, I50-40-49, I20-50-59, I21-50-59, I24-50-59 | E78-40-49, E79-40-49, G45-40-49, I20-40-49, I21-40-49, I44-40-49, I51-40-49, M41-40-49, E79-50-59, M41-50-59 | Diverge |
| I21-40-49, I25-40-49, I42-40-49, I50-40-49, I20-50-59, I21-50-59, I24-50-59 | I25-40-49, I42-40-49, I50-40-49, I42-50-59, I44-50-59, I50-50-59, I51-50-59, I42-60-69, I44-60-69, I44-70-79 | Diverge |
| N80-50-59, N81-50-59, N84-50-59, N85-50-59, N88-50-59, N95-50-59, N80-60-69 | N81-50-59, N81-60-69, N99-60-69 | Diverge |
| D12-40-49, I84-40-49, K64-50-59 | D12-40-49, D12-50-59, C18-60-69, C19-60-69, D12-60-69 | Diverge |
| B02-50-59, D13-50-59, E78-50-59, E83-50-59, G51-50-59, G58-50-59, I07-50-59, I15-50-59, I71-50-59, I72-50-59, I72-60-69 | I72-50-59, I67-60-69, I72-60-69, I67-70-79 | Diverge |
| D17-50-59, D48-50-59, D17-60-69, D48-60-69, D17-70-79 | D17-50-59, N84-60-69, N88-60-69 | Diverge |
| E11-20-29, E66-20-29, G47-20-29, M51-20-29, E65-30-39, L02-30-39 | M51-20-29, G55-30-39, M43-30-39, M48-30-39, M51-30-39, G54-40-49, G55-40-49, G57-40-49, M40-40-49, M51-40-49, M54-40-49, M96-40-49, G54-50-59, M96-50-59, M96-60-69 | Diverge |
| E66-40-49, I89-40-49, K42-40-49, K43-40-49, M16-40-49, E65-50-59, K42-50-59, K43-50-59 | M16-40-49, M16-50-59, M87-50-59 | Diverge |
| I20-40-49, I21-40-49, I25-40-49, I42-40-49, I50-40-49, I20-50-59, I21-50-59, I24-50-59, I25-50-59, I42-50-59, I44-50-59, I50-50-59, I51-50-59 | E78-40-49, E79-40-49, G45-40-49, I20-40-49, I21-40-49, I44-40-49, I51-40-49, M41-40-49, E79-50-59, M41-50-59 | Diverge |
| E11-20-29, E66-20-29, G47-20-29, M51-20-29, E65-30-39, L02-30-39 | M51-20-29, G55-30-39, M43-30-39, M48-30-39, M51-30-39, G55-40-49, G57-40-49, M51-40-49, M96-40-49 | Diverge |
| M51-50-59, M96-50-59, A69-60-69, G54-60-69, G57-60-69, M96-60-69, G57-70-79 | M96-50-59, G55-60-69, G95-60-69, M48-60-69, M50-60-69, M51-60-69, M96-60-69, G54-70-79, G55-70-79, G95-70-79, M47-70-79, M48-70-79, M50-70-79, M51-70-79, M54-70-79, M70-70-79, M93-70-79, M96-70-79 | Diverge |

Table S8: Outcome of diverging trajectories in males.

| Pair | Age | Number of the same patients | Number of patients of the first trajectory exclusively | Number of patients of the first trajectex clusively | Number of the patients from this age group, who are not following these two trajectories | Same diagnoses of trajectories, before they diverge | Exclusive diagnoses of the first trajectory | Exclusive diagnoses of the second trajectory | Average number of diagnosis of patients of the first trajectory | Average number of diagnosis of patients of the second trajectory | Average number of hospital days of patients of the first trajectory | Average number of hospital days of patients of the second trajectory | Average number of hospital stays of patients of the first trajectory | Average number of hospital stays of patients of the second trajectory | Ratio of average number (first vs. second trajectory) of number of hospital diagnoses | days spent in hospital | hospital stays | Mortality of the first trajectory | Mortality of the sencond ttrajectory | Ratio of mortality |
|---|---|---|---|---|---|---|---|---|---|---|---|---|---|---|---|---|---|---|---|---|
| 1 | 5 | 65 | 1496 | 339 | 330106 | N32-40-49 | N35-40-49 N32-50-59 N35-50-59 | N40-40-49 N41-40-49 N21-50-59 N40-50-59 N41-50-59 N42-50-59 N41-60-69 N42-60-69 | 4.806 | 4.932 | 29.478 | 28.091 | 6.084 | 6.758 | 0.974 | 1.049 | 0.9 | 0 | 0.0006400468518 | 0.00 |
| 1 | 6 | 119 | 2215 | 1442 | 325503 | N32-40-49 | N35-40-49 N32-50-59 N35-50-59 | N40-40-49 N41-40-49 N21-50-59 N40-50-59 N41-50-59 N42-50-59 N41-60-69 N42-60-69 | 6.469 | 5.741 | 41.978 | 35.267 | 8.864 | 8.877 | 1.127 | 1.19 | 0.999 | 0.00185890944 | 0.001249571169 | 1.49 |

51

| | | | | | | | | | | | | | | | | | | |
|---|---|---|---|---|---|---|---|---|---|---|---|---|---|---|---|---|---|---|
| 5 | 2 | 6 | 30 | 920 | 160614 | F20-10-19<br>F23-10-19 | E87-10-19<br>F10-10-19<br>F12-10-19<br>F17-10-19<br>F60-10-19<br>F60-20-29<br>F63-20-29 | F12-20-29<br>F15-20-29<br>F20-20-29<br>F23-20-29<br>F25-20-29<br>F31-20-29<br>F20-30-39<br>F23-30-39<br>F25-30-39<br>F20-40-49<br>F25-40-49 | 2.167 | 5.565 | 12.267 | 99.203 | 2.9 | 4.36 | 0.389 | 0.124 | 0.665 | 0.000886943020 5 | 0.00296735905 | 0.30 |
| 5 | 3 | 33 | 3490 | 268 | 224345 | F20-10-19<br>F23-10-19 | E87-10-19<br>F10-10-19<br>F12-10-19<br>F17-10-19<br>F60-10-19<br>F60-20-29<br>F63-20-29 | F12-20-29<br>F15-20-29<br>F20-20-29<br>F23-20-29<br>F25-20-29<br>F31-20-29<br>F20-30-39<br>F23-30-39<br>F25-30-39<br>F20-40-49<br>F25-40-49 | 5.242 | 4.287 | 69.627 | 95.642 | 6.085 | 4.049 | 1.223 | 0.728 | 1.503 | 0 | 0.001162097556 | 0.00 |

52

| | | | | | | | | | | | | | | | | | | |
|---|---|---|---|---|---|---|---|---|---|---|---|---|---|---|---|---|---|---|
| 7 | 5 | 33 | 279 | 371 | 331323 | N32-40-49<br>N40-40-49 | C61-40-49<br>C61-50-59<br>D40-50-59<br>N32-50-59 | N41-40-49<br>N21-50-59<br>N40-50-59<br>N41-50-59<br>N42-50-59<br>N41-60-69<br>N42-60-69 | 6.652 | 5.043 | 32.434 | 28.273 | 7.341 | 6.786 | 1.319 | 1.147 | 1.082 | 0.000370734628 2 | 0.00064000468518 | 0.58 |
| 7 | 6 | 28 | 942 | 1533 | 326775 | N32-40-49<br>N40-40-49 | C61-40-49<br>C61-50-59<br>D40-50-59<br>N32-50-59 | N41-40-49<br>N21-50-59<br>N40-50-59<br>N41-50-59<br>N42-50-59<br>N41-60-69<br>N42-60-69 | 6.224 | 5.879 | 33.037 | 35.95 | 7.055 | 9.01 | 1.059 | 0.919 | 0.783 | 0.004772711511 | 0.001249571169 | 3.82 |
| 8 | 1 | 11 | 192 | 610 | 293266 | N10__0-9 | A41__0-9<br>C91__0-9<br>A41-10-19<br>C91-10-19 | N13__0-9<br>N28__0-9<br>N28-10-19 | 13.99 | 3.831 | 50.161 | 16.319 | 6.104 | 5.356 | 3.652 | 3.074 | 1.14 | 0.000749576592 7 | 0 | Inf |
| 8 | 2 | 2 | 471 | 163 | 160934 | N10__0-9 | A41__0-9<br>C91__0-9<br>A41-10-19<br>C91-10-19 | N13__0-9<br>N28__0-9<br>N28-10-19 | 28.968 | 5.822 | 117.386 | 29.117 | 8.344 | 6.184 | 4.976 | 4.032 | 1.349 | 0.04987296748 | 0.006060606061 | 8.23 |

| | | | | | | | | | | | | | | | | | | |
|---|---|---|---|---|---|---|---|---|---|---|---|---|---|---|---|---|---|---|
| 10 | 4 | 1 | 213 | 226 | 268024 | H36-30-39 H36-40-49 | E10-30-39 E16-30-39 E10-40-49 E16-40-49 | H33-30-39 H35-30-39 H43-40-49 H33-40-49 H34-40-49 H35-40-49 H43-50-59 H33-50-59 H35-50-59 H36-50-59 H43-60-69 H26-60-69 H33-60-69 | 4.056 | 4.509 | 26.925 | 19.279 | 6.216 | 5.301 | 0.9 | 1.397 | 1.173 | 0 | 0 | NA |
| 10 | 5 | 0 | 361 | 116 | 331529 | H36-30-39 H36-40-49 | E10-30-39 E16-30-39 E10-40-49 E16-40-49 | H33-30-39 H35-30-39 H43-40-49 H33-40-49 H34-40-49 H35-40-49 H43-50-59 H33-50-59 H35-50-59 H36-50-59 H43-60-69 H26-60-69 H33-60-69 | 6.675 | 5.517 | 50.764 | 27.181 | 10.314 | 6.819 | 1.21 | 1.868 | 1.513 | 0.003736442512 | 0.00243793546 | 1.53 |

| | | | | | | | | | | | | | | | | | | | |
|---|---|---|---|---|---|---|---|---|---|---|---|---|---|---|---|---|---|---|---|
| 11 | 3 | 22 | 172 | 2372 | 225570 | G47-20-29 G47-30-39 | E11-20-29 E66-20-29 E78-20-29 E79-20-29 | G25-30-39 E66-40-49 G25-40-49 G47-40-49 G25-50-59 H66-50-59 K46-50-59 | 3.343 | 3.274 | 25.326 | 16.366 | 5.878 | 3.797 | 1.021 | 1.547 | 1.548 | 0.000415214532 6 | 0.0004177 10944 | 0.99 |
| 11 | 4 | 6778 | 115 | 316 | 261256 | G47-20-29 G47-30-39 | E11-20-29 E66-20-29 E78-20-29 E79-20-29 | G25-30-39 E66-40-49 G25-40-49 G47-40-49 G25-50-59 H66-50-59 K46-50-59 | 3.87 | 5.095 | 19.383 | 58.595 | 4.383 | 7.959 | 0.76 | 0.331 | 0.551 | 0.00029 0149427 | 0.0002901 49427 | 1.00 |
| 13 | 3 | 107 | 87 | 3422 | 224520 | E78-20-29 E79-20-29 | E11-20-29 E66-20-29 G47-20-29 G47-30-39 | E79-30-39 | 3.379 | 3.836 | 19.115 | 33.534 | 5.632 | 6.128 | 0.881 | 0.57 | 0.919 | 0.000415214532 6 | 0 | Inf |
| 13 | 4 | 107 | 6786 | 2362 | 259210 | E78-20-29 E79-20-29 | E11-20-29 E66-20-29 G47-20-29 G47-30-39 | E79-30-39 | 3.545 | 4.43 | 16.011 | 34.592 | 4.126 | 7.923 | 0.8 | 0.463 | 0.521 | 0.00029 0149427 | 0.0004050 222762 | 0.72 |

| | | | | | | | | | | | | | | | | | | |
|---|---|---|---|---|---|---|---|---|---|---|---|---|---|---|---|---|---|---|
| 17 | 5 | 695 | 436 | 545 | 330330 | L98-40-49 L97-50-59 L98-50-59 | L97-40-49 | M86-40-49 M86-50-59 L97-60-69 L98-60-69 M86-60-69 L97-70-79 L98-70-79 M86-70-79 | 7.312 | 5.512 | 65.617 | 47.108 | 11.161 | 8.327 | 1.327 | 1.393 | 1.34 | 0.005509527755 | 0.004526704995 | 1.22 |
| 17 | 6 | 234 | 1627 | 77 | 327341 | L98-40-49 L97-50-59 L98-50-59 | L97-40-49 | M86-40-49 M86-50-59 L97-60-69 L98-60-69 M86-60-69 L97-70-79 L98-70-79 M86-70-79 | 7.95 | 6.558 | 76.56 | 79.104 | 13.338 | 13.39 | 1.212 | 0.968 | 0.996 | 0.01529484971 | 0.008535476411 | 1.79 |
| 18 | 4 | 40 | 2377 | 214 | 265834 | M25-30-39 | M25-40-49 M66-40-49 M87-40-49 | M19-30-39 M24-30-39 M65-30-39 M67-30-39 M75-30-39 M19-40-49 M24-40-49 M65-40-49 M67-40-49 M75-40-49 M77-40-49 M66-50-59 M75-50-59 M76-50-59 | 3.125 | 2.439 | 17.008 | 10.785 | 5.21 | 3.808 | 1.281 | 1.577 | 1.368 | 0.000413736036 4 | 0 | Inf |

| | | | | | | | | | | | | | | | | | | |
|---|---|---|---|---|---|---|---|---|---|---|---|---|---|---|---|---|---|---|
| 18 | 5 | 1 | 249 | 413 | 331342 | M25-30-39 | M25-40-49<br>M66-40-49<br>M87-40-49 | M77-50-59<br>M19-30-39<br>M24-30-39<br>M65-30-39<br>M67-30-39<br>M75-30-39<br>M19-40-49<br>M24-40-49<br>M65-40-49<br>M67-40-49<br>M75-40-49<br>M77-40-49<br>M66-50-59<br>M75-50-59<br>M76-50-59<br>M77-50-59 | 4.072 | 4.019 | 25.614 | 18.918 | 6.329 | 5.525 | 1.013 | 1.354 | 1.146 | 0.000563380281 7 | 0.0003393 282894 | 1.66 |
| 19 | 2 | 6 | 30 | 920 | 160614 | F20-10-19<br>F23-10-19 | E87-10-19<br>F10-10-19<br>F12-10-19<br>F17-10-19<br>F60-10-19<br>F60-20-29<br>F63-20-29 | F20-20-29<br>F23-20-29<br>F25-20-29<br>F20-30-39<br>F23-30-39<br>F25-30-39 | 2.167 | 5.565 | 12.267 | 99.203 | 2.9 | 4.36 | 0.389 | 0.124 | 0.665 | 0.000886943020 5 | 0.0029673 5905 | 0.30 |

| | | | | | | | | | | | | | | | | | | |
|---|---|---|---|---|---|---|---|---|---|---|---|---|---|---|---|---|---|---|
| 19 | 3 | 79 | 3444 | 683 | 223930 | F20-10-19<br>F23-10-19 | E87-10-19<br>F10-10-19<br>F12-10-19<br>F17-10-19<br>F60-10-19<br>F60-20-29<br>F63-20-29 | F20-20-29<br>F23-20-29<br>F25-20-29<br>F20-30-39<br>F23-30-39<br>F25-30-39 | 5.192 | 4.682 | 68.315 | 84.318 | 6.073 | 3.958 | 1.109 | 0.81 | 1.534 | 0 | 0.0012880<br>53936 | 0.00 |
| 24 | 4 | 1 | 226 | 213 | 268024 | H36-30-39 | H33-30-39<br>H35-30-39<br>H43-40-49<br>H33-40-49<br>H34-40-49<br>H35-40-49<br>H36-40-49<br>H43-50-59<br>H33-50-59<br>H35-50-59<br>H36-50-59<br>H43-60-69<br>H26-60-69<br>H33-60-69 | E10-30-39<br>E16-30-39<br>E16-40-49 | 4.509 | 4.056 | 19.279 | 26.925 | 5.301 | 6.216 | 1.112 | 0.716 | 0.853 | 0 | 0 | NA |

58

| | | | | | | | | | | | | | | | | | | |
|---|---|---|---|---|---|---|---|---|---|---|---|---|---|---|---|---|---|---|
| 24 | 5 | 1 | 115 | 335 | 331555 | H36-30-39 | H33-30-39<br>H35-30-39<br>H43-40-49<br>H33-40-49<br>H34-40-49<br>H35-40-49<br>H36-40-49<br>H43-50-59<br>H33-50-59<br>H35-50-59<br>H36-50-59<br>H43-60-69<br>H26-60-69<br>H33-60-69 | E10-30-39<br>E16-30-39<br>E16-40-49 | 5.296 | 6.934 | 25.548 | 51.332 | 6.583 | 10.617 | 0.764 | 0.498 | 0.62 | 0.002437935468 | 0.02380952381 | 0.10 |
| 26 | 5 | 157 | 247 | 1402 | 330200 | N41-40-49<br>N41-50-59<br>N42-50-59<br>N41-60-69 | N32-40-49<br>N40-40-49<br>N21-50-59<br>N40-50-59<br>N42-60-69 | D40-50-59<br>D40-60-69<br>C61-70-79<br>C64-70-79<br>D40-70-79<br>N41-70-79<br>N42-70-79 | 6.004 | 4.248 | 35.887 | 24.705 | 7.668 | 6.108 | 1.413 | 1.453 | 1.255 | 0.000640046851 8 | 0.0006414368185 | 1.00 |
| 26 | 6 | 56 | 1505 | 420 | 327297 | N41-40-49<br>N41-50-59<br>N42-50-59<br>N41-60-69 | N32-40-49<br>N40-40-49<br>N21-50-59<br>N40-50-59 | D40-50-59<br>D40-60-69<br>C61-70-79<br>C64-70-79<br>D40-70-7 | 5.934 | 5.55 | 36.086 | 30.707 | 9.105 | 7.61 | 1.069 | 1.175 | 1.196 | 0.001249571169 | 0.001252051143 | 1.00 |

| | | | | | | | | | | | | | | | | | | | |
|---|---|---|---|---|---|---|---|---|---|---|---|---|---|---|---|---|---|---|---|
| | | | | | | N42-60-69 | 9<br>N41-70-79<br>N42-70-79 | | | | | | | | | | | | |
| 26 | 7 | 3253 | 1220 | 3601 | 260446 | N41-40-49<br>N41-50-59<br>N42-50-59<br>N41-60-69 | N32-40-49<br>N40-40-49<br>N21-50-59<br>N40-50-59<br>N42-60-69 | D40-50-59<br>D40-60-69<br>C61-70-79<br>C64-70-79<br>D40-70-79<br>N41-70-79<br>N42-70-79 | 6.988 | 7.16 | 41.662 | 38.714 | 10.234 | 9.041 | 0.976 | 1.076 | 1.132 | 0 | 0.001115843341 | 0.00 |
| 27 | 3 | 22 | 172 | 2372 | 225570 | G47-20-29<br>G47-30-39 | E11-20-29<br>E66-20-29<br>E78-20-29<br>E79-20-29 | G25-30-39<br>G25-40-49<br>G47-40-49 | 3.343 | 3.274 | 25.326 | 16.366 | 5.878 | 3.797 | 1.021 | 1.547 | 1.548 | 0.000415214532 6 | 0.000417710944 | 0.99 |
| 27 | 4 | 6778 | 115 | 316 | 261256 | G47-20-29<br>G47-30-39 | E11-20-29<br>E66-20-29<br>E78-20-29<br>E79-20-29 | G25-30-39<br>G25-40-49<br>G47-40-49 | 3.87 | 5.095 | 19.383 | 58.595 | 4.383 | 7.959 | 0.76 | 0.331 | 0.551 | 0.000290149427 | 0.000290149427 | 1.00 |
| 28 | 5 | 836 | 1319 | 5872 | 323979 | K60-40-49<br>K61-40-49<br>K60-50-59<br>K61-50-59 | D12-40-49<br>I84-40-49<br>K64-50-59 | I84-50-59<br>D21-60-69 | 4.087 | 3.929 | 18.363 | 21.105 | 5.647 | 5.123 | 1.04 | 0.87 | 1.102 | 0.001191236991 | 0.00119274659 | 1.00 |
| 28 | 6 | 51 | 620 | 1172 | 327436 | K60-40-49<br>K61-40-49<br>K60-50-59<br>K61-50-59 | D12-40-49<br>I84-40-49<br>K64-50-59 | I84-50-59<br>D21-60-69 | 5.582 | 4.781 | 34.89 | 26.066 | 7.997 | 7.713 | 1.168 | 1.339 | 1.037 | 0.002072950899 | 0.004254518971 | 0.49 |

| | | | | | | | | | | | | | | | | | | | |
|---|---|---|---|---|---|---|---|---|---|---|---|---|---|---|---|---|---|---|---|
| 31 | 2 | 6 | 30 | 920 | 160614 | F20-10-19 F23-10-19 | E87-10-19 F10-10-19 F12-10-19 F17-10-19 F60-10-19 F60-20-29 F63-20-29 | F20-20-29 F23-20-29 F25-20-29 F20-30-39 F23-30-39 F25-30-39 F20-40-49 F25-40-49 | 2.167 | 5.565 | 12.267 | 99.203 | 2.9 | 4.36 | 0.389 | 0.124 | 0.665 | 0.000886943020 5 | 0.00296735905 | 0.30 |
| 31 | 3 | 79 | 3444 | 683 | 223930 | F20-10-19 F23-10-19 | E87-10-19 F10-10-19 F12-10-19 F17-10-19 F60-10-19 F60-20-29 F63-20-29 | F20-20-29 F23-20-29 F25-20-29 F20-30-39 F23-30-39 F25-30-39 F20-40-49 F25-40-49 | 5.192 | 4.682 | 68.315 | 84.318 | 6.073 | 3.958 | 1.109 | 0.81 | 1.534 | 0 | 0.001288053936 | 0.00 |

Table S9: Outcome of diverging trajectories in females.

| | | | | | | | | | | | | | | | Ratio of average number (first vs. second trajectory) of number of | | |
|---|---|---|---|---|---|---|---|---|---|---|---|---|---|---|---|---|---|

| Pair | Age | Number of the same patients | Number of patients of the first trajectory | Number of patients of the first trajectory | Number of the patients from this age group, who are not following these two trajectories | Same diagnoses of trajectories, before they diverge | Exclusive diagnoses of the first trajectory | Exclusive diagnoses of the second trajectory | Average number of diagnosis of patients of the first trajectory | Average number of diagnosis of patients of the second trajectory | Average number of hospital days of patients of the first trajectory | Average number of hospital days of patients of the second trajectory | Average number of hospital stays of patients of the first trajectory | Average number of hospital stays of patients of the second trajectory | hospital diagnoses | days spent in hospital | hospital stays | Mortality of the first trajectory | Mortality of the sencond ttrajectory | Ratio of mortality |
|---|---|---|---|---|---|---|---|---|---|---|---|---|---|---|---|---|---|---|---|---|
| 1 | 5 | 35 | 122 | 1345 | 338227 | F48-40-49 F48-50-59 | G25-40-49 G47-40-49 G25-50-59 G47-50-59 | F06-40-49 F32-50-59 F40-50-59 F44-50-59 D16-60-69 F11-60-69 F22-60-69 F23-60-69 F44-60-69 F51-60-69 F55-60-69 J00-60-69 J37-60-69 L28-60-69 N89-60-69 | 5.82 | 6.779 | 39.885 | 63.533 | 7.926 | 9.541 | 0.859 | 0.628 | 0.831 | 0.000324739769 | 0.0020503280693 | 0.16 |
| 1 | 6 | 2 | 240 | 464 | 293330 | F48-40-49 F48-50-59 | G25-40-49 G47-40-49 G25-50-59 G47-50-59 | F06-40-49 F32-50-59 F40-50-59 F44-50-59 D16-60-69 F11-60-69 F22-60-69 F23-60-69 F44-60-69 F51-60-69 F55-60-69 J00-60-69 J37-60-69 L28-60-69 N89-60-69 | 6.567 | 6.504 | 40.654 | 56.933 | 9.583 | 10.478 | 1.01 | 0.714 | 0.915 | 0.001935872287 | 0.001375372001 | 1.41 |
| 5 | 2 | 2772 | 1366 | 810 | 174853 | F45-10-19 | F41-10-19 F41-20-29 | F44-10-19 F45-20-29 F45-30-39 F45-40-49 | 3.889 | 4.921 | 32.892 | 43.328 | 4.537 | 5.621 | 0.79 | 0.759 | 0.807 | 0 | 0 | NA |
| 5 | 3 | 281 | 3243 | 1876 | 256213 | F45-10-19 | F41-10-19 F41-20-29 | F44-10-19 F45-20-29 F45-30-39 F45-40-49 | 3.953 | 3.778 | 36.607 | 31.165 | 5.075 | 5.201 | 1.046 | 1.175 | 0.976 | 0 | 0 | NA |

| | | | | | | | | | | | | | | | | | | |
|---|---|---|---|---|---|---|---|---|---|---|---|---|---|---|---|---|---|---|
| 6 | 5 | 89 | 1083 | 1020 | 337537 | I42-40-49<br>I50-40-49 | I21-40-49<br>I25-40-49<br>I20-50-59<br>I21-50-59<br>I24-50-59 | I42-50-59<br>I44-50-59<br>I44-60-69<br>I44-70-79 | 5.149 | 6.715 | 33.148 | 50.691 | 8.803 | 11.01 | 0.767 | 0.654 | 0.8 | 0.09892979526 | 0.103745361 | 0.95 |
| 6 | 6 | 34 | 635 | 2008 | 291360 | I42-40-49<br>I50-40-49 | I21-40-49<br>I25-40-49<br>I20-50-59<br>I21-50-59<br>I24-50-59 | I42-50-59<br>I44-50-59<br>I44-60-69<br>I44-70-79 | 5.011 | 6.243 | 30.602 | 46.577 | 9 | 11.405 | 0.803 | 0.657 | 0.789 | 0.02460949692 | 0.005544134826 | 4.44 |
| 7 | 5 | 10 | 50 | 3283 | 336386 | M16-40-49 | E66-40-49<br>I89-40-49<br>K42-40-49<br>K43-40-49<br>E65-50-59<br>K42-50-59<br>K43-50-59 | E66-50-59<br>I89-50-59<br>L30-50-59<br>M16-50-59<br>M87-50-59<br>E65-60-69<br>I89-60-69<br>K45-60-69<br>L27-60-69<br>N62-60-69<br>I89-70-79 | 4.48 | 4.25 | 28.76 | 30.951 | 8.32 | 5.875 | 1.054 | 0.929 | 1.416 | 0 | 0 | NA |
| 7 | 6 | 0 | 250 | 70 | 293717 | M16-40-49 | E66-40-49<br>I89-40-49<br>K42-40-49<br>K43-40-49<br>E65-50-59<br>K42-50-59<br>K43-50-59 | E66-50-59<br>I89-50-59<br>L30-50-59<br>M16-50-59<br>M87-50-59<br>E65-60-69<br>I89-60-69<br>K45-60-69<br>L27-60-69<br>N62-60-69<br>I89-70-79 | 6.647 | 6.057 | 42.462 | 53.6 | 9.353 | 10.743 | 1.097 | 0.792 | 0.871 | 0.003900074178 | 0 | Inf |
| 8 | 6 | 380 | 12579 | 120 | 280958 | M96-50-59<br>M96-60-69 | M51-50-59<br>A69-60-69<br>G54-60-69<br>G57-60-69<br>G57-70-79 | G55-60-69<br>M51-60-69 | 5.748 | 7.333 | 37.359 | 54.067 | 8.693 | 10.458 | 0.784 | 0.691 | 0.831 | 0.001998589563 | 0.002 | 1.00 |
| 8 | 7 | 0 | 29 | 596 | 232659 | M96-50-59<br>M96-60-69 | M51-50-59<br>A69-60-69<br>G54-60-69<br>G57-60-69<br>G57-70-79 | G55-60-69<br>M51-60-69 | 6.586 | 7.653 | 40.483 | 53.196 | 10.586 | 12.201 | 0.861 | 0.761 | 0.868 | 0.002187571886 | 0.002204067934 | 0.99 |
| 11 | 7 | 12 | 241 | 367 | 232664 | D61-60-69<br>D70-60-69 | C38-60-69<br>C78-60-69<br>J90-60-69 | D69-60-69<br>D46-70-79 | 15.614 | 16.777 | 88.859 | 99.747 | 13.519 | 16.033 | 0.931 | 0.891 | 0.843 | 0.06222659619 | 0.03694051803 | 1.68 |

| | | | | | | | | | | | | | | | | | | |
|---|---|---|---|---|---|---|---|---|---|---|---|---|---|---|---|---|---|---|
| | | | | | | J91-60-69<br>C80-70-79<br>D70-70-79 | D61-70-79<br>D69-70-79 | | | | | | | | | | | |
| 11 | 8 | 29 | 2138 | 226 | 217231 | D61-60-69<br>D70-60-69 | C38-60-69<br>C78-60-69<br>J90-60-69<br>J91-60-69<br>C80-70-79<br>D70-70-79 | D69-60-69<br>D46-70-79<br>D61-70-79<br>D69-70-79 | 14.27 | 11.487 | 90.006 | 84.168 | 14.745 | 17.584 | 1.242 | 1.069 | 0.839 | 0.05858531023 | 0.1096844325 | 0.53 |
| 12 | 7 | 27 | 352 | 773 | 232132 | D70-60-69 | D61-60-69<br>D69-60-69<br>D46-70-79<br>D61-70-79<br>D69-70-79 | C77-60-69<br>C80-60-69<br>D63-60-69<br>C23-70-79<br>C77-70-79<br>C78-70-79<br>C79-70-79<br>C80-70-79<br>D48-70-79<br>D63-70-79<br>D70-70-79<br>M84-70-79 | 16.753 | 19.331 | 100.199 | 95.03 | 15.79 | 13.287 | 0.867 | 1.054 | 1.188 | 0.03694051803 | 0.02535817937 | 1.46 |
| 12 | 8 | 0 | 255 | 54 | 219314 | D70-60-69 | D61-60-69<br>D69-60-69<br>D46-70-79<br>D61-70-79<br>D69-70-79 | C77-60-69<br>C80-60-69<br>D63-60-69<br>C23-70-79<br>C77-70-79<br>C78-70-79<br>C79-70-79<br>C80-70-79<br>D48-70-79<br>D63-70-79<br>D70-70-79<br>M84-70-79 | 13.204 | 13.019 | 89.655 | 88.907 | 17.875 | 13.093 | 1.014 | 1.008 | 1.365 | 0.1096844325 | 0.005873087513 | 18.68 |
| 16 | 5 | 10 | 50 | 3283 | 336386 | M16-40-49 | E66-40-49<br>I89-40-49<br>K42-40-49<br>K43-40-49<br>E65-50-59<br>K42-50-59<br>K43-50-59 | I89-50-59<br>M16-50-59<br>M87-50-59<br>I89-60-69<br>I89-70-79 | 4.48 | 4.25 | 28.76 | 30.951 | 8.32 | 5.875 | 1.054 | 0.929 | 1.416 | 0 | 0 | NA |
| 16 | 6 | 1 | 249 | 322 | 293465 | M16-40-49 | E66-40-49<br>I89-40-49<br>K42-40-49<br>K43-40-49<br>E65-50-59<br>K42-50-59<br>K43-50-59 | I89-50-59<br>M16-50-59<br>M87-50-59<br>I89-60-69<br>I89-70-79 | 6.593 | 6.034 | 42.46 | 43.913 | 9.347 | 8.376 | 1.093 | 0.967 | 1.116 | 0.003900074178 | 0 | Inf |

| | | | | | | | | | | | | | | | | | | | |
|---|---|---|---|---|---|---|---|---|---|---|---|---|---|---|---|---|---|---|---|
| 17 | 2 | 19 | 106 | 683 | 178993 | I10-10-19 | E66-10-19<br>E78-10-19<br>F17-10-19<br>K76-10-19<br>K80-10-19<br>K76-20-29 | I10-20-29<br>N18-20-29 | 3.104 | 4.812 | 20.453 | 27.959 | 4.745 | 6.161 | 0.645 | 0.732 | 0.77 | 0 | 0.002849002849 | 0.00 |
| 17 | 3 | 92 | 1027 | 2289 | 258205 | I10-10-19 | E66-10-19<br>E78-10-19<br>F17-10-19<br>K76-10-19<br>K80-10-19<br>K76-20-29 | I10-20-29<br>N18-20-29 | 4.261 | 4.477 | 30.73 | 29.216 | 6.904 | 6.3 | 0.952 | 1.052 | 1.096 | 0.003574620197 | 0.01712988314 | 0.21 |
| 20 | 4 | 60 | 294 | 3442 | 290156 | J44-30-39<br>J43-40-49<br>J44-40-49<br>J96-40-49 | F17-30-39<br>J20-30-39 | J45-30-39<br>J45-40-49<br>J06-50-59 | 5.214 | 4.575 | 39.316 | 29.991 | 7.384 | 6.69 | 1.14 | 1.311 | 1.104 | 0.003339778235 | 0.001725505169 | 1.94 |
| 20 | 5 | 28 | 259 | 710 | 338731 | J44-30-39<br>J43-40-49<br>J44-40-49<br>J96-40-49 | F17-30-39<br>J20-30-39 | J45-30-39<br>J45-40-49<br>J06-50-59 | 6.965 | 6.161 | 51.815 | 40.262 | 10.166 | 9.325 | 1.13 | 1.287 | 1.09 | 0.1687894148 | 0.168070014 | 1.00 |
| 21 | 7 | 0 | 45 | 379 | 232860 | D61-60-69<br>D70-60-69 | C38-60-69<br>C77-60-69<br>C78-60-69<br>C79-60-69<br>C80-60-69<br>D63-60-69<br>J90-60-69<br>J91-60-69<br>M84-60-69<br>C23-70-79<br>C77-70-79<br>C78-70-79<br>C79-70-79<br>C80-70-79<br>D48-70-79<br>D63-70-79<br>D70-70-79<br>M84-70-79 | D69-60-69<br>D46-70-79<br>D61-70-79<br>D69-70-79 | 19.644 | 16.897 | 82.711 | 100.145 | 12.978 | 16.032 | 1.163 | 0.826 | 0.81 | 0.005864607725 | 0.03694051803 | 0.16 |

65

| | | | | | | | | | | | | | | | | | | |
|---|---|---|---|---|---|---|---|---|---|---|---|---|---|---|---|---|---|---|
| 21 | 8 | 0 | 54 | 255 | 219314 | D61-60-69<br>D70-60-69 | C38-60-69<br>C77-60-69<br>C78-60-69<br>C79-60-69<br>C80-60-69<br>D63-60-69<br>J90-60-69<br>J91-60-69<br>M84-60-69<br>C23-70-79<br>C77-70-79<br>C78-70-79<br>C79-70-79<br>C80-70-79<br>D48-70-79<br>D63-70-79<br>D70-70-79<br>M84-70-79 | D69-60-69<br>D46-70-79<br>D61-70-79<br>D69-70-79 | 13.019 | 13.204 | 88.907 | 89.655 | 13.093 | 17.875 | 0.986 | 0.992 | 0.732 | 0.005873087513 | 0.1096844325 | 0.05 |
| | | | | | | | | | | | | | | | | | | |
| 22 | 5 | 21 | 178 | 661 | 338869 | I85-40-49<br>K74-40-49 | D53-40-49<br>F10-40-49<br>G62-40-49<br>K71-40-49<br>D53-50-59<br>F12-50-59<br>G62-50-59<br>K71-50-59 | I85-50-59<br>K74-50-59 | 6.461 | 7.261 | 73.77 | 62.861 | 8.5 | 11.992 | 0.89 | 1.174 | 0.709 | 0.08494135819 | 0.08966753033 | 0.95 |
| 22 | 6 | 6 | 64 | 1194 | 292773 | I85-40-49<br>K74-40-49 | D53-40-49<br>F10-40-49<br>G62-40-49<br>K71-40-49<br>D53-50-59<br>F12-50-59<br>G62-50-59<br>K71-50-59 | I85-50-59<br>K74-50-59 | 10.594 | 7.163 | 81.469 | 62.547 | 14.484 | 12.906 | 1.479 | 1.303 | 1.122 | 0.01490278797 | 0.1246592251 | 0.12 |
| | | | | | | | | | | | | | | | | | | |
| 23 | 5 | 49 | 164 | 1090 | 338426 | F07-40-49<br>G93-40-49<br>G93-50-59 | G40-40-49<br>G41-50-59 | G40-50-59<br>G41-60-69 | 6.726 | 7.009 | 47.86 | 80.524 | 8.689 | 9.997 | 0.96 | 0.594 | 0.869 | 0.2064390873 | 0.2106908247 | 0.98 |
| 23 | 6 | 915 | 142 | 2937 | 290043 | F07-40-49<br>G93-40-49<br>G93-50-59 | G40-40-49<br>G41-50-59 | G40-50-59<br>G41-60-69 | 6.789 | 6.748 | 62.951 | 57.157 | 10.43 | 9.668 | 1.006 | 1.101 | 1.079 | 0.2137521155 | 0.2243905331 | 0.95 |
| | | | | | | | | | | | | | | | | | | |
| 24 | 3 | 127 | 315 | 2837 | 258334 | M51-20-29 | E11-20-29<br>E66-20-29<br>G47-20-29 | G55-30-39<br>M43-30-39 | 3.467 | 3.18 | 22.508 | 19.25 | 4.921 | 4.105 | 1.09 | 1.169 | 1.199 | 0 | 0 | NA |

| | | | | | | | | | | | | | | | | | | | |
|---|---|---|---|---|---|---|---|---|---|---|---|---|---|---|---|---|---|---|---|
| 24 | 4 | 7 | 3904 | 738 | 289303 | M51-20-29 | E11-20-29<br>E66-20-29<br>G47-20-29<br>E65-30-39<br>L02-30-39 | E65-30-39<br>L02-30-39<br>G55-30-39<br>M43-30-39<br>M48-30-39<br>M51-30-39 | 3.83 | 3.449 | 20.633 | 21.692 | 4.649 | 4.585 | 1.11 | 0.951 | 1.014 | 0.0002200 | 0.0006184518 | 0.00 |
| 26 | 5 | 100 | 1072 | 306 | 338251 | I25-40-49<br>I42-40-49<br>I50-40-49 | I21-40-49<br>I20-50-59<br>I21-50-59<br>I24-50-59 | I42-50-59<br>I44-50-59<br>I50-50-59<br>I51-50-59<br>I42-60-69<br>I44-60-69<br>I44-70-79 | 5.349 | 6.634 | 35.91 | 39.02 | 9.243 | 9.703 | 0.806 | 0.92 | 0.953 | 0.09892979526 | 0.1027570063 | 0.96 |
| 26 | 6 | 15 | 654 | 764 | 292604 | I25-40-49<br>I42-40-49<br>I50-40-49 | I21-40-49<br>I20-50-59<br>I21-50-59<br>I24-50-59 | I42-50-59<br>I44-50-59<br>I50-50-59<br>I51-50-59<br>I42-60-69<br>I44-60-69<br>I44-70-79 | 5.057 | 7.379 | 31.115 | 59.641 | 9.089 | 12.743 | 0.685 | 0.522 | 0.713 | 0.02460949692 | 0.02129397016 | 1.16 |
| 27 | 6 | 139 | 451 | 6013 | 287434 | N81-50-59 | N80-50-59<br>N84-50-59<br>N85-50-59<br>N88-50-59<br>N95-50-59<br>N80-60-69 | N81-60-69<br>N99-60-69 | 4.508 | 4.284 | 20.665 | 23.969 | 6.215 | 6.133 | 1.052 | 0.862 | 1.013 | 0.000325311724 3 | 0.0001625487646 | 2.00 |
| 27 | 7 | 94 | 183 | 6538 | 226470 | N81-50-59 | N80-50-59<br>N84-50-59<br>N85-50-59<br>N88-50-59<br>N95-50-59<br>N80-60-69 | N81-60-69<br>N99-60-69 | 6.749 | 5.364 | 42.59 | 33.923 | 10.847 | 7.876 | 1.258 | 1.255 | 1.377 | 0 | 0.02786666857 | 0.00 |
| 28 | 5 | 1571 | 6213 | 0 | 331945 | D12-40-49 | I84-40-49<br>K64-50-59 | D12-50-59<br>C18-60-69<br>C19-60-69<br>D12-60-69 | NA | NA | NA | NA | NA | NA | NA | NA | NA | 0.000298183936 4 | 0.0006365372374 | 0.47 |
| 28 | 6 | 38 | 629 | 2750 | 290620 | D12-40-49 | I84-40-49<br>K64-50-59 | D12-50-59<br>C18-60-69<br>C19-60-69<br>D12-60-69 | 4.308 | 5.666 | 20.44 | 28.875 | 7.323 | 8.235 | 0.76 | 0.708 | 0.889 | 0.001499250375 | 0.00143472023 | 1.04 |

| | | | | | | | | | | | | | | | | | | |
|---|---|---|---|---|---|---|---|---|---|---|---|---|---|---|---|---|---|---|
| 30 | 6 | 3194 | 4600 | 0 | 286243 | D17-50-59 | D48-50-59 D17-60-69 D48-60-69 D17-70-79 | N84-60-69 N88-60-69 | NA | NA | NA | NA | NA | NA | NA | NA | NA | 0.00613 9917014 | 0.000626 1740764 | 9.81 |
| 30 | 7 | 182 | 5446 | 7633 | 220024 | D17-50-59 | D48-50-59 D17-60-69 D48-60-69 D17-70-79 | N84-60-69 N88-60-69 | 8.099 | 5.242 | 45.367 | 28.376 | 9.778 | 7.525 | 1.545 | 1.599 | 1.299 | 0.01113 610784 | 0 | Inf |
| 31 | 3 | 127 | 315 | 2837 | 258334 | M51-20-29 | E11-20-29 E66-20-29 G47-20-29 E65-30-39 L02-30-39 | G55-30-39 M43-30-39 M48-30-39 M51-30-39 G54-40-49 G55-40-49 G57-40-49 M40-40-49 M51-40-49 M54-40-49 M96-40-49 G54-50-59 M96-50-59 M96-60-69 | 3.467 | 3.18 | 22.508 | 19.25 | 4.921 | 4.105 | 1.09 | 1.169 | 1.199 | 0 | 0 | NA |
| 31 | 4 | 7 | 3904 | 738 | 289303 | M51-20-29 | E11-20-29 E66-20-29 G47-20-29 E65-30-39 L02-30-39 | G55-30-39 M43-30-39 M48-30-39 M51-30-39 G54-40-49 G55-40-49 G57-40-49 M40-40-49 M51-40-49 M54-40-49 M96-40-49 G54-50-59 M96-50-59 M96-60-69 | 3.83 | 3.449 | 20.633 | 21.692 | 4.649 | 4.585 | 1.11 | 0.951 | 1.014 | 0 | 0.000220 6184518 | 0.00 |
| 32 | 5 | 10 | 50 | 3283 | 336386 | M16-40-49 | E66-40-49 I89-40-49 K42-40-49 K43-40-49 E65-50-59 K42-50-59 K43-50-59 | M16-50-59 M87-50-59 | 4.48 | 4.25 | 28.76 | 30.951 | 8.32 | 5.875 | 1.054 | 0.929 | 1.416 | 0 | 0 | NA |
| 32 | 6 | 10 | 240 | 8275 | 285512 | M16-40-49 | E66-40-49 I89-40-49 K42-40-49 K43-40-49 | M16-50-59 M87-50-59 | 6.037 | 5.042 | 38.625 | 34.802 | 8.817 | 7.195 | 1.197 | 1.11 | 1.225 | 0.00390 0074178 | 0 | Inf |

68

| | | | | | | | | | | | | | | | | | | | |
|---|---|---|---|---|---|---|---|---|---|---|---|---|---|---|---|---|---|---|---|
| | | | | | | E65-50-59<br>K42-50-59<br>K43-50-59 | | | | | | | | | | | | | |
| | | | | | | | | | | | | | | | | | | | |
| 34 | 3 | 127 | 315 | 2837 | 258334 | M51-20-29 | E11-20-29<br>E66-20-29<br>G47-20-29<br>E65-30-39<br>L02-30-39 | G55-30-39<br>M43-30-39<br>M48-30-39<br>M51-30-39<br>G55-40-49<br>G57-40-49<br>M51-40-49<br>M96-40-49 | 3.467 | 3.18 | 22.508 | 19.25 | 4.921 | 4.105 | 1.09 | 1.169 | 1.199 | 0 | 0 | NA |
| 34 | 4 | 7 | 3904 | 738 | 289303 | M51-20-29 | E11-20-29<br>E66-20-29<br>G47-20-29<br>E65-30-39<br>L02-30-39 | G55-30-39<br>M43-30-39<br>M48-30-39<br>M51-30-39<br>G55-40-49<br>G57-40-49<br>M51-40-49<br>M96-40-49 | 3.83 | 3.449 | 20.633 | 21.692 | 4.649 | 4.585 | 1.11 | 0.951 | 1.014 | 0 | 0.0002206184518 | 0.00 |
| 35 | 6 | 380 | 12579 | 120 | 280958 | M96-50-59<br>M96-60-69 | M51-50-59<br>A69-60-69<br>G54-60-69<br>G57-60-69<br>G57-70-79 | G55-60-69<br>G95-60-69<br>M48-60-69<br>M50-60-69<br>M51-60-69<br>G54-70-79<br>G55-70-79<br>G95-70-79<br>M47-70-79<br>M48-70-79<br>M50-70-79<br>M51-70-79<br>M54-70-79<br>M70-70-79<br>M93-70-79<br>M96-70-79 | 5.748 | 7.333 | 37.359 | 54.067 | 8.693 | 10.458 | 0.784 | 0.691 | 0.831 | 0.001998589563 | 0.002 | 1.00 |
| 35 | 7 | 0 | 29 | 352 | 232903 | M96-50-59<br>M96-60-69 | M51-50-59<br>A69-60-69<br>G54-60-69<br>G57-60-69<br>G57-70-79 | G55-60-69<br>G95-60-69<br>M48-60-69<br>M50-60-69<br>M51-60-69<br>G54-70-79<br>G55-70-79<br>G95-70-79<br>M47-70-79<br>M48-70-79<br>M50-70-79<br>M51-70-79<br>M54-70-79 | 6.586 | 8.065 | 40.483 | 57.426 | 10.586 | 12.716 | 0.817 | 0.705 | 0.832 | 0.002187571886 | 0.002127753604 | 1.03 |

| | | | | | | | M70-70-79 M93-70-79 M96-70-79 | | | | | | | | | | | |
|---|---|---|---|---|---|---|---|---|---|---|---|---|---|---|---|---|---|---|
| 35 | 8 | 1 | 584 | 41 | 218998 | M96-50-59 M96-60-69 | G55-60-69 G95-60-69 M48-60-69 M50-60-69 M51-60-69 G54-70-79 G55-70-79 G95-70-79 M47-70-79 M48-70-79 M50-70-79 M51-50-59 M51-70-79 A69-60-69 M54-70-79 G54-60-69 M70-70-79 G57-60-69 M93-70-79 G57-70-79 M96-70-79 | 8.691 | 8.439 | 82.165 | 67.195 | 16.407 | 15.902 | 1.03 | 1.223 | 1.0 32 | 0.01025 641026 | 0.006390 357284 | 1.60 |



\end{document}